 \documentclass[preprint,12pt]{elsarticle}
\usepackage{amssymb,amsmath,latexsym}
\usepackage[dvips]{color}
\usepackage[headings]{fullpage}
\usepackage{epic,epsfig,graphicx}
\usepackage{enumerate}
\usepackage{multirow}
\usepackage{setspace}
\usepackage{keyval}
\usepackage{float}

\usepackage{amssymb}

 \newtheorem{thm}{Theorem}[section]

 \newtheorem{lemma}{Lemma}[section]

 \newtheorem{defn}{Definition}[section]

 \newtheorem{rem}{Remark}[section]
  \newtheorem{assumption}{Assumption}[section]

 \numberwithin{equation}{section}

\journal{arXiv}

\begin{document}
\begin{frontmatter}
\title{The stochastic permanence of malaria, and the existence of a stationary distribution for a class of malaria models}
\author{Divine Wanduku }
\address{Department of Mathematical Sciences,
Georgia Southern University, 65 Georgia Ave, Room 3042, Statesboro,
Georgia, 30460, U.S.A. E-mail:dwanduku@georgiasouthern.edu;wandukudivine@yahoo.com\footnote{Corresponding author. Tel: +14073009605.
} }
\begin{abstract}
This paper investigates the stochastic permanence of malaria and the existence of a stationary distribution for the stochastic process describing the disease dynamics over sufficiently longtime. The malaria system is highly  random with fluctuations from the disease transmission and natural deathrates, which are expressed as independent white noise processes in a family of stochastic differential equation epidemic models. Other sources of variability in the malaria dynamics are the random incubation and naturally acquired immunity periods of malaria. Improved analytical techniques and local martingale characterizations are applied to describe the character of the sample paths of the solution process of the system in the neighborhood of an endemic equilibrium. Emphasis of this study is laid on examination of the impacts of (1) the sources of variability- disease transmission and natural death rates, and (2) the intensities of the white noise processes in the system on the stochastic permanence of malaria, and also on the existence of the stationary distribution for the solution process over sufficiently long time. Numerical simulation examples are presented to illuminate the persistence and stochastic permanence of malaria, and also to numerically approximate the stationary distribution of the states of the solution process.
\end{abstract}

\end{frontmatter}
\section{Introduction\label{ch1.sec0}}
According to the WHO estimates released in December $2016$,  about 212 million cases of malaria occurred in $2015$ resulting in about 429 thousand deaths. In addition, the highest mortality rates were recorded for the sub-Saharan African countries,  where about $90\%$ of the global malaria cases occurred, and resulted to about $75\%$ of the global malaria related deaths.  Moreover, more than two third of these global malaria related deaths were children younger than or exactly five years old. Furthermore, in spite of the fact that malaria is a curable and preventable disease, and despite all technological advances to control and contain the disease, malaria imposes serious menace to human health and the welfare of many economies in the world.

In fact, WHO has reported in $2015$ that nearly half of the world's population was at risk to malaria, and the disease was actively and continuously transmitted in about $91$ countries in the world. Moreover, the most severely affected economies are the sub-Saharan countries, and the most vulnerable sub-human populations include children younger than five years old, pregnant women, people suffering from HIV/AIDS, and travellers from regions with low malaria transmission to malaria endemic zones\cite{WHO,CDC}. These facts serve as motivation to foster research about malaria and understand all aspects of the disease that lead to its containment, and amelioration of the burdens of the disease.

Mathematical modeling is one special way of understanding malaria, and malaria models go as far back as 1911 with Ross\cite{ross} who studied mosquito control. Several other authors such as \cite{wanduku-biomath,macdonald,ngwa-shu,hyun,may,kazeem,gungala,anita,tabo} have also made strides in the understanding of malaria mathematically.

 Malaria is a vector-borne disease  caused by protozoa (a micro-parasitic organism) of the genus \textit{Plasmodium}. There are several different species of the parasite that cause disease in humans namely: \textit{P. falciparum, P. viviax, P. ovale} and \textit{P. malariae}. However, the species that causes the most severe and fatal disease is the \textit{P. falciparum}.  Malaria is transmitted between humans by the infectious bite of a female mosquito of the genus \textit{Anopheles}. The complete life cycle of the  malaria  plasmodium entails two-hosts: (1) the female anopheles mosquito vector, and (2) the susceptible or infectious human being\cite{malaria,WHO,CDC}.

 The stages of maturation of the plasmodium within the human host is called the \textit{exo-erythrocytic cycle}. Moreover, the total duration of the \textit{exo-erythrocytic cycle} is estimated  between 7-30 days depending on the species of plasmodium, with the exceptions of the plasmodia- \textit{P. vivax} and \textit{P. ovale} that may be delayed for as long as 1 to 2 years. See for example \cite{malaria,WHO,CDC}. Also, the stages of development of the plasmodium within the mosquito host is called the  \textit{sporogonic cycle}. It is estimated that the duration of  the \textit{sporogonic cycle} is over 2 to 3 weeks\cite{malaria,WHO,CDC}.  The delay between infection of the mosquito and maturation of the parasite inside the mosquito suggests that the mosquito must survive a minimum of the 2 to 3 weeks to be able to transmit malaria\cite{malaria}. These facts are important in deriving a mathematical model to represent the dynamics of malaria.
   More details about the mosquito biting habit, the life cycle of malaria and the key issues related to the mathematical model for malaria in this study  are located in Wanduku\cite{wanduku-biomath}, and also in  \cite{malaria,WHO,CDC}.

    It is also important to note that malaria confers natural immunity\cite{CDC,lars,denise} after recovery from the disease. The strength and effectiveness of the natural immunity against the disease depends primarily on the frequency of  exposure to the parasites and other biological factors such as age, pregnancy, and genetic structure of red blood cells of people with malaria. The natural immunity against malaria has been studied mathematically by several different authors, for example, \cite{wanduku-biomath,hyun}. The duration of the naturally acquired immunity period is random with a range of possible values from zero for individuals with almost no history of the disease (for instance, young babies etc.), to sufficiently long time for people with genetic resistance against the disease( for instance, people with sickle cell anemia, and duffy negative blood type conditions etc.). All of these facts related to the naturally acquired immunity against malaria, and development of the acquired immunity into a mathematical expression are discussed in  Wanduku\cite{wanduku-biomath}, and  \cite{CDC,lars,denise,hyun}.

As with every other infectious disease dynamics, there is inevitable presence of noise in the dynamics of malaria. Mathematically, the noise in an infectious system over continuous time can be expressed in one way as a Wierner or Brownian motion process obtained as an approximation of a random walk process over an infinitesimally small time interval. Moreover, the central limit theorem is applied to obtain  this approximation.

There are several different ways to introduce white noise into the infectious system, for example, by considering the variation of the driving parameters of the infectious system, or considering the random perturbation of the density of the system etc. Regardless of the method of introducing the noise into the system, the mathematical systems obtained from the approximation process above are called  stochastic differential equation systems.

 Stochastic systems offer a better representation of reality, and a better fit for most dynamic processes that occur in real life. This is because of the inevitable occurrence of random fluctuations in the dynamic real life systems. Whilst several deterministic systems for malaria dynamics have been studied \cite{wanduku-biomath,macdonald,ngwa-shu,hyun,may,kazeem,gungala,anita,tabo}, to the best of the author's knowledge, little or no mathematical studies authored by other experts exist about malaria in the framework of Ito-Doob type stochastic differential equations. The study by Wanduku\cite{wanduku-comparative} would appear to lead as the first attempt to understand the impacts of white noise on various aspects of malaria dynamics. And the mode of adding white noise into the malaria dynamics in this study is similar to the earlier studies \cite{Wanduku-2017,wanduku-fundamental,wanduku-delay}.

An important investigation in the study of infectious population dynamic systems influenced by white noise is the permanence of the disease, and the existence of a stationary distribution for the infectious system. Several papers in the literature\cite{aadil,yanli,yongli,yongli-2,yzhang,mao-2}  have addressed these topics. Investigations about the permanence of the disease in the population seek to find conditions that negatively favor the survival of the endemic population classes (such as the exposed, infectious and removal classes) in the far future time. Moreover, in a white noise influenced infectious system, the permanence of the disease requires the existence of a nonzero average population size for the infectious classes over sufficiently long time.

The existence of a stationary distribution for a stochastic infectious system implies that in the far future the statistical properties of the different states of the system can be determined accurately by knowing the distribution of a single random variable, which is the limit of convergence in distribution of the random process describing the dynamics of the disease. Since most realistic stochastic models formulated in terms of stochastic differential equations are nonlinear, and explicit solutions are nontrivial, numerical methods can be used to approximate the stationary distribution for the random process. See for example \cite{aadil,yongli-2,mao-2}

Along with the stationary behavior of the stochastic infectious system over sufficiently long time, another topic of  investigation concerns the ergodic character of the sample paths of the disease system. The ergodicity of the stochastic disease system ensures that the statistical properties of the disease in the system in the far future time can be understood, and estimated by the sample realizations of disease over sufficiently long time. That is, while insights about the ensemble nature of the disease are difficult to obtain directly from the explicit solutions of the stochastic system because of the nonlinear structure of the system, the stationary and ergodic properties of the stochastic system ensure that sufficient information about the disease is obtained from the sample paths of the disease over sufficiently long time. See for example \cite{aadil,yongli-2,mao-2}

Several different studies suggest that the strength or the intensity of the white noise in the infectious system plays a major role on the permanence of the disease, and also on the existence of a stationary distribution for the stochastic system\cite{aadil,yanli,yongli,yongli-2, yzhang}. In most of these studies, one can deduce that low intensity of the white noise  is associated with the permanence of the disease in the far future time, and consequently lead to the existence of a stationary distribution for the stochastic infectious system.

The primary objective of this study is to characterize the role of the intensities of the white noises from different sources in a malaria dynamics on the overall behavior of the disease, and in particular on the permanence of the disease. Furthermore, another objective is to also understand the existence of an endemic stationary distribution, which numerically can be approximated for a given set of parameters corresponding to a malaria scenario.

Recently, Wanduku\cite{wanduku-biomath} presented a class of deterministic models for malaria, where the class type is determined by a generalized nonlinear incidence rate of the disease. The class of epidemic dynamic models incorporates the three delays in the dynamics of malaria from the incubation of the disease inside the mosquito (\textit{sporogonic cycle}), the incubation of the plasmodium inside the human being (\textit{exo-erythrocytic cycle}), and also the period of effective naturally acquired immunity against malaria. Moreover, the delay periods are all random and arbitrarily distributed.

Some special cases of the generalized nonlinear incidence rate include (1) a malaria scenario where the response rate of the disease transmission from infectives to susceptible individuals increases initially for small number of infectious individuals, and then saturates with a horizontal asymptote for large and larger number of infectious individuals, (2) a malaria scenario where the response rate of the disease transmission from infectives to susceptible individuals initially decreases, and saturates at a lower horizontal asymptote as the infected population increases, and (3) a malaria scenario where the response rate of the disease transmission from infectives to  susceptible individuals initially increases, attains a maximum level and decreases as the number of infected individuals increases etc.

Some extensions of Wanduku\cite{wanduku-biomath} will appear in the context of a general class of  vector-borne disease models such as dengue fever and malaria in  Wanduku\cite{wanduku-theorBio},  where the role of the different sources of variability on vector-borne diseases are investigated, and their intensities are classified. The focus of \cite{wanduku-theorBio} is on disease eradication in the steady state population. In the extension Wanduku\cite{wanduku-comparative}, a general class of malaria stochastic models is investigated, and the emphasis is to examine the extend to which the different sources of noise in the system deviate the stochastic dynamics of malaria from its ideal dynamics in the absence of noise. Note that Wanduku\cite{wanduku-comparative} is a comparative study.
%

The current study extends Wanduku\cite{wanduku-biomath} by introducing the sources of variability in Wanduku\cite{wanduku-comparative} with a more detailed formulation of the white noise processes in the malaria dynamics,
 and provides detailed analytical techniques, biological interpretation and numerical simulation results to comprehend (1) the behavior of the stochastic system in the neighborhood of a potential endemic equilibrium, (2) the stochastic permanence of the disease, and the fundamental role of the intensities of the noises in the system in determining the persistence of the disease, and (3) the existence of a stationary endemic distribution to fully characterize the statistical properties of the states in the system in the long-term.

   This work is presented as follows:- in Section~\ref{ch1.sec0}, the fundamentals in the derivation of the class of deterministic models for malaria in Wanduku \cite{wanduku-biomath} are discussed, and essential information to this study is presented. In Section~\ref{ch1.sec0.sec1}, the new class of stochastic models is extensively formulated. In Section~\ref{ch1.sec1}, the model validation results are presented for the stochastic system. 
   In Section~\ref{ch1.sec3}, the persistence of the disease in the human population is exhibited. The permanence of malaria in the mean in the human population is also exhibited in Section~\ref{ch1.sec5}. Furthermore, the existence of a stationary distribution  for the class of stochastic models is presented in Section~\ref{ch1.sec3.sec1}. Moreover, the ergodicity of the stochastic system is also exhibited in this section. Finally, numerical results are given to test the permanence of malaria, and approximate the stationary distribution of malaria in Section~\ref{ch1.sec4}.
\section{The derivation of the model and some preliminary results}\label{ch1.sec0}
In the recent study by Wanduku \cite{wanduku-biomath}, a class of SEIRS epidemic dynamic models  for malaria with three random delays is presented.
 The delays represent the incubation periods of the infectious agent (plasmodium) inside the vector(mosquito) denoted $T_{1}$, and inside the human host denoted $T_{2}$. The third delay represents the naturally acquired immunity period of the disease $T_{3}$, where the delays are random variables with density functions $f_{T_{1}}, t_{0}\leq T_{1}\leq h_{1}, h_{1}>0$, and $f_{T_{2}}, t_{0}\leq T_{2}\leq h_{2}, h_{2}>0$ and $f_{T_{3}}, t_{0}\leq T_{3}<\infty$. Furthermore, the joint density of $T_{1}$ and $T_{2}$  given by $f_{T_{1},T_{2}}, t_{0}\leq T_{1}\leq h_{1} , t_{0}\leq T_{2}\leq h_{2}$, is also expressed as $f_{T_{1},T_{2}}=f_{T_{1}}.f_{T_{2}}, t_{0}\leq T_{1}\leq h_{1} , t_{0}\leq T_{2}\leq h_{2}$, since  it is assumed that the random variables $T_{1}$ and $T_{2}$ are independent (see \cite{wanduku-biomath}). The independence between $T_{1}$ and $T_{2}$ is justified from the understanding that the  incubation of  the infectious agent for the vector-borne disease depends on the suitable  biological environmental requirements for incubation inside the vector and the human body which are unrelated. Furthermore, the independence between $T_{1}$ and $T_{3}$ follows from the lack of any real biological evidence to justify the connection between the incubation of  the infectious agent inside the vector and the acquired natural immunity conferred to the human being. But $T_{2}$ and $T_{3}$ may be dependent as biological evidence suggests that the naturally acquired immunity is induced by exposure to the infectious agent.

By employing similar reasoning in \cite{cooke,qun,capasso,huo}, the expected incidence rate of the disease or force of infection of the disease at time $t$ due to the disease transmission process between the infectious vectors and susceptible humans, $S(t)$, is given by the expression $\beta \int^{h_{1}}_{t_{0}}f_{T_{1}}(s) e^{-\mu s}S(t)G(I(t-s))ds$, where $\mu$ is the natural death rate of individuals in the population, and it is assumed for simplicity that the natural death rate for the vectors and human beings are the same. Assuming exponential lifetimes for the people and vectors in the population, the term $0<e^{-\mu s}\leq 1, s\in [t_{0}, h_{1}], h_{1}>0$ represents  the survival probability rate of  exposed vectors over the incubation period, $T_{1}$, of the infectious agent inside the vectors with the length of the period given as $T_{1}=s, \forall s \in [t_{0}, h_{1}]$, where the vectors acquired infection at the earlier time $t-s$ from an infectious human via a successful infected blood meal, and  become infectious at time $t$.  Furthermore, it is assumed that the survival of the vectors over the incubation period of length  $s\in [t_{0}, h_{1}]$ is independent of the age of the vectors. In addition, $I(t-s)$, is the infectious human population at earlier time $t-s$, $G$ is a nonlinear incidence function of the disease dynamics,  and $\beta$ is the average number of effective contacts per infectious individual per unit time. Indeed, the force of infection,  $\beta \int^{h_{1}}_{t_{0}}f_{T_{1}}(s) e^{-\mu s}S(t)G(I(t-s))ds$  signifies the expected rate of new infections at time $t$ between the infectious vectors and the susceptible human population $S(t)$ at time $t$, where the infectious agent is transmitted per infectious vector per unit time at the rate $\beta$. Furthermore, it is assumed that the number of infectious vectors at time $t$ is proportional to the infectious human population at earlier time $t-s$. Moreover, it is further assumed that the interaction between the infectious vectors and  susceptible humans exhibits nonlinear behavior, for instance, psychological and overcrowding effects,  which is characterized by the nonlinear incidence  function $G$. Therefore, the force of infection given by
 \begin{equation}\label{ch1.sec0.eqn0}
   \beta \int^{h_{1}}_{t_{0}}f_{T_{1}}(s) e^{-\mu s}S(t)G(I(t-s))ds,
 \end{equation}
  represents the expected rate at which infected individuals leave the susceptible state and become exposed at time $t$.

 The susceptible individuals who have acquired infection from infectious vectors but are non infectious form the exposed class $E$. The population of exposed individuals at time $t$ is denoted $E(t)$. After the incubation period, $T_{2}=u\in [t_{0}, h_{2}]$, of the infectious agent in the exposed human host, the individual becomes infectious, $I(t)$, at time $t$. Applying similar reasoning in  \cite{cooke-driessche},
 the exposed population, $E(t)$, at time $t$ can be written as follows
  \begin{equation}\label{ch1.sec0.eqn1a}
    E(t)=E(t_{0})e^{-\mu (t-t_{0})}p_{1}(t-t_{0})+\int^{t}_{t_{0}}\beta S(\xi)e^{-\mu T_{1}}G(I(\xi-T_{1}))e^{-\mu(t-\xi)}p_{1}(t-\xi)d \xi,
   \end{equation}
   where
   \begin{equation}\label{ch1.seco.eqn1b}
     p_{1}(t-\xi)=\left\{\begin{array}{l}0,t-\xi\geq T_{2},\\
 1, t-\xi< T_{2} \end{array}\right.
   \end{equation}
   represents the probability that an individual remains exposed over the time interval $[\xi, t]$.
   It is easy to see from (\ref{ch1.sec0.eqn1a}) that under the assumption that the disease has been in the population for at least a time $t>\max_{t_{0}\leq T_{1}\leq h_{1}, t_{0}\leq T_{2}\leq h_{2}} {( T_{1}+ T_{2})}$, in fact, $t>h_{1}+h_{2}$, so that all initial perturbations have died out, the expected number of exposed individuals at time $t$ is given  by
\begin{equation}\label{ch1.sec0.eqn1}
E(t)=\int_{t_{0}}^{h_{2}}f_{T_{2}}(u)\int_{t-u}^{t}\beta \int^{h_{1}}_{t_{0}} f_{T_{1}}(s) e^{-\mu s}S(v)G(I(v-s))e^{-\mu(t-u)}dsdvdu.
\end{equation}
    Similarly, for the removal population, $R(t)$, at time $t$, individuals recover from the infectious state $I(t)$  at the per capita rate $\alpha$  and acquire natural immunity.  The natural immunity wanes after the varying immunity period $T_{3}=r\in [ t_{0},\infty]$, and removed individuals become susceptible again to the disease. Therefore, at time $t$, individuals leave the infectious state at the rate $\alpha I(t)$  and become part of the removal population $R(t)$. Thus, at time $t$ the removed population is given by the following equation
  \begin{equation}\label{ch1.sec0.eqn2a}
    R(t)=R(t_{0})e^{-\mu (t-t_{0})}p_{2}(t-t_{0})+\int^{t}_{t_{0}}\alpha I(\xi)e^{-\mu(t-\xi)}p_{2}(t-\xi)d \xi,
  \end{equation}
    where
    \begin{equation}\label{ch1.sec0.eqn2b}
      p_{2}(t-\xi)=\left\{\begin{array}{l}0,t-\xi\geq T_{3},\\
 1, t-\xi< T_{3} \end{array}\right.
    \end{equation}
 represents the probability that an individual remains naturally immune to the disease over the time interval $[\xi, t]$.
 But it follows from  (\ref{ch1.sec0.eqn2a}) that under the assumption that the disease has been in the population for at least a time $t> \max_{t_{0}\leq T_{1}\leq h_{1}, t_{0}\leq T_{2}\leq h_{2}, T_{3}\geq t_{0}}{(T_{1}+ T_{2}, T_{3})}\geq \max_{t_{0}\leq T_{3}}{(T_{3})}$, in fact, the disease has been in the population for sufficiently large  amount of time so that all initial perturbations have died out,  then the expected number of removal individuals at time $t$ can be written as
  \begin{equation}\label{ch1.sec0.eqn2}
R(t)=\int_{t_{0}}^{\infty}f_{T_{3}}(r)\int_{t-r}^{t}\alpha I(v)e^{-\mu (t-v)}dvdr.
\end{equation}
There is also constant birth rate $B$ of susceptible individuals in the population. Furthermore, individuals die additionally due to disease related causes at the rate $d$. Moreover, $B>$, $d>0$, and all the other parameters are nonnegative.  A compartmental framework illustrating the transition rates between the different states in the system and also showing the delays in the disease dynamics is given in Figure~\ref{ch1.sec4.figure 1}.
\begin{figure}[H]
\includegraphics[height=8cm]{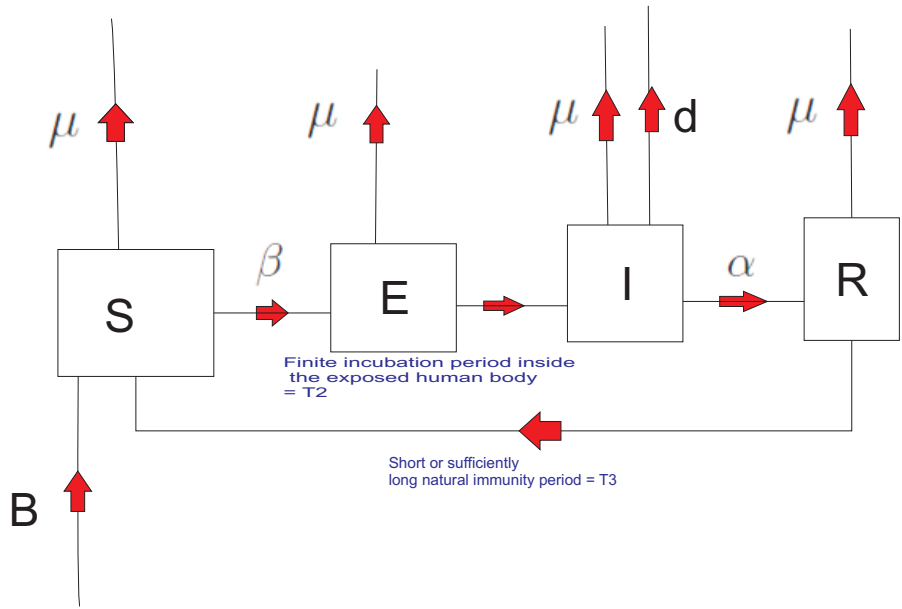}
\caption{The compartmental framework illustrates the transition rates between the states $S,E,I,R$ of the system. It also shows the incubation delay $T_{2}$ and the naturally acquired immunity $T_{3}$ periods. \label{ch1.sec4.figure 1}}
\end{figure}
It follows from (\ref{ch1.sec0.eqn0}), (\ref{ch1.sec0.eqn1}), (\ref{ch1.sec0.eqn2}) and the transition rates illustrated in the compartmental framework in Figure~\ref{ch1.sec4.figure 1} above that the family of SEIRS epidemic dynamic models for a malaria and vector-borne diseases in general in the absence of any random environmental fluctuations in the disease dynamics can be written as follows:
\begin{eqnarray}
dS(t)&=&\left[ B-\beta S(t)\int^{h_{1}}_{t_{0}}f_{T_{1}}(s) e^{-\mu s}G(I(t-s))ds - \mu S(t)+ \alpha \int_{t_{0}}^{\infty}f_{T_{3}}(r)I(t-r)e^{-\mu r}dr \right]dt,\nonumber\\
&&\label{ch1.sec0.eq3}\\
dE(t)&=& \left[ \beta S(t)\int^{h_{1}}_{t_{0}}f_{T_{1}}(s) e^{-\mu s}G(I(t-s))ds - \mu E(t)\right.\nonumber\\
&&\left.-\beta \int_{t_{0}}^{h_{2}}f_{T_{2}}(u)S(t-u)\int^{h_{1}}_{t_{0}}f_{T_{1}}(s) e^{-\mu s-\mu u}G(I(t-s-u))dsdu \right]dt,\label{ch1.sec0.eq4}\\
&&\nonumber\\
dI(t)&=& \left[\beta \int_{t_{0}}^{h_{2}}f_{T_{2}}(u)S(t-u)\int^{h_{1}}_{t_{0}}f_{T_{1}}(s) e^{-\mu s-\mu u}G(I(t-s-u))dsdu- (\mu +d+ \alpha) I(t) \right]dt,\nonumber\\
&&\label{ch1.sec0.eq5}\\
dR(t)&=&\left[ \alpha I(t) - \mu R(t)- \alpha \int_{t_{0}}^{\infty}f_{T_{3}}(r)I(t-r)e^{-\mu s}dr \right]dt,\label{ch1.sec0.eq6}
\end{eqnarray}
where the initial conditions are given in the following: Let $h= h_{1}+ h_{2}$ and define
\begin{eqnarray}
&&\left(S(t),E(t), I(t), R(t)\right)
=\left(\varphi_{1}(t),\varphi_{2}(t), \varphi_{3}(t),\varphi_{4}(t)\right), t\in (-\infty,t_{0}],\nonumber\\
&&\varphi_{k}\in \mathcal{C}((-\infty,t_{0}],\mathbb{R}_{+}),\forall k=1,2,3,4, \nonumber\\
&&\varphi_{k}(t_{0})>0,\forall k=1,2,3,4,\nonumber\\
 \label{ch1.sec0.eq06a}
\end{eqnarray}
where $\mathcal{C}((-\infty,t_{0}],\mathbb{R}_{+})$ is the space of continuous functions with  the supremum norm
\begin{equation}\label{ch1.sec0.eq06b}
||\varphi||_{\infty}=\sup_{ t\leq t_{0}}{|\varphi(t)|}.
\end{equation}
The following general properties of the incidence function $G$ studied in \cite{wanduku-biomath} are given as follows:
\begin{assumption}\label{ch1.sec0.assum1}
\begin{enumerate}
  \item [$A1$]$G(0)=0$.
  \item [$A2$]$G(I)$ is strictly monotonic on $[0,\infty)$.
  \item [$A3$] $G''(I)<0$ $\Leftrightarrow$ $G(I)$ is differentiable concave on $[0,\infty)$.
  \item [$A4$] $\lim_{I\rightarrow \infty}G(I)=C, 0\leq C<\infty$ $\Leftrightarrow$  $G(I)$ has a horizontal asymptote $0\leq C<\infty$.
  \item [$A5$] $G(I)\leq I, \forall I>0$ $\Leftrightarrow$ $G(I)$ is at most as large as the identity function $f:I\mapsto I$ over the positive all $I\in (0,\infty)$.
\end{enumerate}
\end{assumption}
Note that an incidence function $G$ that satisfies  Assumption~\ref{ch1.sec0.assum1} $A1$-$A5$ can be used to describe the disease transmission rate of a vector-borne disease scenario, where the disease dynamics is represented by the system (\ref{ch1.sec0.eq3})-(\ref{ch1.sec0.eq6}), and the disease transmission rate between the vectors and the human beings initially increases or decreases for small values of the infectious population size, and is bounded or steady for sufficiently large size of the infectious  individuals in the population.  It is noted that Assumption~\ref{ch1.sec0.assum1} is a generalization of some subcases of the assumptions $A1$-$A5$ investigated in \cite{gumel,zhica,kyrychko, qun}.
Some examples of frequently used incidence functions in the literature that  satisfy Assumption~\ref{ch1.sec0.assum1}$A1$-$A5$ include:  $G(I(t))=\frac{I(t)}{1+\alpha I(t)}, \alpha>0$, $G(I(t))=\frac{I(t)}{1+\alpha I^{2}(t)}, \alpha>0$, $G(I(t))=I^{p}(t),0<p<1$ and $G(I)=1-e^{-aI}, a>0$.

 In the analysis of the deterministic malaria model (\ref{ch1.sec0.eq3})-(\ref{ch1.sec0.eq6}) with initial conditions in (\ref{ch1.sec0.eq06a})-(\ref{ch1.sec0.eq06b})  in Wanduku\cite{wanduku-biomath}, the threshold values for disease eradication such as the basic reproduction number for the disease when the system is in steady state are obtained in both cases where the delays in the system $T_{1}, T_{2}$ and $T_{3}$ are constant and  also arbitrarily distributed.
  It should be noted that the assumption  of constant delay times representing the incubation period of the disease in the vector, $T_{1}$, incubation period of the disease in the host, $T_{2}$, and immunity period of the disease in the human population, $T_{3}$  in (\ref{ch1.sec0.eq3})-(\ref{ch1.sec0.eq6}) is equivalent to the special case of letting the probability density functions $f_{T_{i}}, i=1,2,3$ of the random variables $T_{1}, T_{2}$ and $T_{3}$ be the dirac-delta function. That is,
\begin{equation}\label{ch1.sec2.eq4}
f_{T_{i}}(s)=\delta(s-T_{i})=\left\{\begin{array}{l}+\infty, s=T_{i},\\
0, otherwise,
\end{array}\right.
, i=1, 2, 3.
\end{equation}
Moreover, under the assumption that $T_{1}\geq 0, T_{2}\geq 0$ and $T_{3}\geq 0$ are constant, the following expectations can be written as  $E(e^{-2\mu (T_{1}+T_{2})})=e^{-2\mu (T_{1}+T_{2})} $, $E(e^{-2\mu T_{1}})=e^{-2\mu T_{1}} $ and $E(e^{-2\mu T_{3}})=e^{-2\mu T_{3}} $.
%

When the delays in the system are all constant, the basic reproduction number of the disease is given by
\begin{equation} \label{ch1.sec2.lemma2a.corrolary1.eq4}
\hat{R}^{*}_{0}=\frac{\beta S^{*}_{0} }{(\mu+d+\alpha)}.
\end{equation}
This threshold value $\hat{R}^{*}_{0}=\frac{\beta S^{*}_{0} }{(\mu+d+\alpha)}$ from (\ref{ch1.sec2.lemma2a.corrolary1.eq4}), represents the total number of infectious cases  that result from one malaria infectious individual present in a completely disease free population with state given by $S^{*}_{0}=\frac{B}{\mu}$, over the average lifetime given by $\frac{1 }{(\mu+d+\alpha)}$ of a person who has survived from disease related death $d$,  natural death $\mu$  and recovered at rate $\alpha$ from infection. Hence,  $\hat{R}^{*}_{0}$ is also the noise-free basic reproduction number of the disease, whenever the incubation periods of the malaria parasite inside the human and mosquito hosts given by $T_{i}, i=1,2$, and also the period of effective naturally acquired immunity against malaria given by $T_{3}$, are all positive constants. Furthermore, the threshold condition $\hat{R}^{*}_{0}<1$ is required for the disease to be eradicated from the noise free human population, whenever the constant delays in the system also satisfy the following:
  \begin{equation} \label{ch1.sec2.lemma2a.corrolary1.eq7}
      T_{max}\geq \frac{1}{2\mu}\log{\frac{\hat{R}^{*}_{1}}{1-\hat{R}^{*}_{0}}},
    \end{equation}
    where
    \begin{equation} \label{ch1.sec2.lemma2a.corrolary1.eq8}
      T_{max}=\max{(T_{1}+T_{2}, T_{3})},
    \end{equation}
    and
    \begin{equation} \label{ch1.sec2.lemma2a.corrolary1.eq3}
\hat{R}^{*}_{1}=\frac{\beta S^{*}_{0} \hat{K}^{*}_{0}+\alpha}{(\mu+d+\alpha)},
\end{equation}
with some constant $\hat{K}^{*}_{0}>0$ (in fact, $\hat{K}^{*}_{0}=4 $).

On the other hand, when the delays in the system  $T_{i}, i=1,2$  are random, and arbitrarily distributed, the basic reproduction number is given by
\begin{equation}\label{ch1.sec2.theorem1.corollary1.eq3}
R_{0}=\frac{\beta S^{*}_{0} \hat{K}_{0}}{(\mu+d+\alpha)}+\frac{\alpha}{(\mu+d+\alpha)},
\end{equation}
where, $\hat{K}_{0}>0$ is a constant that depends only on $S^{*}_{0}$  (in fact, $\hat{K}_{0}=4+ S^{*}_{0} $). In addition, malaria is eradicated from the system in the steady state, whenever $R_{0}\leq 1$,  $\hat{U}_{0}\leq 1$ and $\hat{V}_{0}\leq 1$, where
\begin{equation}\label{ch1.sec2.theorem1.corollary1.eq4}
\hat{U}_{0}=\frac{2\beta S^{*}_{0}+\beta +\alpha + 2\frac{\mu}{\tilde{K}(\mu)^{2}}}{2\mu},
\end{equation}
 and
 \begin{equation}\label{ch1.sec2.theorem1.corollary1.eq5}
\hat{V}_{0}=\frac{(2\mu \tilde{K}(\mu)^{2} + \alpha + \beta (2S^{*}_{0}+1 ) )}{2\mu},
\end{equation}
are other threshold values for the stability of the disease-free steady state $E_{0}=(S^{*}_{0},0,0),S^{*}_{0}=\frac{B}{\mu}$.

Note that the threshold value $R_{0}$ in (\ref{ch1.sec2.theorem1.corollary1.eq3}) is a modification of the basic reproduction number $\hat{R}^{*}_{0}$  defined in (\ref{ch1.sec2.lemma2a.corrolary1.eq4}), and it is therefore the corresponding noise-free basic reproduction number for the disease dynamics described by deterministic system (\ref{ch1.sec0.eq3})-(\ref{ch1.sec0.eq6}), whenever the delays  $T_{i}, i=1,2,3$ in the system are random variables. See  Wanduku\cite{wanduku-biomath} for more conceptual and biological interpretation of the threshold values for disease eradication.
 The stochastic extension of the deterministic model (\ref{ch1.sec0.eq3})-(\ref{ch1.sec0.eq6}) is derived and studied in the following section.
\section{The stochastic model}\label{ch1.sec0.sec1}
 Stochastic models are more realistic because of the inevitable occurrence of random fluctuations in the dynamics of diseases, and in addition, stochastic models provide a  better fit for these disease scenarios than their deterministic counterparts.

 There are several different techniques to add gaussian noise  processes into a dynamic system. One method involves adding the noise into the system as direct influence to the state of the system, where the random fluctuations in the system are, for instance, (1) proportional to the state of the system, or (2) proportional to the deviation of the state of the system from a nonzero steady state etc. Another approach to adding white noise into a dynamic system involves (3) incorporating the random fluctuations in the driving parameters of the system such as the birth, death, recovery and disease transmission rates etc. of an infectious system. See for example \cite{imf}.

 In this study, the third approach is utilized to model the  random environmental fluctuations in the disease transmission rate $\beta$, and also in the natural death rates $\mu$ of the different states $S(t)$, $E(t)$ , $I(t)$ and $R(t)$ of the  human population.  This approach entails the construction of a random walk process for the rates $\beta$, and $\mu$ over an infinitesimally small interval $[t, t+dt]$ and applying the central limit theorem. See for example \cite{wanduku-bookchapter}

 For $t\geq t_{0}$, let $(\Omega, \mathfrak{F}, P)$ be a complete probability space, and $\mathfrak{F}_{t}$ be a filtration (that is, sub $\sigma$- algebra $\mathfrak{F}_{t}$ that satisfies the following: given $t_{1}\leq t_{2} \Rightarrow \mathfrak{F}_{t_{1}}\subset \mathfrak{F}_{t_{2}}; E\in \mathfrak{F}_{t}$ and $P(E)=0 \Rightarrow E\in \mathfrak{F}_{0} $ ).
 The variability in the disease transmission and natural death rates are represented by independent white noise or Wiener processes with drifts, and the rates are expressed as follows:
 \begin{equation}\label{ch1.sec0.eq7}
 \mu  \rightarrow \mu  + \sigma_{i}\xi_{i}(t),\quad \xi_{i}(t) dt= dw_{i}(t),i=S,E,I,R, \quad  \beta \rightarrow \beta + \sigma_{\beta}\xi_{\beta}(t),\quad \xi_{\beta}(t)dt=dw_{\beta}(t),
 \end{equation}
 where $\xi_{i}(t)$ and $w_{i}(t)$ represent the  standard white noise and normalized wiener processes  for the  $i^{th}$ state at time $t$, with the following properties: $w(0)=0, E(w(t))=0, var(w(t))=t$.  Furthermore,  $\sigma_{i},i=S,E,I,R $, represents the intensity of the white noise process due to the random natural death rate of the $i^{th}$ state, and $\sigma_{\beta}$ is the intensity of the white noise process due to the random disease transmission rate. Moreover, the $ w_{i}(t),i=S,E,I,R,\beta,\forall t\geq t_{0}$, are all independent.
 The detailed formulation of the expressions in (\ref{ch1.sec0.eq7}) will appear in the book chapter by Wanduku\cite{wanduku-bookchapter}.

The ideas behind the formulation of the expressions in (\ref{ch1.sec0.eq7}) are
   given in the following. The constant parameters $\mu$ and $\beta$ represent the natural death and disease transmission rates per unit time, respectively. In reality, random environmental fluctuations impact these rates turning them into random variables $\tilde{\mu}$ and $\tilde{\beta}$. Thus, the natural death and disease transmission rates over an infinitesimally small interval of time $[t, t+dt]$  with length $dt$ is given by the expressions  $\tilde{\mu}(t)=\tilde{\mu}dt$ and $\tilde{\beta}(t)=\tilde{\beta}dt$, respectively. It is assumed that there are independent and identical random impacts acting upon these rates at times $t_{j+1}$ over $n$ subintervals $[t_{j}, t_{j+1}]$ of length $\triangle t=\frac{dt}{n}$, where $t_{j}=t_{0}+j\triangle t, j=0,1,\cdots,n$, and  $t_{0}=t$. Furthermore,  it is assumed that $\tilde{\mu}(t_{0})=\tilde{\mu}(t)=\mu dt$ is  constant or deterministic, and $\tilde{\beta}(t_{0})=\tilde{\beta}(t)=\beta dt$ is also a constant. It follows that by letting the independent identically distributed random variables $Z_{i},i=1,\cdots,n $ represent the random effects acting on the natural death rate, then it follows further that the rate at time $t_{n}=t+dt$, that is,
   \begin{equation}\label{ch1.sec0.eq7.eq1}
    \tilde{\mu}(t+dt)=\tilde{\mu}(t)+\sum_{j=1}^{n}Z_{j},
  \end{equation}
  where $E(Z_{j})=0$, and $Var(Z_{j})=\sigma^{2}_{i}\triangle t, i\in \{S, E, I, R\}$.
    Note that $\tilde{\beta}(t+dt)$ can similarly be expressed as (\ref{ch1.sec0.eq7.eq1}). And for sufficient large value of $n$, the summation in (\ref{ch1.sec0.eq7.eq1}) converges in distribution by the central limit theorem to a random variable which is identically distributed as the wiener process $\sigma_{i}(w_{i}(t+dt)-w_{i}(t))=\sigma_{i}dw_{i}(t)$, with mean $0$ and variance $\sigma^{2}_{i}dt, i\in \{S, E, I, R\}$. It follows easily from (\ref{ch1.sec0.eq7.eq1}) that
  \begin{equation}\label{ch1.sec0.eq7.eq2}
   \tilde{\mu}dt =\mu dt+ \sigma_{i}dw_{i}(t), i\in \{S, E, I, R\}.
  \end{equation}
  Similarly, it can be easily seen that
  \begin{equation}\label{ch1.sec0.eq7.eq2}
   \tilde{\beta}dt =\beta dt+ \sigma_{\beta}dw_{\beta}(t).
  \end{equation}
    Note that the intensities $\sigma^{2}_{i},i=S,E,I,R, \beta $ of the independent white noise processes in the expressions   $\tilde{\mu}(t)=\mu dt  + \sigma_{i}\xi_{i}(t)$ and $\tilde{\beta} (t)=\beta dt + \sigma_{\beta}\xi_{\beta}(t)$ that represent the  natural death rate, $\tilde{\mu}(t)$, and disease transmission rate, $\tilde{\beta} (t)$,  at time $t$,  measure the average deviation of the random variable disease transmission rate, $\tilde{\beta}$,  and natural death rate, $\tilde{\mu}$, about their constant mean values $\beta$ and $\mu$, respectively, over the infinitesimally small time interval $[t, t+dt]$. These measures reflect the force of the random fluctuations that occur during the disease outbreak at anytime, and which lead to oscillations in the natural death and disease transmission rates overtime, and consequently lead to oscillations of the susceptible, exposed, infectious and removal states of the system over time during the disease outbreak. Thus, in this study the words "strength" and "intensity" of the white noise are used synonymously. Also, the constructions "strong noise" and "weak noise" are used to refer to white noise with high and low intensities, respectively.

    It is easy to see from (\ref{ch1.sec0.eq7}) that the random natural death rate per unit time denoted $\tilde{\mu}$ is given by  $\tilde{\mu}=\mu   + \sigma_{i}\xi_{i}(t),\quad \xi_{i}(t) dt= dw_{i}(t),i=S,E,I,R,$. It follows further that under the assumption of independent deaths in the human population, so that the number of natural deaths $N(t)$ over an interval $[t_{0}, t_{0}+t]$ of length $t$ follows a Poisson process $\{N(t),t\geq t_{0}\}$ with intensity of the process $E(\tilde{\mu})=\mu$, and mean $E(N(t))=E(\tilde{\mu}t)=\mu t$, then the time until death is exponentially distributed with mean $\frac{1}{\mu}$. Moreover, the survival function is given by
    \begin{equation}\label{ch1.sec0.eq7.eq3}
      \mathfrak{S}(t)=e^{-\mu t},t>0.
    \end{equation}
 Substituting (\ref{ch1.sec0.eq7})-(\ref{ch1.sec0.eq7.eq3}) into the deterministic system (\ref{ch1.sec0.eq3})-(\ref{ch1.sec0.eq6}) leads to the following generalized system of Ito-Doob stochastic differential equations describing the dynamics of  vector-borne diseases in the human population.
 \begin{eqnarray}
dS(t)&=&\left[ B-\beta S(t)\int^{h_{1}}_{t_{0}}f_{T_{1}}(s) e^{-\mu s}G(I(t-s))ds - \mu S(t)+ \alpha \int_{t_{0}}^{\infty}f_{T_{3}}(r)I(t-r)e^{-\mu r}dr \right]dt\nonumber\\
&&-\sigma_{S}S(t)dw_{S}(t)-\sigma_{\beta} S(t)\int^{h_{1}}_{t_{0}}f_{T_{1}}(s) e^{-\mu s}G(I(t-s))dsdw_{\beta}(t) \label{ch1.sec0.eq8}\\
dE(t)&=& \left[ \beta S(t)\int^{h_{1}}_{t_{0}}f_{T_{1}}(s) e^{-\mu s}G(I(t-s))ds - \mu E(t)\right.\nonumber\\
&&\left.-\beta \int_{t_{0}}^{h_{2}}f_{T_{2}}(u)S(t-u)\int^{h_{1}}_{t_{0}}f_{T_{1}}(s) e^{-\mu s-\mu u}G(I(t-s-u))dsdu \right]dt\nonumber\\
&&-\sigma_{E}E(t)dw_{E}(t)+\sigma_{\beta} S(t)\int^{h_{1}}_{t_{0}}f_{T_{1}}(s) e^{-\mu s}G(I(t-s))dsdw_{\beta}(t)\nonumber\\
&&-\sigma_{\beta} \int_{t_{0}}^{h_{2}}f_{T_{2}}(u)S(t-u)\int^{h_{1}}_{t_{0}}f_{T_{1}}(s) e^{-\mu s-\mu u}G(I(t-s-u))dsdudw_{\beta}(t)\label{ch1.sec0.eq9}\\
dI(t)&=& \left[\beta \int_{t_{0}}^{h_{2}}f_{T_{2}}(u)S(t-u)\int^{h_{1}}_{t_{0}}f_{T_{1}}(s) e^{-\mu s-\mu u}G(I(t-s-u))dsdu- (\mu +d+ \alpha) I(t) \right]dt\nonumber\\
&&-\sigma_{I}I(t)dw_{I}(t)+\sigma_{\beta} \int_{t_{0}}^{h_{2}}f_{T_{2}}(u)S(t-u)\int^{h_{1}}_{t_{0}}f_{T_{1}}(s) e^{-\mu s-\mu u}G(I(t-s-u))dsdudw_{\beta}(t)\nonumber\\
&&\label{ch1.sec0.eq10}\\
dR(t)&=&\left[ \alpha I(t) - \mu R(t)- \alpha \int_{t_{0}}^{\infty}f_{T_{3}}(r)I(t-r)e^{-\mu s}dr \right]dt-\sigma_{R}R(t)dw_{R}(t),\label{ch1.sec0.eq11}
\end{eqnarray}
where the initial conditions are given in the following: Let $h= h_{1}+ h_{2}$ and define
\begin{eqnarray}
&&\left(S(t),E(t), I(t), R(t)\right)
=\left(\varphi_{1}(t),\varphi_{2}(t), \varphi_{3}(t),\varphi_{4}(t)\right), t\in (-\infty,t_{0}],\nonumber\\
&&\varphi_{k}\in \mathcal{C}((-\infty,t_{0}],\mathbb{R}_{+}),\forall k=1,2,3,4, \nonumber\\
&&\varphi_{k}(t_{0})>0,\forall k=1,2,3,4,\nonumber\\
 \label{ch1.sec0.eq12}
\end{eqnarray}
where $\mathcal{C}((-\infty,t_{0}],\mathbb{R}_{+})$ is the space of continuous functions with  the supremum norm
\begin{equation}\label{ch1.sec0.eq13}
||\varphi||_{\infty}=\sup_{ t\leq t_{0}}{|\varphi(t)|}.
\end{equation}
Furthermore, the random continuous functions $\varphi_{k},k=1,2,3,4$ are
$\mathfrak{F}_{0}-measurable$, or  independent of $w(t)$
for all $t\geq t_{0}$.

In a similar structure to the study \cite{cooke-driessche}, one major advantage of the family of  vector-borne disease models (\ref{ch1.sec0.eq8})-(\ref{ch1.sec0.eq11}) is the sufficiency in its simplistic nature to provide insights about the vector-borne disease in the human population with limited characterization or limited knowledge of the life cycle of the disease vector. This model provides a suitable platform to study control strategies against the disease with primary focus directed to the human population, whenever there is limited information about the vector life cycle, for instance, in an emergency situation where there is a sudden deadly vector-borne disease outbreak, and there is limited time to investigate the biting habits and life cycles of the vectors.
Observe that (\ref{ch1.sec0.eq9}) and (\ref{ch1.sec0.eq11}) decouple from the other two equations in the system (\ref{ch1.sec0.eq8})-(\ref{ch1.sec0.eq11}). Nevertheless, for mathematical convenience the results in this paper  will be shown for the vector $X(t)=(S(t), E(t), I(t))^{T}$. The following notations are utilized:
\begin{equation}\label{ch1.sec0.eq13b}
\left\{
  \begin{array}{lll}
    Y(t)&=&(S(t), E(t), I(t), R(t))^{T} \\
   X(t)&=&(S(t), E(t), I(t))^{T} \\
   N(t)&=&S(t)+ E(t)+ I(t)+ R(t).
  \end{array}
  \right.
\end{equation}
\section{Model validation \label{ch1.sec1}}
 The existence and uniqueness of solution of the stochastic system  (\ref{ch1.sec0.eq8})-(\ref{ch1.sec0.eq11}) is exhibited in the following theorem. Moreover, the feasibility region of the the solution process $\{X(t), t\geq t_{0}\}$ of the system (\ref{ch1.sec0.eq8})-(\ref{ch1.sec0.eq11}) is defined.  The standard methods  utilized in the earlier studies\cite{wanduku-determ,Wanduku-2017,wanduku-delay,divine5} are applied to establish the results.

  It should be noted that the existence and qualitative behavior of the positive solution process of the system (\ref{ch1.sec0.eq8})-(\ref{ch1.sec0.eq11}) depend on the sources (natural death or disease transmission rates) of variability in the system. As it is shown below, certain sources of variability lead to very complex uncontrolled behavior of the sample paths of the system.

  The following lemma describes the behavior of the positive local solution process for the system (\ref{ch1.sec0.eq8})-(\ref{ch1.sec0.eq11}). This result will be useful in   establishing the existence and uniqueness results for the global solutions of the stochastic system (\ref{ch1.sec0.eq8})-(\ref{ch1.sec0.eq11}).
\begin{lemma}\label{ch1.sec1.lemma1}
Suppose for some $\tau_{e}>t_{0}\geq 0$ the system (\ref{ch1.sec0.eq8})-(\ref{ch1.sec0.eq11}) with initial condition in (\ref{ch1.sec0.eq12}) has a unique positive solution process $Y(t)\in \mathbb{R}^{4}_{+}$, for all $t\in (-\infty, \tau_{e}]$, then  it follows that
\item[(a.)] if $N(t_{0})\leq \frac{B}{\mu}$, and the intensities of the independent white noise processes in the system satisfy  $\sigma_{i}=0$, $i\in \{S, E, I\}$ and $\sigma_{\beta}\geq 0$, then $N(t)\leq \frac{B}{\mu}$, and in addition, the set denoted by
\begin{equation}\label{ch1.sec1.lemma1.eq1}
  D(\tau_{e})=\left\{Y(t)\in \mathbb{R}^{4}_{+}: N(t)=S(t)+ E(t)+ I(t)+ R(t)\leq \frac{B}{\mu}, \forall t\in (-\infty, \tau_{e}] \right\}=\bar{B}^{(-\infty, \tau_{e}]}_{\mathbb{R}^{4}_{+},}\left(0,\frac{B}{\mu}\right),
\end{equation}
is locally self-invariant with respect to the system (\ref{ch1.sec0.eq8})-(\ref{ch1.sec0.eq11}), where $\bar{B}^{(-\infty, \tau_{e}]}_{\mathbb{R}^{4}_{+},}\left(0,\frac{B}{\mu}\right)$ is the closed ball in $\mathbb{R}^{4}_{+}$ centered at the origin with radius $\frac{B}{\mu}$ containing the local positive solutions defined over $(-\infty, \tau_{e}]$.
\item[(b.)] If the intensities of the independent white noise processes in the system satisfy  $\sigma_{i}>0$, $i\in \{S, E, I\}$ and $\sigma_{\beta}\geq 0$, then $Y(t)\in \mathbb{R}^{4}_{+}$ and $N(t)\geq 0$, for all $t\in (-\infty, \tau_{e}]$.
\end{lemma}
Proof:\\
 It follows directly from (\ref{ch1.sec0.eq8})-(\ref{ch1.sec0.eq11}) that when $\sigma_{i}=0$, $i\in \{S, E, I\}$ and $\sigma_{\beta}\geq 0$, then
\begin{equation}\label{ch1.sec1.lemma1.eq2}
dN(t)=[B-\mu N(t)-dI(t)]dt
\end{equation}
The result in (a.) follows easily by observing that for $Y(t)\in \mathbb{R}^{4}_{+}$, the equation (\ref{ch1.sec1.lemma1.eq2}) leads to  $N(t)\leq \frac{B}{\mu}-\frac{B}{\mu}e^{-\mu(t-t_{0})}+N(t_{0})e^{-\mu(t-t_{0})}$. And under the assumption that $N(t_{0})\leq \frac{B}{\mu}$, the result follows immediately. The result in (b.) follows directly from Theorem~\ref{ch1.sec1.thm1}.\\
 The following theorem presents the existence and uniqueness results for the global solutions  of the stochastic system (\ref{ch1.sec0.eq8})-(\ref{ch1.sec0.eq11}). 
\begin{thm}\label{ch1.sec1.thm1}
  Given the initial conditions (\ref{ch1.sec0.eq12}) and (\ref{ch1.sec0.eq13}), there exists a unique solution process $X(t,w)=(S(t,w),E(t,w), I(t,w))^{T}$ satisfying (\ref{ch1.sec0.eq8})-(\ref{ch1.sec0.eq11}), for all $t\geq t_{0}$. Moreover,
  \item[(a.)] the solution process is positive for all $t\geq t_{0}$ a.s. and lies in $D(\infty)$, whenever  the intensities of the independent white noise processes in the system satisfy  $\sigma_{i}=0$, $i\in \{S, E, I\}$ and $\sigma_{\beta}\geq 0$.
        That is, $S(t,w)>0,E(t,w)>0,  I(t,w)>0, \forall t\geq t_{0}$ a.s. and $X(t,w)\in D(\infty)=\bar{B}^{(-\infty, \infty)}_{\mathbb{R}^{4}_{+},}\left(0,\frac{B}{\mu}\right)$, where $D(\infty)$ is defined in Lemma~\ref{ch1.sec1.lemma1}, (\ref{ch1.sec1.lemma1.eq1}).
        \item[(b.)] Also, the solution process is positive for all $t\geq t_{0}$ a.s. and lies in $\mathbb{R}^{4}_{+}$, whenever  the intensities of the independent white noise processes in the system satisfy  $\sigma_{i}>0$, $i\in \{S, E, I\}$ and $\sigma_{\beta}\geq 0$.
        That is, $S(t,w)>0,E(t,w)>0,  I(t,w)>0, \forall t\geq t_{0}$ a.s. and $X(t,w)\in \mathbb{R}^{4}_{+}$.
\end{thm}
Proof:\\
 A similar proof of this result appears in a more general study of vector-borne diseases in Wanduku\cite{wanduku-theorBio}, nevertheless it is added here for completion. It is easy to see that the coefficients of (\ref{ch1.sec0.eq8})-(\ref{ch1.sec0.eq11}) satisfy the local Lipschitz condition for the given initial data (\ref{ch1.sec0.eq12}). Therefore there exist a unique maximal local solution $X(t,w)=(S(t,w), E(t,w), I(t,w))$ on $t\in (-\infty,\tau_{e}(w)]$, where $\tau_{e}(w)$ is the first hitting time or the explosion time of the process\cite{mao}.
 The following shows that $X(t,w)\in D(\tau_{e})$ almost surely, whenever $\sigma_{i}=0$, $i\in \{S, E, I\}$ and $\sigma_{\beta}\geq 0$,  where $D(\tau_{e}(w))$ is defined in Lemma~\ref{ch1.sec1.lemma1} (\ref{ch1.sec1.lemma1.eq1}), and also that $X(t,w)\in \mathbb{R}^{4}_{+}$, whenever  $\sigma_{i}>0$, $i\in \{S, E, I\}$ and $\sigma_{\beta}\geq 0$.
Define the following stopping time
\begin{equation}\label{ch1.sec1.thm1.eq1}
\left\{
\begin{array}{lll}
\tau_{+}&=&sup\{t\in (t_{0},\tau_{e}(w)): S|_{[t_{0},t]}>0,\quad E|_{[t_{0},t]}>0,\quad and\quad I|_{[t_{0},t]}>0 \},\\
\tau_{+}(t)&=&\min(t,\tau_{+}),\quad for\quad t\geq t_{0}.\\
\end{array}
\right.
\end{equation}
and lets show that $\tau_{+}(t)=\tau_{e}(w)$ a.s. Suppose on the contrary that $P(\tau_{+}(t)<\tau_{e}(w))>0$. Let $w\in \{\tau_{+}(t)<\tau_{e}(w)\}$, and $t\in [t_{0},\tau_{+}(t))$. Define
\begin{equation}\label{ch1.sec1.thm1.eq2}
\left\{
\begin{array}{ll}
V(X(t))=V_{1}(X(t))+V_{2}(X(t))+V_{3}(X(t)),\\
V_{1}(X(t))=\ln(S(t)),\quad V_{2}(X(t))=\ln(E(t)),\quad V_{3}(X(t))=\ln(I(t)),\forall t\leq\tau_{+}(t).
\end{array}
\right.
\end{equation}
It follows from (\ref{ch1.sec1.thm1.eq2}) that
\begin{equation}\label{ch1.sec1.thm1.eq3}
  dV(X(t))=dV_{1}(X(t))+dV_{2}(X(t))+dV_{3}(X(t)),
\end{equation}
where
\begin{eqnarray}
  dV_{1}(X(t)) &=& \frac{1}{S(t)}dS(t)-\frac{1}{2}\frac{1}{S^{2}(t)}(dS(t))^{2}\nonumber \\
   &=&\left[ \frac{B}{S(t)}-\beta \int^{h_{1}}_{t_{0}}f_{T_{1}}(s) e^{-\mu s}G(I(t-s))ds - \mu + \frac{\alpha}{S(t)} \int_{t_{0}}^{\infty}f_{T_{3}}(r)I(t-r)e^{-\mu r}dr \right.\nonumber\\
   &&\left.-\frac{1}{2}\sigma^{2}_{S}-\frac{1}{2}\sigma^{2}_{\beta}\left(\int^{h_{1}}_{t_{0}}f_{T_{1}}(s) e^{-\mu s}G(I(t-s))ds\right)^{2}\right]dt\nonumber\\
&&-\sigma_{S}dw_{S}(t)-\sigma_{\beta} \int^{h_{1}}_{t_{0}}f_{T_{1}}(s) e^{-\mu s}G(I(t-s))dsdw_{\beta}(t), \label{ch1.sec1.thm1.eq4}
\end{eqnarray}
\begin{eqnarray}
  dV_{2}(X(t)) &=& \frac{1}{E(t)}dE(t)-\frac{1}{2}\frac{1}{E^{2}(t)}(dE(t))^{2} \nonumber\\
  &=& \left[ \beta \frac{S(t)}{E(t)}\int^{h_{1}}_{t_{0}}f_{T_{1}}(s) e^{-\mu s}G(I(t-s))ds - \mu \right.\nonumber\\
&&\left.-\beta\frac{1}{E(t)} \int_{t_{0}}^{h_{2}}f_{T_{2}}(u)S(t-u)\int^{h_{1}}_{t_{0}}f_{T_{1}}(s) e^{-\mu s-\mu u}G(I(t-s-u))dsdu \right.\nonumber\\
&&\left.-\frac{1}{2}\sigma^{2}_{E}-\frac{1}{2}\sigma^{2}_{\beta}\frac{S^{2}(t)}{E^{2}(t)}\left(\int^{h_{1}}_{t_{0}}f_{T_{1}}(s) e^{-\mu s}G(I(t-s))ds\right)^{2}\right.\nonumber\\
&&\left.-\frac{1}{2}\sigma^{2}_{\beta}\frac{1}{E^{2}(t)}\left(\int_{t_{0}}^{h_{2}}f_{T_{2}}(u)S(t-u)\int^{h_{1}}_{t_{0}}f_{T_{1}}(s) e^{-\mu s-\mu u}G(I(t-s-u))dsdu \right)^{2}\right]dt\nonumber\\
&&-\sigma_{E}dw_{E}(t)+\sigma_{\beta} \frac{S(t)}{E(t)}\int^{h_{1}}_{t_{0}}f_{T_{1}}(s) e^{-\mu s}G(I(t-s))dsdw_{\beta}(t)\nonumber\\
&&-\sigma_{\beta}\frac{1}{E(t)} \int_{t_{0}}^{h_{2}}f_{T_{2}}(u)S(t-u)\int^{h_{1}}_{t_{0}}f_{T_{1}}(s) e^{-\mu s-\mu u}G(I(t-s-u))dsdudw_{\beta}(t),\nonumber\\
\label{ch1.sec1.thm1.eq5}
\end{eqnarray}
and
\begin{eqnarray}
  dV_{3}(X(t)) &=& \frac{1}{I(t)}dI(t)-\frac{1}{2}\frac{1}{I^{2}(t)}(dI(t))^{2}\nonumber \\
  &=& \left[\beta \frac{1}{I(t)}\int_{t_{0}}^{h_{2}}f_{T_{2}}(u)S(t-u)\int^{h_{1}}_{t_{0}}f_{T_{1}}(s) e^{-\mu s-\mu u}G(I(t-s-u))dsdu- (\mu +d+ \alpha)\right.  \nonumber\\
  &&\left.-\frac{1}{2}\sigma^{2}_{I}-\frac{1}{2}\sigma^{2}_{\beta}\left(\int_{t_{0}}^{h_{2}}f_{T_{2}}(u)S(t-u)\int^{h_{1}}_{t_{0}}f_{T_{1}}(s) e^{-\mu s-\mu u}G(I(t-s-u))dsdu \right)^{2}\right]dt\nonumber\\
&&-\sigma_{I}dw_{I}(t)+\sigma_{\beta}\frac{1}{I(t)} \int_{t_{0}}^{h_{2}}f_{T_{2}}(u)S(t-u)\int^{h_{1}}_{t_{0}}f_{T_{1}}(s) e^{-\mu s-\mu u}G(I(t-s-u))dsdudw_{\beta}(t)\nonumber\\
&&\label{ch1.sec1.thm1.eq6}
\end{eqnarray}
It follows from (\ref{ch1.sec1.thm1.eq3})-(\ref{ch1.sec1.thm1.eq6}) that for $t<\tau_{+}(t)$,
\begin{eqnarray}
  V(X(t))-V(X(t_{0})) &\geq& \int^{t}_{t_{0}}\left[-\beta \int^{h_{1}}_{t_{0}}f_{T_{1}}(s) e^{-\mu s}G(I(\xi-s))ds-\frac{1}{2}\sigma^{2}_{S}\right.\nonumber\\
   &&\left.-\frac{1}{2}\sigma^{2}_{\beta}\left(\int^{h_{1}}_{t_{0}}f_{T_{1}}(s) e^{-\mu s}G(I(\xi-s))ds\right)^{2}\right]d\xi\nonumber\\
   &&+ \int_{t}^{t_{0}}\left[-\beta\frac{1}{E(\xi)} \int_{t_{0}}^{h_{2}}f_{T_{2}}(u)S(\xi-u)\int^{h_{1}}_{t_{0}}f_{T_{1}}(s) e^{-\mu s-\mu u}G(I(\xi-s-u))dsdu
\right.\nonumber\\
&&\left.-\frac{1}{2}\sigma^{2}_{E}-\frac{1}{2}\sigma^{2}_{\beta}\frac{S^{2}(\xi)}{E^{2}(\xi)}\left(\int^{h_{1}}_{t_{0}}f_{T_{1}}(s) e^{-\mu s}G(I(\xi-s))ds\right)^{2}\right.\nonumber\\
&&\left.-\frac{1}{2}\sigma^{2}_{\beta}\frac{1}{E^{2}(\xi)}\left(\int_{t_{0}}^{h_{2}}f_{T_{2}}(u)S(\xi-u)\int^{h_{1}}_{t_{0}}f_{T_{1}}(s) e^{-\mu s-\mu u}G(I(\xi-s-u))dsdu \right)^{2}\right]d\xi\nonumber\\
   &&+ \int_{t}^{t_{0}}\left[- (3\mu +d+ \alpha)-\frac{1}{2}\sigma^{2}_{I}\right.  \nonumber\\
  &&\left.-\frac{1}{2}\sigma^{2}_{\beta}\left(\int_{t_{0}}^{h_{2}}f_{T_{2}}(u)S(\xi-u)\int^{h_{1}}_{t_{0}}f_{T_{1}}(s) e^{-\mu s-\mu u}G(I(\xi-s-u))dsdu \right)^{2}\right]d\xi\nonumber\\
&&+\int_{t}^{t_{0}}\left[-\sigma_{S}dw_{S}(\xi)-\sigma_{\beta} \int^{h_{1}}_{t_{0}}f_{T_{1}}(s) e^{-\mu s}G(I(\xi-s))dsdw_{\beta}(\xi)\right]\nonumber \\
  &&+\int_{t}^{t_{0}}\left[-\sigma_{E}dw_{E}(\xi)+\sigma_{\beta} \frac{S(\xi)}{E(\xi)}\int^{h_{1}}_{t_{0}}f_{T_{1}}(s) e^{-\mu s}G(I(\xi-s))dsdw_{\beta}(\xi)\right]\nonumber\\
&&-\int_{t}^{t_{0}}\left[\sigma_{\beta}\frac{1}{E(\xi)} \int_{t_{0}}^{h_{2}}f_{T_{2}}(u)S(\xi-u)\int^{h_{1}}_{t_{0}}f_{T_{1}}(s) e^{-\mu s-\mu u}G(I(\xi-s-u))dsdudw_{\beta}(\xi)\right]\nonumber\\
&&+\int_{t_{0}}^{t}\left[-\sigma_{I}dw_{I}(\xi)\right.\nonumber\\
&&\left.+\sigma_{\beta}\frac{1}{I(\xi)} \int_{t_{0}}^{h_{2}}f_{T_{2}}(u)S(\xi-u)\int^{h_{1}}_{t_{0}}f_{T_{1}}(s) e^{-\mu s-\mu u}G(I(\xi-s-u))dsdudw_{\beta}(\xi)\right].\nonumber\\
&&\label{ch1.sec1.thm1.eq7}
\end{eqnarray}
Taking the limit on (\ref{ch1.sec1.thm1.eq7}) as $t\rightarrow \tau_{+}(t)$, it follows from (\ref{ch1.sec1.thm1.eq1})-(\ref{ch1.sec1.thm1.eq2}) that the left-hand side $V(X(t))-V(X(t_{0}))\leq -\infty$. This contradicts the finiteness of the right-handside of the inequality (\ref{ch1.sec1.thm1.eq7}). Hence $\tau_{+}(t)=\tau_{e}(w)$ a.s., that is, $X(t,w)\in D(\tau_{e})$,   whenever  $\sigma_{i}=0$, $i\in \{S, E, I\}$ and $\sigma_{\beta}\geq 0$, and $X(t,w)\in \mathbb{R}^{4}_{+}$, whenever  $\sigma_{i}>0$, $i\in \{S, E, I\}$ and $\sigma_{\beta}\geq 0$.

The following shows that $\tau_{e}(w)=\infty$. Let $k>0$ be a positive integer such that $||\vec{\varphi}||_{1}\leq k$, where $\vec{\varphi}=\left(\varphi_{1}(t),\varphi_{2}(t), \varphi_{3}(t)\right), t\in (-\infty,t_{0}]$ defined in (\ref{ch1.sec0.eq12}), and $||.||_{1}$ is the $p-sum$ norm defined on $\mathbb{R}^{3}$, when $p=1$. Define the stopping time
\begin{equation}\label{ch1.sec1.thm1.eq8}
\left\{
\begin{array}{ll}
\tau_{k}=sup\{t\in [t_{0},\tau_{e}): ||X(s)||_{1}=S(s)+E(s)+I(s)\leq k, s\in[t_{0},t] \}\\
\tau_{k}(t)=\min(t,\tau_{k}).
\end{array}
\right.
\end{equation}
It is easy to see that as $k\rightarrow \infty$, $\tau_{k}$ increases. Set $\lim_{k\rightarrow \infty}\tau_{k}(t)=\tau_{\infty}$. Then it follows that $\tau_{\infty}\leq \tau_{e}$ a.s.
 We show in the following that: (1.) $\tau_{e}=\tau_{\infty}\quad a.s.\Leftrightarrow P(\tau_{e}\neq \tau_{\infty})=0$, (2.)  $\tau_{\infty}=\infty\quad a.s.\Leftrightarrow P(\tau_{\infty}=\infty)=1$.

Suppose on the contrary that $P(\tau_{\infty}<\tau_{e})>0$. Let $w\in \{\tau_{\infty}<\tau_{e}\}$ and $t\leq \tau_{\infty}$.
 Define
\begin{equation}\label{ch1.sec1.thm1.eq9}
\left\{
\begin{array}{ll}
\hat{V}_{1}(X(t))=e^{\mu t}(S(t)+E(t)+I(t)),\\
\forall t\leq\tau_{k}(t).
\end{array}
\right.
\end{equation}
The Ito-Doob differential $d\hat{V}_{1}$ of (\ref{ch1.sec1.thm1.eq9}) with respect to the system (\ref{ch1.sec0.eq8})-(\ref{ch1.sec0.eq11}) is given as follows:
\begin{eqnarray}
 d\hat{V}_{1} &=& \mu e^{\mu t}(S(t)+E(t)+I(t)) dt + e^{\mu t}(dS(t)+dE(t)+dI(t))  \\
   &=& e^{\mu t}\left[B+\alpha \int_{t_{0}}^{\infty}f_{T_{3}}(r)I(t-r)e^{-\mu r}dr-(\alpha + d)I(t)\right]dt\nonumber\\
   &&-\sigma_{S}e^{\mu t}S(t)dw_{S}(t)-\sigma_{E}e^{\mu t}E(t)dw_{E}(t)-\sigma_{I}e^{\mu t}I(t)dw_{I}(t)\label{ch1.sec1.thm1.eq10}
\end{eqnarray}
Integrating (\ref{ch1.sec1.thm1.eq9}) over the interval $[t_{0}, \tau]$, and applying some algebraic manipulations and  simplifications  it follows that
\begin{eqnarray}
  V_{1}(X(\tau)) &=& V_{1}(X(t_{0}))+\frac{B}{\mu}\left(e^{\mu \tau}-e^{\mu t_{0}}\right)\nonumber\\
  &&+\int_{t_{0}}^{\infty}f_{T_{3}}(r)e^{-\mu r}\left(\int_{t_{0}-r}^{t_{0}}\alpha I(\xi)d\xi-\int_{\tau-r}^{\tau}\alpha I(\xi)d\xi\right)dr-\int_{t_{0}}^{\tau}d I(\xi)d\xi \nonumber\\
  &&+\int^{\tau}_{t_{0}}\left[-\sigma_{S}e^{\mu \xi}S(\xi)dw_{S}(\xi)-\sigma_{E}e^{\mu \xi}E(\xi)dw_{E}(\xi)-\sigma_{I}e^{\mu \xi}I(\xi)dw_{I}(\xi)\right]\label{ch1.sec1.thm1.eq11}
\end{eqnarray}
Removing negative terms from (\ref{ch1.sec1.thm1.eq11}), it implies from (\ref{ch1.sec0.eq12}) that
\begin{eqnarray}
  V_{1}(X(\tau)) &\leq& V_{1}(X(t_{0}))+\frac{B}{\mu}e^{\mu \tau}\nonumber\\
  &&+\int_{t_{0}}^{\infty}f_{T_{3}}(r)e^{-\mu r}\left(\int_{t_{0}-r}^{t_{0}}\alpha \varphi_{3}(\xi)d\xi\right)dr \nonumber\\
  &&+\int^{\tau}_{t_{0}}\left[-\sigma_{S}e^{\mu \xi}S(\xi)dw_{S}(\xi)-\sigma_{E}e^{\mu \xi}E(\xi)dw_{E}(\xi)-\sigma_{I}e^{\mu \xi}I(\xi)dw_{I}(\xi)\right]\label{ch1.sec1.thm1.eq12}
\end{eqnarray}
But from (\ref{ch1.sec1.thm1.eq9}) it is easy to see that for $\forall t\leq\tau_{k}(t)$,
\begin{equation}\label{ch1.sec1.thm1.eq12a}
  ||X(t)||_{1}=S(t)+E(t)+I(t)\leq V(X(t)).
\end{equation}
 Thus setting $\tau=\tau_{k}(t)$, then it follows from
(\ref{ch1.sec1.thm1.eq8}), (\ref{ch1.sec1.thm1.eq12}) and  (\ref{ch1.sec1.thm1.eq12a}) that
\begin{equation}\label{ch1.sec1.thm1.eq13}
  k=||X(\tau_{k}(t))||_{1}\leq V_{1}(X(\tau_{k}(t)))
\end{equation}
Taking the limit on (\ref{ch1.sec1.thm1.eq13}) as $k\rightarrow \infty$ leads to a contradiction because the left-hand-side of the inequality (\ref{ch1.sec1.thm1.eq13}) is infinite, but following the right-hand-side  from (\ref{ch1.sec1.thm1.eq12}) leads to a finite value. Hence $\tau_{e}=\tau_{\infty}$ a.s. The following shows that $\tau_{e}=\tau_{\infty}=\infty$ a.s.

  Let $\ w\in \{\tau_{e}<\infty\}$. It follows from (\ref{ch1.sec1.thm1.eq11})-(\ref{ch1.sec1.thm1.eq12}) that
  \begin{eqnarray}
  I_{\{\tau_{e}<\infty\}}V_{1}(X(\tau)) &\leq& I_{\{\tau_{e}<\infty\}}V_{1}(X(t_{0}))+I_{\{\tau_{e}<\infty\}}\frac{B}{\mu}e^{\mu \tau}\nonumber\\
  &&+I_{\{\tau_{e}<\infty\}}\int_{t_{0}}^{\infty}f_{T_{3}}(r)e^{-\mu r}\left(\int_{t_{0}-r}^{t_{0}}\alpha \varphi_{3}(\xi)d\xi\right)dr\nonumber\\
  &&+I_{\{\tau_{e}<\infty\}}\int^{\tau}_{t_{0}}\left[-\sigma_{S}e^{\mu \xi}S(\xi)dw_{S}(\xi)-\sigma_{E}e^{\mu \xi}E(\xi)dw_{E}(\xi)-\sigma_{I}e^{\mu \xi}I(\xi)dw_{I}(\xi)\right].
  \nonumber\\
  \label{ch1.sec1.thm1.eq14}
\end{eqnarray}
Suppose $\tau=\tau_{k}(t)\wedge T$, where $ T>0$ is arbitrary, then taking the expected value of (\ref{ch1.sec1.thm1.eq14}) follows that
\begin{equation}\label{ch1.sec1.thm1.eq14a}
  E(I_{\{\tau_{e}<\infty\}}V_{1}(X(\tau_{k}(t)\wedge T))) \leq V_{1}(X(t_{0}))+\frac{B}{\mu}e^{\mu T}
\end{equation}
But from (\ref{ch1.sec1.thm1.eq12a}) it is easy to see that
\begin{equation}\label{ch1.sec1.thm1.eq15}
 I_{\{\tau_{e}<\infty,\tau_{k}(t)\leq T\}}||X(\tau_{k}(t))||_{1}\leq I_{\{\tau_{e}<\infty\}}V_{1}(X(\tau_{k}(t)\wedge T))
\end{equation}
It follows from (\ref{ch1.sec1.thm1.eq14})-(\ref{ch1.sec1.thm1.eq15}) and
   (\ref{ch1.sec1.thm1.eq8}) that
 \begin{eqnarray}
 P(\{\tau_{e}<\infty,\tau_{k}(t)\leq T\})k&=&E\left[I_{\{\tau_{e}<\infty,\tau_{k}(t)\leq T\}}||X(\tau_{k}(t))||_{1}\right]\nonumber\\
 &\leq& E\left[I_{\{\tau_{e}<\infty\}}V_{1}(X(\tau_{k}(t)\wedge T))\right]\nonumber\\
 &\leq& V_{1}(X(t_{0}))+\frac{B}{\mu}e^{\mu T}.
\label{ch1.sec1.thm1.eq16}
 \end{eqnarray}
  It follows immediately from (\ref{ch1.sec1.thm1.eq16}) that
 $P(\{\tau_{e}<\infty,\tau_{\infty}\leq T\})\rightarrow 0$ as $k\rightarrow \infty$. Furthermore, since $T<\infty$ is arbitrary, we conclude that $P(\{\tau_{e}<\infty,\tau_{\infty}< \infty\})= 0$.
Finally,  by the total probability principle,
 \begin{eqnarray}
 P(\{\tau_{e}<\infty\})&=&P(\{\tau_{e}<\infty,\tau_{\infty}=\infty\})+P(\{\tau_{e}<\infty,\tau_{\infty}<\infty\})\nonumber\\
 &\leq&P(\{\tau_{e}\neq\tau_{\infty}\})+P(\{\tau_{e}<\infty,\tau_{\infty}<\infty\})\nonumber\\
 &=&0.\label{ch1.sec1.thm1.eq17}
 \end{eqnarray}
 Thus from (\ref{ch1.sec1.thm1.eq17}), $\tau_{e}=\tau_{\infty}=\infty$ a.s.. In addition, $X(t)\in D(\infty)$, whenever  $\sigma_{i}=0$, $i\in \{S, E, I\}$ and $\sigma_{\beta}\geq 0$, and $X(t,w)\in \mathbb{R}^{4}_{+}$, whenever  $\sigma_{i}>0$, $i\in \{S, E, I\}$ and $\sigma_{\beta}\geq 0$.
\begin{rem}\label{ch1.sec0.remark1}
Theorem~\ref{ch1.sec1.thm1} and Lemma~\ref{ch1.sec1.lemma1} signify that the stochastic system (\ref{ch1.sec0.eq8})-(\ref{ch1.sec0.eq11}) has a unique positive solution process  $Y(t)\in \mathbb{R}^{4}_{+}$ globally for all $t\in (-\infty, \infty)$. Furthermore, it follows that every positive sample path for the stochastic system that starts in the closed ball centered at the origin with a radius of $\frac{B}{\mu}$, $D(\infty)=\bar{B}^{(-\infty, \infty)}_{\mathbb{R}^{4}_{+},}\left(0,\frac{B}{\mu}\right)$, will continue to oscillate and remain bounded in the closed ball for all time $t\geq t_{0}$, whenever the intensities of the independent white noise processes in the system satisfy  $\sigma_{i}=0$, $i\in \{S, E, I\}$ and $\sigma_{\beta}\geq 0$. Hence, the set $D(\infty)=\bar{B}^{(-\infty, \infty)}_{\mathbb{R}^{4}_{+},}\left(0,\frac{B}{\mu}\right)$ is a positive self-invariant set, or the feasibility region  for the stochastic system (\ref{ch1.sec0.eq8})-(\ref{ch1.sec0.eq11}), whenever $\sigma_{i}=0$, $i\in \{S, E, I\}$ and $\sigma_{\beta}\geq 0$.
 In the case where the intensities of the independent white noise processes in the system satisfy  $\sigma_{i}>0$, $i\in \{S, E, I\}$ and $\sigma_{\beta}\geq 0$, the sample path solutions are positive and unique, and continue to oscillate in the unbounded space of positive real numbers $\mathbb{R}^{4}_{+}$. In other words, all positive sample path solutions of the system that start in the bounded region $D(\infty)$, remain bounded for all time, whenever $\sigma_{i}=0$, $i\in \{S, E, I\}$ and $\sigma_{\beta}\geq 0$, and the positive sample paths  may become unbounded,  whenever $\sigma_{i}>0$, $i\in \{S, E, I\}$ and $\sigma_{\beta}\geq 0$.

The implication of this result to the disease dynamics represented by (\ref{ch1.sec0.eq8})-(\ref{ch1.sec0.eq11}) is that the occurrence of noise exclusively from the disease transmission rate allows a controlled situation for the disease dynamics, since the positive solutions exist within a well-defined positive self invariant space. The additional source of variability from the natural death rate of any of the different disease classes (susceptible, exposed, infectious or removed), may lead to more complex and uncontrolled situations for the disease dynamics, since it is obvious that the intensities of the white noise processes from the natural death rates of the different states in the system may drive the positive sample path solutions of the system unbounded. Some examples of uncontrolled disease situations that can occur when the positive solutions are unbounded include:-  (1) extinction of the population, (2) failure of existence of an infection-free steady population state, wherein the disease can be controlled into the state, and (3)  a sudden significant random rise or drop of a given state, such as the infectious state, from a low to high value, or vice versa over a short time period etc.

It is shown in the subsequent sections that the stronger noise in the system tends to enhance the persistence of the disease, and possible eventual extinction of the human population.
\end{rem}
\section{Stochasticity of the endemic equilibrium and persistence of disease\label{ch1.sec3}}
From a probabilistic perspective, the stochastic asymptotic stability (in the sense of Lyapunov) of an endemic equilibrium $E_{1}$, whenever it exists, ensures that every sample path for the stochastic system that starts in the neighborhood of the steady state $E_{1}$, has a high probability of oscillating in the neighborhood of the state, and eventually converges to that steady state, almost surely.

The biological significance of the stochastic stability of the endemic equilibrium $E_{1}$, whenever it exists is that, there exists a steady state for all disease related states in the population (exposed, infectious and removed), denoted $E_{1}$, where all sample paths for the disease related states that start in the neighborhood of the state $E_{1}$, have a high probability of oscillating in the neighborhood of the state $E_{1}$, and eventually converge to that steady state in the definite sense. In other words, in the long term, there is certainty of an endemic population, which persists the disease. Epidemiologists require the basic reproduction numbers $R^{*}_{0}$ or $R_{0}$ defined in (\ref{ch1.sec2.lemma2a.corrolary1.eq4}) and (\ref{ch1.sec2.theorem1.corollary1.eq3}), respectively, to satisfy the conditions  $R^{*}_{0}>1$ or $R_{0}\geq 1$ for the disease to persist.These facts are discussed in this section, and examples provided to substantiate the results.

It is easy to see that the stochastic system (\ref{ch1.sec0.eq8})-(\ref{ch1.sec0.eq11}) does not have a nontrivial or endemic steady state, whenever at least one of the intensities of the independent white noise processes in the system $\sigma_{i}>0, i=S, E, I, R, \beta$.  Nevertheless, when the intensities of the noises of the system are infinitesimally small, that is,  $\sigma_{i}=0, i=S, E, I, R, \beta$, the resulting system behaves approximately in the same manner as the deterministic system  (\ref{ch1.sec0.eq3})-(\ref{ch1.sec0.eq6}), which has an endemic equilibrium $E_{1}$ studied in \cite{wanduku-biomath}. Thus, in this section, the asymptotic behavior of the sample paths of the stochastic system  (\ref{ch1.sec0.eq8})-(\ref{ch1.sec0.eq11}) in the neighborhood of the potential endemic steady state, denoted $E_{1}=(S^{*}_{1}, E^{*}_{1}, I^{*}_{1})$, is exhibited.
%

 The following results are quoted from \cite{wanduku-biomath} about  sufficient conditions for the existence of the endemic equilibrium of the deterministic system (\ref{ch1.sec0.eq3})-(\ref{ch1.sec0.eq6}).
\begin{thm}\label{ch1.sec3.thm1}
  Suppose the threshold condition $R_{0}>1$ is satisfied, where $R_{0}$ is  defined in (\ref{ch1.sec2.theorem1.corollary1.eq3}). It follows that when the delays in the system namely $T_{i}, i=1, 2, 3$ are random, and arbitrarily distributed, then the deterministic system (\ref{ch1.sec0.eq3})-(\ref{ch1.sec0.eq6}) has a unique positive equilibrium state denoted by $E_{1}=(S^{*}_{1}, E^{*}_{1}, I^{*}_{1})$,  whenever
\begin{equation}\label{ch1.sec3.thm1.eq1}
E(e^{-\mu(T_{1}+T_{2})})\geq \frac{\hat{K}_{0}+\frac{\alpha}{\beta \frac{B}{\mu}}}{G'(0)},
\end{equation}
where $\hat{K}_{0}$ is also defined in  (\ref{ch1.sec2.theorem1.corollary1.eq3}).
\end{thm}
Proof:\\
See\cite{wanduku-biomath}.
\begin{thm}\label{ch1.sec3.thm1.corrolary1}
Suppose the incubation periods of the malaria plasmodium inside the mosquito and human hosts $T_{1}$ and $T_{2}$, and also the period of effective natural immunity against malaria inside the human being $T_{3}$ are constant.  Let the threshold condition $R^{*}_{0}>1$ be satisfied, where $R^{*}_{0}$ is  defined in (\ref{ch1.sec2.lemma2a.corrolary1.eq4}). It follows that the deterministic system (\ref{ch1.sec0.eq3})-(\ref{ch1.sec0.eq6}) has a unique positive equilibrium state denoted by $E_{1}=(S^{*}_{1}, E^{*}_{1}, I^{*}_{1})$,  whenever
\begin{equation}\label{ch1.sec3.thm1.corrolary1.eq1}
 T_{1}+T_{2}\leq\frac{1}{\mu}\log{(G^{'}(0))}.
\end{equation}
\end{thm}
Proof:\\
See\cite{wanduku-biomath}.

It should be noted from Assumption~\ref{ch1.sec0.assum1} that the nonlinear function $G$ is  bounded.  Therefore, suppose
\begin{equation}\label{ch1.sec3.rem1.eqn1}
G^{*}=\sup_{x>0}{G(x)},
\end{equation}
 then it is easy to see that $0\leq  G(x)\leq G^{*},\forall x>0$. It follows further from Assumption~\ref{ch1.sec0.assum1} that given $\lim _{I\rightarrow \infty}{G(I)}=C$, if $G$ is strictly monotonic increasing then $G^{*}\leq C$. Also, if $G$ is strictly monotonic decreasing then $G^{*}\geq C$.

It easy to see that when the deterministic system (\ref{ch1.sec0.eq3})-(\ref{ch1.sec0.eq6}) is perturbed by the noise in the system from at least one of the different sources- natural  death or disease transmission rates, that is,  whenever at least one of $\sigma_{i}> 0,i=S,E,I,R,\beta$, then the nontrivial steady state $E_{1}=(S^{*}_{1}, E^{*}_{1}, I^{*}_{1})$ ceases to exist for the resulting perturbed system from (\ref{ch1.sec0.eq3})-(\ref{ch1.sec0.eq6}). It is important to understand the extend to which the  sample paths are deviated from the endemic steady state $E_{1}$, under the influence of the noises in the system.

The following lemma will be utilized to prove the results that characterize the asymptotic behavior of the sample paths of the stochastic system (\ref{ch1.sec0.eq8})-(\ref{ch1.sec0.eq11}) in the neighborhood of the nontrivial steady state $E_{1}=(S^{*}_{1}, E^{*}_{1}, I^{*}_{1})$,  whenever at least one of $\sigma_{i} \neq 0,i=S,E,I,R,\beta$.
\begin{lemma}\label{ch1.sec3.lemma1}
Let the hypothesis of Theorem~\ref{ch1.sec3.thm1} be satisfied and define the $\mathcal{C}^{2,1}-$ function $V:\mathbb{R}^{3}_{+}\times \mathbb{R}_{+}\rightarrow \mathbb{R}_{+}$ where
\begin{equation}\label{ch1.sec3.lemma1.eq1}
  V(t)=V_{1}(t)+V_{2}(t)+V_{3}(t) +V_{4}(t),
\end{equation}
where,
\begin{equation}\label{ch1.sec3.lemma1.eq2}
  V_{1}(t)=\frac{1}{2}\left(S(t)-S^{*}_{1}+E(t)-E^{*}_{1}+I(t)-I^{*}_{1}\right)^{2},
\end{equation}
\begin{equation}\label{ch1.sec3.lemma1.eq3}
  V_{2}(t)=\frac{1}{2}\left(S(t)-S^{*}_{1}\right)^{2}
\end{equation}
\begin{equation}\label{ch1.sec3.lemma1.eq4}
  V_{3}(t)=\frac{1}{2}\left(S(t)-S^{*}_{1}+E(t)-E^{*}_{1}\right)^{2}.
\end{equation}
 and
 \begin{eqnarray}
  V_{4}(t)&=&\frac{3}{2}\frac{\alpha}{\lambda(\mu)}\int_{t_{0}}^{\infty}f_{T_{3}}(r)e^{-2\mu r}dr(I(\theta)-I^{*})^{2}d\theta dr\nonumber\\
  &&+\frac{\beta S^{*}_{1}}{\lambda(\mu)}(G'(I^{*}_{1}))^{2} \int_{t_{0}}^{h_{2}}\int_{t_{0}}^{h_{1}}f_{T_{2}}(u)f_{T_{1}}(s)e^{-2\mu (s+u)}\int^{t}_{t-(s+u)}(I(\theta)-I^{*}_{1})^{2}d\theta dsdu\nonumber\\
  &&+[\frac{\beta \lambda(\mu)}{2} \int_{t_{0}}^{h_{2}}\int_{t_{0}}^{h_{1}}f_{T_{2}}(u)f_{T_{1}}(s)e^{-2\mu (s+u)}\int^{t}_{t-(s+u)}G^{2}(I(\theta))(S(\theta)-S^{*}_{1})^{2}d\theta dsdu\nonumber\\
  &&+\sigma^{2}_{\beta} \int_{t_{0}}^{h_{2}}\int_{t_{0}}^{h_{1}}f_{T_{2}}(u)f_{T_{1}}(s)e^{-2\mu (s+u)}\int^{t}_{t-(s+u)}G^{2}(I(\theta))(S(\theta)-S^{*}_{1})^{2}d\theta dsdu,\nonumber\\\label{ch1.sec3.lemma1.eq4b}
\end{eqnarray}
 where $\lambda(\mu)>0$ is a real valued function of $\mu$.
Suppose  $\tilde{\phi}_{1}$, $\tilde{\psi}_{1}$ and $\tilde{\varphi}_{1}$ are defined as follows
\begin{eqnarray}
\tilde{\phi}_{1}&=&3\mu-\left[2\mu\lambda{(\mu)}+(2\mu+d+\alpha)\frac{\lambda{(\mu)}}{2}+\alpha\lambda{(\mu)}+\frac{\beta S^{*}_{1}\lambda{(\mu)}}{2}+3\sigma ^{2}_{S}+\left(\frac{\beta (G^{*})^{2}}{2\lambda{(\mu)}}+\sigma^{2}_{\beta}(G^{*})^{2}\right)E(e^{-2\mu T_{1}})\right.\nonumber\\
&&\left.+\left(\frac{\beta \lambda{(\mu)}(G^{*})^{2}}{2}+\sigma^{2}_{\beta}(G^{*})^{2}\right)E(e^{-2\mu (T_{1}+T_{2})})\right]\label{ch1.sec3.lemma1.eq5a}\\
  \tilde{\psi}_{1} &=& 2\mu-\left[\frac{\beta }{2\lambda{(\mu)}}+\frac{\beta S^{*}_{1}\lambda{(\mu)}}{2}+ \frac{2\mu}{\lambda{(\mu)}}+(2\mu+d+\alpha)\frac{\lambda{(\mu)}}{2}+\alpha \lambda{(\mu)}+ 2 \sigma^{2}_{E} \right]\label{ch1.sec3.lemma1.eq5b}\\
    \tilde{\varphi}_{1}&=& (\mu + d+\alpha)-\left[(2\mu+d+\alpha)\frac{1}{\lambda{(\mu)}}+ \frac{\alpha\lambda{(\mu)}}{2} + \sigma^{2}_{I}+\frac{3\alpha}{2\lambda{(\mu)}}E(e^{-2\mu T_{3}})\right.\nonumber\\
    &&\left. +\left(\frac{\beta S^{*}_{1}(G'(I^{*}_{1}))^{2}}{\lambda{(\mu)}}
\right)E(e^{-2\mu (T_{1}+T_{2})})\right].\label{ch1.sec3.lemma1.eq5c}
   \end{eqnarray}
  The differential operator $dV$ applied to $V(t)$ with respect to the stochastic system (\ref{ch1.sec0.eq8})-(\ref{ch1.sec0.eq11}) can be written as follows:
 \begin{equation}\label{ch1.sec3.lemma1.eq5}
   dV=LV(t)dt + \overrightarrow{g}(S(t), E(t), I(t))d\overrightarrow{w(t)},
 \end{equation}
 where for $\overrightarrow{w(t)}=(w_{S},w_{E}, w_{I}, w_{\beta})^{T}$ and the  function $(S(t), E(t), I(t))\mapsto g(S(t), E(t), I(t))$, is defined as follows:
 \begin{eqnarray}
   &&\overrightarrow{g}(S(t), E(t), I(t))d\overrightarrow{w(t)}= -\sigma_{S}(3(S(t)-S^{*}_{1})+2(E(t)-E^{*}_{1})+I(t)-I^{*}_{1})S(t)dw_{S}(t)\nonumber\\
   &&-\sigma_{E}(2(S(t)-S^{*}_{1})+2(E(t)-E^{*}_{1})+I(t)-I^{*}_{1})E(t)dw_{E}(t)\nonumber\\
   &&-\sigma_{I}((S(t)-S^{*}_{1})+(E(t)-E^{*}_{1})+I(t)-I^{*}_{1})I(t)dw_{I}(t)\nonumber\\
   &&-\sigma_{\beta}(S(t)-S^{*}_{1})S(t)\int_{t_{0}}^{h_{1}}f_{T_{1}}(s)e^{-\mu s}G(I(t-s))dsdw_{\beta}(t)\nonumber\\
   &&-\sigma_{\beta}((S(t)-S^{*}_{1})+(E(t)-E^{*}_{1}))\int_{t_{0}}^{h_{2}}\int_{t_{0}}^{h_{1}}f_{T_{2}}(u)f_{T_{1}}(s)e^{-\mu (s+u)}S(t-u)G(I(t-s-u))dsdudw_{\beta}(t),\nonumber\\\label{ch1.sec3.lemma1.eq6}
 \end{eqnarray}
  and $LV$ satisfies the following inequality
  \begin{eqnarray}
    LV(t)&\leq& -\left\{\tilde{\phi}_{1}(S(t)-S^{*}_{1})^{2}+\tilde{\psi}_{1}(E(t)-E^{*}_{1})^{2}+\tilde{\varphi}_{1}(I(t)-I^{*}_{1})^{2}\right\}\nonumber\\
    &&+3\sigma^{2}_{S}(S^{*}_{1})^{2}+ 2\sigma^{2}_{E}(E^{*}_{1})^{2}+\sigma^{2}_{I}(I^{*}_{1})^{2}+\sigma^{2}_{\beta}(S^{*}_{1})^{2}(G^{*})^{2}E(e^{-2\mu (T_{1}+T_{2})})+\sigma^{2}_{\beta}(S^{*}_{1})^{2}(G^{*})^{2}E(e^{-\mu T_{1}}).\nonumber\\\label{ch1.sec3.lemma1.eq7}
  \end{eqnarray}
\end{lemma}
Proof\\
From (\ref{ch1.sec3.lemma1.eq2})-(\ref{ch1.sec3.lemma1.eq4}) the derivative of $V_{1}$, $V_{2}$ and $V_{3}$ with respect to the system (\ref{ch1.sec0.eq8})-(\ref{ch1.sec0.eq11}) can be written in the form:
\begin{eqnarray}
dV_{1}&=&LV_{1}dt -\sigma_{S}((S(t)-S^{*}_{1})+(E(t)-E^{*}_{1})+I(t)-I^{*}_{1})S(t)dw_{S}(t)\nonumber\\
   &&-\sigma_{E}((S(t)-S^{*}_{1})+(E(t)-E^{*}_{1})+I(t)-I^{*}_{1})E(t)dw_{E}(t)\nonumber\\
   &&-\sigma_{I}((S(t)-S^{*}_{1})+(E(t)-E^{*}_{1})+I(t)-I^{*}_{1})I(t)dw_{I}(t),\nonumber\\\label{ch1.sec3.lemma1.proof.eq1}
\end{eqnarray}
\begin{eqnarray}
dV_{2}&=&LV_{2}dt -\sigma_{S}((S(t)-S^{*}_{1}))S(t)dw_{S}(t)\nonumber\\
   &&-\sigma_{\beta}(S(t)-S^{*}_{1})S(t)\int_{t_{0}}^{h_{1}}f_{T_{1}}(s)e^{-\mu s}G(I(t-s))dsdw_{\beta}(t)\nonumber\\\label{ch1.sec3.lemma1.proof.eq2}
\end{eqnarray}
and
\begin{eqnarray}
dV_{3}&=&LV_{3}dt -\sigma_{S}((S(t)-S^{*}_{1})+(E(t)-E^{*}_{1}))S(t)dw_{S}(t)-\sigma_{S}((S(t)-S^{*}_{1})+(E(t)-E^{*}_{1}))E(t)dw_{E}(t)\nonumber\\
   &&-\sigma_{\beta}((S(t)-S^{*}_{1})+(E(t)-E^{*}_{1}))\int_{t_{0}}^{h_{2}}\int_{t_{0}}^{h_{1}}f_{T_{2}}(u)f_{T_{1}}(s)e^{-\mu (s+u)}S(t-u)G(I(t-s-u))dsdudw_{\beta}(t),\nonumber\\\label{ch1.sec3.lemma1.proof.eq3}
\end{eqnarray}
where utilizing (\ref{ch1.sec0.eq3})-(\ref{ch1.sec0.eq6}),   $LV_{1}$, $LV_{2}$ and $LV_{3}$ can be written as follows:
\begin{eqnarray}
LV_{1}(t)&=&-\mu (S(t)-S^{*}_{1})^{2}-\mu (E(t)-E^{*}_{1})^{2}-(\mu + d+ \alpha) (I(t)-I^{*}_{1})^{2}\nonumber\\
&&-2\mu (S(t)-S^{*}_{1})(E(t)-E^{*}_{1})-(2\mu + d+ \alpha) (S(t)-S^{*}_{1})(I(t)-I^{*}_{1})\nonumber\\
&&-(2\mu + d+ \alpha) (E(t)-E^{*}_{1})(I(t)-I^{*}_{1})\nonumber\\
&&+\alpha((S(t)-S^{*}_{1})+(E(t)-E^{*}_{1})+(I(t)-I^{*}_{1}))\int_{t_{0}}^{\infty}f_{T_{3}}(r)e^{-\mu r}(I(t-r)-I^{*}_{1})dr\nonumber\\
&&+\frac{1}{2}\sigma^{2}_{S}S^{2}(t)+\frac{1}{2}\sigma^{2}_{E}E^{2}(t)+\frac{1}{2}\sigma^{2}_{I}I^{2}(t),\label{ch1.sec3.lemma1.proof.eq4}
\end{eqnarray}
\begin{eqnarray}
LV_{2}(t)&=&-\mu (S(t)-S^{*}_{1})^{2}+\alpha(S(t)-S^{*}_{1})\int_{t_{0}}^{\infty}f_{T_{3}}(r)e^{-\mu r}(I(t-r)-I^{*}_{1})dr\nonumber\\
&&-\beta(S(t)-S^{*}_{1})^{2}\int_{t_{0}}^{h_{1}}f_{T_{1}}(s)e^{-\mu s}G(I(t-s))ds\nonumber\\
&&-\beta S^{*}_{1}(S(t)-S^{*}_{1})\int_{t_{0}}^{h_{1}}f_{T_{1}}(s)e^{-\mu s}(G(I(t-s))-G(I^{*}_{1}))ds\nonumber\\
&&+\frac{1}{2}\sigma^{2}_{S}S^{2}(t)+\frac{1}{2}\sigma^{2}_{\beta}S^{2}(t)\left(\int_{t_{0}}^{h_{1}}f_{T_{1}}(s)e^{-\mu s}G(I(t-s))ds\right)^{2},\label{ch1.sec3.lemma1.proof.eq5}
\end{eqnarray}
and
\begin{eqnarray}
LV_{3}(t)&=&-\mu (S(t)-S^{*}_{1})^{2}-\mu (E(t)-E^{*}_{1})^{2}-2\mu (S(t)-S^{*}_{1})(E(t)-E^{*}_{1})\nonumber\\
&&+\alpha((S(t)-S^{*}_{1})+(E(t)-E^{*}_{1}))\int_{t_{0}}^{\infty}f_{T_{3}}(r)e^{-\mu r}(I(t-r)-I^{*}_{1})dr\nonumber\\
&&-\beta(S(t)-S^{*}_{1})\int_{t_{0}}^{h_{2}}\int_{t_{0}}^{h_{1}}f_{T_{2}}(u)f_{T_{1}}(s)e^{-\mu (s+u)}(S(t-u)-S^{*}_{1})G(I(t-s-u))dsdu\nonumber\\
&&-\beta(E(t)-E^{*}_{1})\int_{t_{0}}^{h_{2}}\int_{t_{0}}^{h_{1}}f_{T_{2}}(u)f_{T_{1}}(s)e^{-\mu (s+u)}(S(t-u)-S^{*}_{1})G(I(t-s-u))dsdu\nonumber\\
&&-\beta S^{*}_{1}(S(t)-S^{*}_{1})\int_{t_{0}}^{h_{2}}\int_{t_{0}}^{h_{1}}f_{T_{2}}(u)f_{T_{1}}(s)e^{-\mu (s+u)}(G(I(t-s-u))-G(I^{*}_{1}))dsdu\nonumber\\
&&-\beta S^{*}_{1}(E(t)-E^{*}_{1})\int_{t_{0}}^{h_{2}}\int_{t_{0}}^{h_{1}}f_{T_{2}}(u)f_{T_{1}}(s)e^{-\mu (s+u)}(G(I(t-s-u))-G(I^{*}_{1}))dsdu\nonumber\\
&&+\frac{1}{2}\sigma^{2}_{S}S^{2}(t)+\frac{1}{2}\sigma^{2}_{E}E^{2}(t)\nonumber\\
&&+\frac{1}{2}\sigma^{2}_{\beta}\left(\int_{t_{0}}^{h_{2}}\int_{t_{0}}^{h_{1}}f_{T_{2}}(u)f_{T_{1}}(s)e^{-\mu (s+u)}S(t-u)G(I(t-s-u))dsdu\right)^{2}.\label{ch1.sec3.lemma1.proof.eq6}
\end{eqnarray}
 From (\ref{ch1.sec3.lemma1.proof.eq4})-(\ref{ch1.sec3.lemma1.proof.eq6}), the set of inequalities that follow will be used to estimate the sum $LV_{1}(t)+LV_{2}(t)+LV_{3}(t)$. That is,
  applying  $Cauchy-Swartz$ and  $H\ddot{o}lder$ inequalities,  and also applying the algebraic inequality 
\begin{equation}\label{ch2.sec2.thm2.proof.eq2*}
2ab\leq \frac{a^{2}}{g(c)}+b^{2}g(c)
\end{equation}
where $a,b,c\in \mathbb{R}$,  and the function $g$ is such that $g(c)> 0$,
 the terms associated with the integral term (sign)    $\int_{t_{0}}^{\infty}f_{T_{3}}(r)e^{-\mu r}(I(t-r)-I^{*}_{1})dr$ are estimated as follows:
 \begin{equation}\label{ch1.sec3.lemma1.proof.eq7}
  (a(t)-a^{*})\int_{t_{0}}^{\infty}f_{T_{3}}(r)e^{-\mu r}(I(t-r)-I^{*}_{1})dr\leq  \frac{\lambda(\mu)}{2}(a-a^{*})^{2} + \frac{1}{2\lambda(\mu)}\int_{t_{0}}^{\infty}f_{T_{3}}(r)e^{-2\mu r}(I(t-r)-I^{*}_{1})^{2}dr,
 \end{equation}
 where $a(t)\in \{S(t), E(t), I(t)\}$ and $a^{*}\in \{S^{*}_{1}, E^{*}_{1}, I^{*}_{1}\}$. Furthermore,  the terms with the integral sign that depend on $G(I(t-s))$ and $G(I(t-s-u))$ are estimated as follows:
 \begin{eqnarray}
 &&-\beta(S(t)-S^{*}_{1})^{2}\int_{t_{0}}^{h_{1}}f_{T_{1}}(s)e^{-\mu s}G(I(t-s))ds \leq  \frac{\beta\lambda(\mu)}{2}(S(t)-S^{*}_{1})^{2}\nonumber\\
&&+\frac{\beta}{2\lambda(\mu)}(S(t)-S^{*}_{1})^{2}\int_{t_{0}}^{h_{1}}f_{T_{1}}(s)e^{-2\mu s}G^{2}(I(t-s))ds.\nonumber\\
&& -\beta(E(t)-E^{*}_{1})\int_{t_{0}}^{h_{2}}\int_{t_{0}}^{h_{1}}f_{T_{2}}(u)f_{T_{1}}(s)e^{-\mu (s+u)}(S(t-u)-S^{*}_{1})G(I(t-s-u))dsdu\leq  \frac{\beta}{2\lambda(\mu)}(E(t)-E^{*}_{1})^{2} \nonumber\\
&&+\frac{\beta\lambda(\mu)}{2}\int_{t_{0}}^{h_{2}}\int_{t_{0}}^{h_{1}}f_{T_{2}}(u)f_{T_{1}}(s)e^{-2\mu (s+u)}(S(t-u)-S^{*}_{1})^{2}G^{2}(I(t-s-u))dsdu. \label{ch1.sec3.lemma1.proof.eq8}
 \end{eqnarray}
The terms with the integral sign that depend on $G(I(t-s))-G(I^{*}_{1})$ and $G(I(t-s-u))-G(I^{*}_{1})$ are estimated as follows:
\begin{eqnarray}
&&-\beta S^{*}_{1}(S(t)-S^{*}_{1})\int_{t_{0}}^{h_{1}}f_{T_{1}}(s)e^{-\mu s}(G(I(t-s))-G(I^{*}_{1}))ds\leq \frac{\beta S^{*}_{1}\lambda(\mu)}{2}(S(t)-S^{*}_{1})^{2}\nonumber\\
&& +\frac{\beta S^{*}_{1}}{2\lambda(\mu)}\int_{t_{0}}^{h_{1}}f_{T_{1}}(s)e^{-2\mu s}(I(t-s)-I^{*}_{1})^{2}\left(\frac{G(I(t-s))-G(I^{*}_{1})}{I(t-s)-I^{*}_{1}}\right)^{2}ds\nonumber\\
&&\leq \frac{\beta S^{*}_{1}\lambda(\mu)}{2}(S(t)-S^{*}_{1})^{2}\nonumber\\
&& +\frac{\beta S^{*}_{1}}{2\lambda(\mu)}\int_{t_{0}}^{h_{1}}f_{T_{1}}(s)e^{-2\mu s}(I(t-s)-I^{*}_{1})^{2}\left(G'(I^{*}_{1})\right)^{2}ds.\nonumber\\
&&-\beta S^{*}_{1}(E(t)-E^{*}_{1})\int_{t_{0}}^{h_{2}}\int_{t_{0}}^{h_{1}}f_{T_{2}}(u)f_{T_{1}}(s)e^{-\mu (s+u)}(G(I(t-s-u))-G(I^{*}_{1}))dsdu\leq \frac{\beta S^{*}_{1}\lambda(\mu)}{2}(E(t)-E^{*}_{1})^{2}\nonumber\\
&& +\frac{\beta S^{*}_{1}}{2\lambda(\mu)}\int_{t_{0}}^{h_{2}}\int_{t_{0}}^{h_{1}}f_{T_{2}}(u)f_{T_{1}}(s)e^{-2\mu s}(I(t-s-u)-I^{*}_{1})^{2}\left(\frac{G(I(t-s-u))-G(I^{*}_{1})}{I(t-s-u)-I^{*}_{1}}\right)^{2}ds\nonumber\\
&&\leq \frac{\beta S^{*}_{1}\lambda(\mu)}{2}(E(t)-E^{*}_{1})^{2}\nonumber\\
&& +\frac{\beta S^{*}_{1}}{2\lambda(\mu)}\int_{t_{0}}^{h_{2}}\int_{t_{0}}^{h_{1}}f_{T_{2}}(u)f_{T_{1}}(s)e^{-2\mu s}(I(t-s-u)-I^{*}_{1})^{2}\left(G'(I^{*}_{1})\right)^{2}ds,\nonumber\\\label{ch1.sec3.lemma1.proof.eq9}
\end{eqnarray}
where the inequality in (\ref{ch1.sec3.lemma1.proof.eq9}) follows from Assumption~\ref{ch1.sec0.assum1}. That is, $G$ is a differentiable monotonic function with $G''(I)<0$, and consequently,  $0< \frac{G(I)-G(I^{*}_{1})}{(I-I^{*}_{1})}\leq G'(I^{*}_{1}), \forall I>0$.

By employing the  $Cauchy-Swartz$ and $H\ddot{o}lder$ inequalities, and also applying the following algebraic inequality $(a+b)^{2}\leq 2a^{2}+ 2b^{2}$, the last set of terms with integral signs on (\ref{ch1.sec3.lemma1.proof.eq5})-(\ref{ch1.sec3.lemma1.proof.eq6}) are estimated as follows:
\begin{eqnarray}
&&\frac{1}{2}\sigma^{2}_{\beta}S^{2}(t)\left(\int_{t_{0}}^{h_{1}}f_{T_{1}}(s)e^{-\mu s}G(I(t-s))ds\right)^{2}\leq \sigma^{2}_{\beta}\left((S(t)-S^{*}_{1})^{2}+(S^{*}_{1})^{2}\right)\times\nonumber\\
&&\times\int_{t_{0}}^{h_{1}}f_{T_{1}}(s)e^{-2\mu s}G^{2}(I(t-s))ds.\nonumber\\
&&\frac{1}{2}\sigma^{2}_{\beta}\left(\int_{t_{0}}^{h_{2}}\int_{t_{0}}^{h_{1}}f_{T_{2}}(u)f_{T_{1}}(s)e^{-\mu (s+u)}S(t-u)G(I(t-s-u))dsdu\right)^{2}\leq \sigma^{2}_{\beta}\times\nonumber\\
&&\times\int_{t_{0}}^{h_{2}}\int_{t_{0}}^{h_{1}}f_{T_{2}}(u)f_{T_{1}}(s)e^{-2\mu (s+u)}\left((S(t-u)-S^{*}_{1})^{2}+(S^{*}_{1})^{2}\right)G^{2}(I(t-s-u))dsdu.\nonumber\\\label{ch1.sec3.lemma1.proof.eq10}
\end{eqnarray}
By further applying the algebraic inequality (\ref{ch2.sec2.thm2.proof.eq2*}) and the inequalities (\ref{ch1.sec3.lemma1.proof.eq7})-(\ref{ch1.sec3.lemma1.proof.eq10}) on the sum $LV_{1}(t)+LV_{2}(t)+LV_{3}(t)$, it is easy to see from (\ref{ch1.sec3.lemma1.proof.eq4})-(\ref{ch1.sec3.lemma1.proof.eq6}) that
\begin{eqnarray}
&&LV_{1}(t)+LV_{2}(t)+LV_{3}(t)\leq (S(t)-S^{*}_{1})^{2}\left[-3\mu +2\mu \lambda(\mu) +(2\mu+ d+\alpha)\frac{\lambda(\mu)}{2}+\alpha\lambda(\mu)\right.\nonumber\\
&&\left.+ \frac{\beta\lambda(\mu)}{2}+\frac{\beta S^{*}_{1}\lambda(\mu)}{2}+ \frac{\beta}{2\lambda(\mu)}(G^{*})^{2}E(e^{-2\mu T_{1}})+ 3\sigma^{2}_{S}+\sigma^{2}_{S} (G^{*})^{2}E(e^{-2\mu T_{1}})\right]\nonumber\\
&&(E(t)-E^{*}_{1})^{2}\left[-2\mu +\frac{2\mu }{\lambda(\mu)} +(2\mu+ d+\alpha)\frac{\lambda(\mu)}{2}+\alpha\lambda(\mu)+ \frac{\beta}{2\lambda(\mu)}+\frac{\beta S^{*}_{1}\lambda(\mu)}{2}+ 2\sigma^{2}_{E}\right]\nonumber\\
&&+(I(t)-I^{*}_{1})^{2}\left[-(\mu+d+\alpha)  +(2\mu+ d+\alpha)\frac{1}{\lambda(\mu)}+\frac{\alpha\lambda(\mu)}{2}+ \frac{\beta}{2\lambda(\mu)}+\frac{\beta S^{*}_{1}\lambda(\mu)}{2}+ \sigma^{2}_{I}\right]\nonumber\\
&&+\frac{3\alpha}{2\lambda(\mu)}\int_{t_{0}}^{\infty}f_{T_{3}}(r)e^{-2\mu r}(I(t-r)-I^{*}_{1})^{2}dr\nonumber\\
&&+\frac{\beta S^{*}_{1}}{\lambda(\mu)}(G'(I^{*}_{1}))^{2}\int_{t_{0}}^{h_{2}}\int_{t_{0}}^{h_{1}}f_{T_{2}}(u)f_{T_{1}}(s)e^{-2\mu (s+u)}(I(t-s-u)-I^{*}_{1})^{2}dsdu\nonumber\\
&&+\frac{\beta \lambda(\mu)}{2}\int_{t_{0}}^{h_{2}}\int_{t_{0}}^{h_{1}}f_{T_{2}}(u)f_{T_{1}}(s)e^{-2\mu (s+u)}G^{2}(I(t-s-u))(S(t-s)-S^{*}_{1})^{2}dsdu\nonumber\\
&&+3\sigma^{2}_{S}(S^{*}_{1})^{2}+2\sigma^{2}_{E}(E^{*}_{1})^{2}+ \sigma^{2}_{I}(I^{*}_{1})^{2}\nonumber\\
&&+\sigma^{2}_{\beta}(S^{*}_{1})^{2}\int_{t_{0}}^{h_{1}}f_{T_{1}}(r)e^{-2\mu }f_{T_{1}}(s)e^{-2\mu s}G^{2}(I(t-s))ds\nonumber\\
&&+\sigma^{2}_{\beta}\int_{t_{0}}^{h_{2}}\int_{t_{0}}^{h_{1}}f_{T_{2}}(u)f_{T_{1}}(s)e^{-2\mu (s+u)}G^{2}(I(t-s-u))(S(t-s-u)-S^{*}_{1})^{2}dsdu\nonumber\\\label{ch1.sec3.lemma1.proof.eq11}
\end{eqnarray}
But $V(t)=V_{1}(t)+V_{2}(t)+V_{3}(t)+V_{4}(t)$, therefore from (\ref{ch1.sec3.lemma1.proof.eq11}),  (\ref{ch1.sec3.lemma1.eq4b}) and (\ref{ch1.sec3.lemma1.proof.eq4})-(\ref{ch1.sec3.lemma1.proof.eq6}), the results in (\ref{ch1.sec3.lemma1.eq5})-(\ref{ch1.sec3.lemma1.eq7}) follow directly.

Theorems~[\ref{ch1.sec3.thm1}, \ref{ch1.sec3.thm1.corrolary1}] assert that the deterministic system (\ref{ch1.sec0.eq3})-(\ref{ch1.sec0.eq6}) has an endemic equilibrium denoted $E_{1}=(S^{*}_{1}, E^{*}_{1}, I^{*}_{1})$, whenever the basic reproduction numbers $R_{0}$ and $R^{*}_{0}$ for the disease in the absence of noise in the system satisfy $R_{0}>1$ and $R^{*}_{0}>1$, respectively. One common technique to obtain insight about the asymptotic behavior of the sample paths of the stochastic system (\ref{ch1.sec0.eq8})-(\ref{ch1.sec0.eq11}) near the potential  endemic equilibrium $E_{1}=(S^{*}_{1}, E^{*}_{1}, I^{*}_{1})$ for the stochastic system, is to characterize the long-term average distance of the paths of the stochastic system (\ref{ch1.sec0.eq8})-(\ref{ch1.sec0.eq11}) from the endemic equilibrium $E_{1}$.

Indeed, justification for this technique is the fact that for the second order stochastic solution process $\{X(t),t\geq t_{0}\}$ of the system (\ref{ch1.sec0.eq8})-(\ref{ch1.sec0.eq11}) defined in Theorem~\ref{ch1.sec1.thm1}, the long-term average distance of the sample paths from the endemic equilibrium $E_{1}$, denoted $\limsup_{t\rightarrow \infty}{\frac{1}{t}\int^{t}_{0}||X(s)-E_{1}||ds}$ estimates the long-term ensemble mean denoted\\
 $\limsup_{t\rightarrow \infty}{E||X(t)-E_{1}||}$, almost surely. Moreover, if the solution process $\{X(t),t\geq t_{0}\}$ of the system (\ref{ch1.sec0.eq8})-(\ref{ch1.sec0.eq11}) is stationary and ergodic, then the long-term ensemble mean $\limsup_{t\rightarrow \infty}{E||X(t)-E_{1}||}=E||X-E_{1}||$, where $X$ is the limit of convergence in distribution of the solution process $\{X(t),t\geq t_{0}\}$. That is, $\limsup_{t\rightarrow \infty}{\frac{1}{t}\int^{t}_{0}||X(s)-E_{1}||ds}=E||X-E_{1}||$, almost surely.  These stationary and ergodic properties of the solution process $\{X(t),t\geq t_{0}\}$ are discussed in details in Section~\ref{ch1.sec3.sec1}.

For convenience, the following notations are introduced and used in the rest of the results that follow in the subsequent sections.  Let $a_{1}(\mu, d, \alpha, \beta, B, \sigma ^{2}_{S}, \sigma^{2}_{\beta})$, $a_{2}(\mu, d, \alpha, \beta, B, \sigma ^{2}_{I})$,  $a_{3}(\mu, d, \alpha, \beta, B)$, and $a_{3}(\mu, d, \alpha, \beta, B, \sigma ^{2}_{E})$ represent the following set of parameters
\begin{equation}\label{ch1.sec3.lemma1.proof.eq13a}
\left\{
\begin{array}{lll}
a_{1}(\mu, d, \alpha, \beta, B, \sigma ^{2}_{S}, \sigma^{2}_{\beta})&=&2\mu\lambda{(\mu)}+(2\mu+d+\alpha)\frac{\lambda{(\mu)}}{2}+\alpha\lambda{(\mu)}+\frac{\beta S^{*}_{1}\lambda{(\mu)}}{2}+3\sigma ^{2}_{S}\\
&&+\left(\frac{\beta \lambda{(\mu)}(G^{*})^{2}}{2}+\sigma^{2}_{\beta}(G^{*})^{2}\right)\left(\frac{1}{G'(0)}\right)^{2}\\
a_{1}(\mu, d, \alpha, \beta, B)&=&2\mu\lambda{(\mu)}+(2\mu+d+\alpha)\frac{\lambda{(\mu)}}{2}+\alpha\lambda{(\mu)}+\frac{\beta S^{*}_{1}\lambda{(\mu)}}{2}\\
&&+\left(\frac{\beta \lambda{(\mu)}(G^{*})^{2}}{2}\right)\left(\frac{1}{G'(0)}\right)^{2}\\
a_{2}(\mu, d, \alpha, \beta, B, \sigma ^{2}_{I})&=&(2\mu+d+\alpha)\frac{1}{\lambda{(\mu)}}+ \frac{\alpha\lambda{(\mu)}}{2} + \sigma^{2}_{I}\\
&&+\left(\frac{\beta S^{*}_{1}(G'(I^{*}_{1}))^{2}}{\lambda{(\mu)}}
\right)\left(\frac{1}{G'(0)}\right)^{2}\\
a_{2}(\mu, d, \alpha, \beta, B)&=&(2\mu+d+\alpha)\frac{1}{\lambda{(\mu)}}+ \frac{\alpha\lambda{(\mu)}}{2}\\
&&+\left(\frac{\beta S^{*}_{1}(G'(I^{*}_{1}))^{2}}{\lambda{(\mu)}}
\right)\left(\frac{1}{G'(0)}\right)^{2}\\
a_{3}(\mu, d, \alpha, \beta, B, \sigma ^{2}_{E})&=&\frac{\beta }{2\lambda{(\mu)}}+\frac{\beta S^{*}_{1}\lambda{(\mu)}}{2}+ \frac{2\mu}{\lambda{(\mu)}}+(2\mu+d+\alpha)\frac{\lambda{(\mu)}}{2}+\alpha \lambda{(\mu)}+ 2 \sigma^{2}_{E}\\
a_{3}(\mu, d, \alpha, \beta, B)&=&\frac{\beta }{2\lambda{(\mu)}}+\frac{\beta S^{*}_{1}\lambda{(\mu)}}{2}+ \frac{2\mu}{\lambda{(\mu)}}+(2\mu+d+\alpha)\frac{\lambda{(\mu)}}{2}+\alpha \lambda{(\mu)}
\end{array}
\right.
\end{equation}
Also let $\tilde{a}_{1}(\mu, d, \alpha, \beta, B, \sigma ^{2}_{S}, \sigma^{2}_{\beta})$, $\tilde{a}_{1}(\mu, d, \alpha, \beta, B)$, $\tilde{a}_{2}(\mu, d, \alpha, \beta, B, \sigma ^{2}_{I})$, $\tilde{a}_{2}(\mu, d, \alpha, \beta, B)$  represent the following set of parameters
\begin{equation}\label{ch1.sec3.thm2.proof.eq4a}
\left\{
\begin{array}{lll}
\tilde{a}_{1}(\mu, d, \alpha, \beta, B,\sigma ^{2}_{S}, \sigma^{2}_{\beta})&=&2\mu\lambda{(\mu)}+(2\mu+d+\alpha)\frac{\lambda{(\mu)}}{2}+\alpha\lambda{(\mu)}+\frac{\beta S^{*}_{1}\lambda{(\mu)}}{2}+3\sigma ^{2}_{S}\\
&&+\left(\frac{\beta \lambda{(\mu)}(G^{*})^{2}}{2}+\sigma^{2}_{\beta}(G^{*})^{2}\right)+\left(\frac{\beta (G^{*})^{2}}{2\lambda{(\mu)}}+\sigma^{2}_{\beta}(G^{*})^{2}\right)\\
\tilde{a}_{1}(\mu, d, \alpha, \beta, B)&=&2\mu\lambda{(\mu)}+(2\mu+d+\alpha)\frac{\lambda{(\mu)}}{2}+\alpha\lambda{(\mu)}+\frac{\beta S^{*}_{1}\lambda{(\mu)}}{2}\\
&&+\left(\frac{\beta \lambda{(\mu)}(G^{*})^{2}}{2}\right)+\left(\frac{\beta (G^{*})^{2}}{2\lambda{(\mu)}}\right)\\
\tilde{a}_{2}(\mu, d, \alpha, \beta, B, \sigma^{2}_{I})&=&(2\mu+d+\alpha)\frac{1}{\lambda{(\mu)}}+ \frac{\alpha\lambda{(\mu)}}{2} + \sigma^{2}_{I}\\
&&+\left(\frac{\beta S^{*}_{1}(G'(I^{*}_{1}))^{2}}{\lambda{(\mu)}}
\right)+\frac{3\alpha}{2\lambda(\mu)},\\
\tilde{a}_{2}(\mu, d, \alpha, \beta, B)&=&(2\mu+d+\alpha)\frac{1}{\lambda{(\mu)}}+ \frac{\alpha\lambda{(\mu)}}{2} +\left(\frac{\beta S^{*}_{1}(G'(I^{*}_{1}))^{2}}{\lambda{(\mu)}}
\right)+\frac{3\alpha}{2\lambda(\mu)}.
\end{array}
\right.
\end{equation}
 The result in Theorem~\ref{ch1.sec3.thm2} characterizes the behavior of the sample paths of the stochastic system (\ref{ch1.sec0.eq8})-(\ref{ch1.sec0.eq11}) in the neighborhood of the nontrivial steady states $E_{1}=(S^{*}_{1}, E^{*}_{1}, I^{*}_{1})$ defined in Theorem~\ref{ch1.sec3.thm1.corrolary1}, whenever the incubation and natural immunity delay periods of the disease denoted by $T_{1}$,  $T_{2}$, and  $T_{3}$ are constant for all individuals in the population, and Theorem~\ref{ch1.sec1.thm1}[a.] holds. The following partial result from [\cite{maobook}, Theorem~3.4] called the strong law of large number for local martingales will be used to establish the result.
\begin{lemma}\label{ch1.sec3.thm2.lemma1.lemma1}
Let $M=\{M_{t}\}_{t\geq 0}$ be a real valued continuous local martingale vanishing at $t=0$. Then
\[\lim_{t\rightarrow \infty}{<M(t),M(t)>}=\infty, \quad a.s.\quad\Rightarrow\quad \lim_{t\rightarrow\infty} \frac{M(t)}{<M(t), M(t)>}=0,\quad a.s.\]
and also
\[\limsup_{t\rightarrow \infty}{\frac{<M(t),M(t)>}{t}}<\infty, \quad a.s.\quad\Rightarrow\quad \lim_{t\rightarrow\infty} \frac{M(t)}{t}=0,\quad a.s.\]
\end{lemma}
The notation $<M(t),M(t)>$ is used to denote the quadratic variation of the local martingale $M=\{M(t),\forall t\geq t_{0}\}$.

Recall, the assumptions that $T_{1}, T_{2}$ and $T_{3}$ are constant, is also equivalent to the special case of letting the probability density functions of $T_{1}, T_{2}$ and $T_{3}$ to be the dirac-delta function defined in (\ref{ch1.sec2.eq4}).
Moreover, under the assumption that $T_{1}\geq 0, T_{2}\geq 0$ and $T_{3}\geq 0$ are constant, it follows  from (\ref{ch1.sec3.lemma1.eq5a})-(\ref{ch1.sec3.lemma1.eq5c}),  that  $E(e^{-2\mu (T_{1}+T_{2})})=e^{-2\mu (T_{1}+T_{2})} $, $E(e^{-2\mu T_{1}})=e^{-2\mu T_{1}} $ and $E(e^{-2\mu T_{3}})=e^{-2\mu T_{3}} $.
\begin{thm}\label{ch1.sec3.thm2}
Let the hypotheses of Theorem~\ref{ch1.sec1.thm1}[a.], Theorem~\ref{ch1.sec3.thm1.corrolary1} and  Lemma~\ref{ch1.sec3.lemma1} be satisfied and let
 \begin{eqnarray}
 \mu>\max{ \left(\frac{1}{3}a_{1}(\mu, d, \alpha, \beta, B, \sigma^{2}_{S}=0,\sigma^{2}_{\beta}),\frac{1}{2}a_{3}(\mu, d, \alpha, \beta, B)\right)},\quad and\nonumber\\
 (\mu+d+\alpha)>a_{2}(\mu, d, \alpha, \beta, B ).\label{ch1.sec3.thm2.eq1}
 \end{eqnarray}
 Also let the delay times $T_{1}, T_{2}$ and $T_{3}$ be constant, that is, the probability density functions of $T_{1}, T_{2}$ and $T_{3}$ respectively denoted by $f_{T_{i}}, i=1, 2, 3$ are the  dirac-delta functions defined in (\ref{ch1.sec2.eq4}). Furthermore, let the constants $T_{1}, T_{2}$ and $T_{3}$ satisfy the following set of inequalities:
\begin{equation}\label{ch1.sec3.thm2.eq2}
T_{1}>\frac{1}{2\mu}\log{\left(\frac{\left(\frac{\beta (G^{*})^{2}}{2\lambda{(\mu)}}+\sigma^{2}_{\beta}(G^{*})^{2}\right)}{(3\mu-a_{1}(\mu, d, \alpha, \beta, B, \sigma^{2}_{S}=0,\sigma^{2}_{\beta}))}\right)},
\end{equation}
\begin{equation}\label{ch1.sec3.thm2.eq3}
T_{2}<\frac{1}{2\mu}\log{\left(\frac{(3\mu-a_{1}(\mu, d, \alpha, \beta, B, \sigma^{2}_{S}=0,\sigma^{2}_{\beta}))}{\left(\frac{\beta (G^{*})^{2}}{2\lambda{(\mu)}}+\sigma^{2}_{\beta}(G^{*})^{2}\right)\left(\frac{1}{G'(0)}\right)^{2}}\right)},
\end{equation}
and
\begin{equation}\label{ch1.sec3.thm2.eq3b}
T_{3}>\frac{1}{2\mu}\log{\left(\frac{\frac{3\alpha}{2\lambda(\mu)}}{(\mu+d+\alpha)-a_{2}(\mu, d, \alpha, \beta, B)}\right)}.
\end{equation}
There exists a positive real number $\mathfrak{m}_{1}>0$, such that
\begin{eqnarray}
  &&\limsup_{t\rightarrow \infty}\frac{1}{t}\int^{t}_{0}\left[ ||X(v)-E_{1}||_{2}\right]^{2}dv\nonumber\\
  &&\leq \frac{\sigma^{2}_{\beta}(S^{*}_{1})^{2}(G^{*})^{2}e^{-2\mu (T_{1}+T_{2})}+\sigma^{2}_{\beta}(S^{*}_{1})^{2}(G^{*})^{2}e^{-\mu T_{1}}}{\mathfrak{m}_{1}},\nonumber\\
  \label{ch2.sec3.thm2.eq4}
\end{eqnarray}
almost surely, where $X(t)$ is defined in (\ref{ch1.sec0.eq13b}), and $||.||_{2}$ is the natural Euclidean norm in $\mathbb{R}^{2}$.
\end{thm}
Proof:\\
From Lemma~\ref{ch1.sec3.lemma1},  (\ref{ch1.sec3.lemma1.eq5a})-(\ref{ch1.sec3.lemma1.eq5c}), it is easy to see that under the assumptions in (\ref{ch1.sec3.thm1.corrolary1.eq1}) and (\ref{ch1.sec3.thm2.eq1})-(\ref{ch1.sec3.thm2.eq3b}), then $\tilde{\phi}_{1}>0$, $\tilde{\psi}_{1}>0$ and $\tilde{\varphi}_{1}>0$. Therefore,  from (\ref{ch1.sec3.lemma1.eq5})-(\ref{ch1.sec3.lemma1.eq7}) it is also easy to see that
\begin{eqnarray}
   dV&=&LV(t)dt + \overrightarrow{g}(S(t), E(t), I(t))d\overrightarrow{w(t)},\nonumber\\
   &\leq&-\min\{\tilde{\phi}_{1}, \tilde{\psi}_{1}, \tilde{\varphi}_{1}\}\left[ (S(t)-S^{*}_{1})^{2}+ (E(t)-E^{*}_{1})^{2}+  (I(t)-I^{*}_{1})^{2}\right]\nonumber\\
   && +3\sigma^{2}_{S}(S^{*}_{1})^{2}+ 2\sigma^{2}_{E}(E^{*}_{1})^{2}+\sigma^{2}_{I}(I^{*}_{1})^{2}+\sigma^{2}_{\beta}(S^{*}_{1})^{2}(G^{*})^{2}e^{-2\mu (T_{1}+T_{2})}+\sigma^{2}_{\beta}(S^{*}_{1})^{2}(G^{*})^{2}e^{-\mu T_{1}}\nonumber\\
   &&+ \overrightarrow{g}(S(t), E(t), I(t))d\overrightarrow{w(t)},\label{ch1.sec3.thm2.proof.eq1}
 \end{eqnarray}
 Integrating both sides of (\ref{ch1.sec3.thm2.proof.eq1}) from 0 to $t$, it follows that
 \begin{eqnarray}
 &&(V(t)-V(0))\leq -\mathfrak{m_{1}}\int^{t}_{0}\left[ (S(v)-S^{*}_{1})^{2}+ (E(v)-E^{*}_{1})^{2}+  (I(v)-I^{*}_{1})^{2}\right]dv\nonumber\\
   && +\left(3\sigma^{2}_{S}(S^{*}_{1})^{2}+ 2\sigma^{2}_{E}(E^{*}_{1})^{2}+\sigma^{2}_{I}(I^{*}_{1})^{2}+\sigma^{2}_{\beta}(S^{*}_{1})^{2}(G^{*})^{2}e^{-2\mu (T_{1}+T_{2})}+\sigma^{2}_{\beta}(S^{*}_{1})^{2}(G^{*})^{2}e^{-\mu T_{1}}\right)t,\nonumber\\
   &&+\int^{t}_{0}\overrightarrow{g}(S(v), E(v), I(v))d\overrightarrow{w(v)},\label{ch2.sec3.thm2.proof.eq2}
\end{eqnarray}
where $V(0)\geq 0$  and
\begin{equation}\label{ch2.sec3.thm2.proof.eq3}
\mathfrak{m}_{1}=min(\tilde{\phi},\tilde{ \psi},\tilde{\varphi})>0.
\end{equation}
are constants and
\begin{eqnarray}
&&\overrightarrow{g}(S(t), E(t), I(t))d\overrightarrow{w(t)}=
   -\sigma_{\beta}(S(t)-S^{*}_{1})S(t)e^{-\mu T_{1}}G(I(t-T_{1}))dw_{\beta}(t)\nonumber\\
   &&-\sigma_{\beta}((S(t)-S^{*}_{1})+(E(t)-E^{*}_{1}))e^{-\mu (T_{1}+T_{2})}S(t-T_{2})G(I(t-T_{1}-T_{2}))dw_{\beta}(t),\nonumber\\\label{ch2.sec3.thm2.proof.eq4}
 \end{eqnarray}
 since Theorem~\ref{ch1.sec1.thm1}[a.] holds and $T_{1}$ and $T_{2}$ are constants. Now, define
 \begin{eqnarray}
 N_{1}(t)=-\int^{t}_{0}\sigma_{\beta}(S(v)-S^{*}_{1})S(v)e^{-\mu T_{1}}G(I(v-T_{1}))dw_{\beta}(v),\quad and\quad \nonumber\\ N_{2}(t)=-\int^{t}_{0}\sigma_{\beta}((S(v)-S^{*}_{1})+(E(v)-E^{*}_{1}))e^{-\mu (T_{1}+T_{2})}S(v-T_{2})G(I(v-T_{1}-T_{2}))dw_{\beta}(v).\nonumber\\
 \label{ch2.sec3.thm2.proof.eq5}
 \end{eqnarray}
 Also,  the quadratic variations of $N_{1}(t)$ and $N_{2}(t)$  in (\ref{ch2.sec3.thm2.proof.eq5}) are given by
 \begin{eqnarray}
 <N(t)_{1}(t), N_{1}(t)>&=&\int^{t}_{0}\sigma^{2}_{\beta}(S(v)-S^{*}_{1})^{2}S^{2}(v)e^{-2\mu T_{1}}G^{2}(I(v-T_{1}))dv,\nonumber\\
 <N(t)_{2}(t), N_{2}(t)>&=&\int^{t}_{0}\sigma^{2}_{\beta}((S(v)-S^{*}_{1})+(E(v)-E^{*}_{1}))^{2}e^{-2\mu (T_{1}+T_{2})}S^{2}(v-T_{2})G^{2}(I(v-T_{1}-T_{2}))dv.\nonumber\\\label{ch2.sec3.thm2.proof.eq6}
 \end{eqnarray}
It follows that when Theorem~\ref{ch1.sec1.thm1}[a.] holds, then from Assumption~\ref{ch1.sec0.assum1} and (\ref{ch2.sec3.thm2.proof.eq6}),
\begin{eqnarray}
<N(t)_{1}, N_{1}(t)>&\leq& \int^{t}_{0}\sigma^{2}_{\beta}\left(\frac{\beta}{\mu}+S^{*}_{1}\right)^{2}\left(\frac{\beta}{\mu}\right)^{2}e^{-2\mu T_{1}}\left(\frac{\beta}{\mu}\right)^{2}dv\nonumber\\
&=&\sigma^{2}_{\beta}\left(\frac{\beta}{\mu}+S^{*}_{1}\right)^{2}\left(\frac{\beta}{\mu}\right)^{4}e^{-2\mu T_{1}}t.\label{ch2.sec3.thm2.proof.eq7}
\end{eqnarray}
From (\ref{ch2.sec3.thm2.proof.eq7}), it is easy to see that $\limsup_{t\rightarrow \infty }{\frac{1}{t}<N(t)_{1}, N_{1}(t)>}<\infty$, a.s.. Therefore by the strong law of large numbers for local martingales in Lemma~\ref{ch1.sec3.thm2.lemma1.lemma1}, it follows that $\limsup_{t\rightarrow \infty }\frac{1}{t}N_{1}(t)=0$, a.s. By the same reasoning, it can be shown that $\limsup_{t\rightarrow \infty }\frac{1}{t}N_{2}(t)=0$, a.s. Moreover, from (\ref{ch2.sec3.thm2.proof.eq4}), it can be seen that $\limsup_{t\rightarrow \infty }\int^{t}_{0}\overrightarrow{g}(S(v), E(v), I(v))d\overrightarrow{w(v)}=0$, a.s.
 Hence, diving both sides of (\ref{ch2.sec3.thm2.proof.eq2}) by $t$ and $\mathfrak{m}_{1}$, and taking the $\limsup_{t\rightarrow \infty}$, then (\ref{ch2.sec3.thm2.eq4}) follows directly.
\begin{rem}\label{ch2.sec3.thm2.rem1}
Theorem~\ref{ch1.sec3.thm2} asserts that when the basic reproduction number $R^{*}_{0}$ defined in (\ref{ch1.sec2.lemma2a.corrolary1.eq4}) satisfies $R^{*}_{0}>1$, and  the disease dynamics is perturbed by random fluctuations exclusively in the disease transmission rate, that is, the  intensity $\sigma_{\beta}>0$, it is noted earlier that the stochastic system (\ref{ch1.sec0.eq8})-(\ref{ch1.sec0.eq11}) does not have an endemic equilibrium state. Nevertheless, the conditions in Theorem~\ref{ch1.sec3.thm2} provide estimates for the constant delay times $T_{1}, T_{2}$ and $T_{3}$ in (\ref{ch1.sec3.thm2.eq2})-(\ref{ch1.sec3.thm2.eq3b}) in addition to other parametric restrictions in (\ref{ch1.sec3.thm2.eq1}) that are sufficient for the solution paths of the perturbed stochastic system (\ref{ch1.sec0.eq8})-(\ref{ch1.sec0.eq11}) to oscillate near the nontrivial steady state, $E_{1}$, of the deterministic system (\ref{ch1.sec0.eq3})-(\ref{ch1.sec0.eq6}) found in Theorem~\ref{ch1.sec3.thm1.corrolary1}. The result in (\ref{ch2.sec3.thm2.eq4}) estimates the average distance between the sample paths of the stochastic system (\ref{ch1.sec0.eq8})-(\ref{ch1.sec0.eq11}), and the nontrivial steady state $E_{1}$. Moreover, (\ref{ch2.sec3.thm2.eq4}) depicts the size of the oscillations of the paths of the stochastic system relative to $E_{1}$, where smaller values for the intensity ( $\sigma_{\beta}> 0$) lead to oscillations of the paths in close proximity to the steady state $E_{1}$, and vice versa.

In a biological context, the result of this theorem signifies that when the basic reproduction number exceeds one, and the other parametric restrictions in (\ref{ch1.sec3.thm2.eq1})-(\ref{ch1.sec3.thm2.eq3b}) are satisfied,  then the disease related classes ($E, I, R$), and consequently the disease in a whole will persist in state near the endemic equilibrium state $E_{1}$. Moreover, stronger noise in the system from the disease transmission rate of the disease tends to persist the disease in state further away from the endemic equilibrium state $E_{1}$, and vice versa. Nevertheless, in this case of variability exclusively from the disease transmission rate, the numerical simulation results in Example~\ref{ch1.sec4.subsec1.1} suggest that continuous decrease in size of some of the subpopulation classes- susceptible, exposed, infectious and removal may occur, as the intensity of the noise from the disease transmission rate increases, but there is no definite indication of extinction of the entire human population over time. Note that, comparing to the simulation results in Example~\ref{ch1.sec4.subsec1.2}, there is some evidence that the strength of the noise from the disease transmission rate persists the disease, but not to the point of extinction of the entire population (or at least not at the same rate as the case of the strength of the noises from the natural deathrates).

These observations in Example~\ref{ch1.sec4.subsec1.1} also support the remark for Theorem~\ref{ch1.sec1.thm1}[a] that the sample paths for the stochastic system exhibit non-complex behaviors such as extinction of the entire human population, when the disease dynamics is perturbed exclusively by noise from the disease transmission rate, compared to the complex behavior observed when the disease dynamics is perturbed by noise from the natural death rates.

 These facts suggest that malaria control policies in the event where the disease is persistent, should focus on reducing the intensity of the fluctuations in the disease transmission  rate, perhaps through vector control and better care of the people in the population to keep the transmission rate constant, in order to reduce the number of malaria cases which lead to the persistence of the disease.
\end{rem}
The subsequent result provides more general conditions irrespective of the probability distribution of the random variables $T_{1}, T_{2}$ and $T_{3}$, that are sufficient for the trajectories of the stochastic system (\ref{ch1.sec0.eq8})-(\ref{ch1.sec0.eq11}) to oscillate near the nontrivial steady state $E_{1}$ of the deterministic system (\ref{ch1.sec0.eq3})-(\ref{ch1.sec0.eq6}), whenever the intensities of the gaussian noises  in the system are positive, that is whenever $\sigma_{i}>0, i=S, E, I, R, \beta$. The following result from [Lemma 2.2  \cite{maobook}] and [lemma 15,\cite{Li2008}] will be used to achieve this result.
\begin{lemma}\label{ch1.sec3.thm3.lemma1}
Let $M(t); t\geq 0$ be a continuous local martingale with initial value $M(0)=0$. Let $<M(t),M(t)>$ be its quadratic  variation. Let $\delta >1$ be a number, and let $\nu_{k}$ and $\tau_{k}$ be two sequences of positive  numbers. Then, for almost all $w\in\Omega$, there exists a random integer $k_{0}=k_{0}(w)$ such that, for all $k\geq k_{0}$,
\[M(t)\leq \frac{1}{2}\nu_{k}<M(t), M(t)>+\frac{\delta ln k}{\nu_{k}}, 0\leq t\leq \tau_{k}.\]
\end{lemma}
Proof:\\
See \cite{maobook, Li2008}.
\begin{thm}\label{ch2.sec3.thm3}
Suppose the hypotheses of Theorem~\ref{ch1.sec1.thm1}[b.], Theorem~\ref{ch1.sec3.thm1} and  Lemma~\ref{ch1.sec3.lemma1} are satisfied, and let
 \begin{equation}\label{ch1.sec3.thm3.eq1}
 \mu> \max\left\{\frac{1}{3}\tilde{a}_{1}(\mu, d, \alpha, \beta, B, \sigma ^{2}_{S}, \sigma^{2}_{\beta}),\frac{1}{2}a_{3}(\mu, d, \alpha, \beta, B, \sigma ^{2}_{E})\right\}\quad and\quad(\mu+d+\alpha)>\tilde{a}_{2}(\mu, d, \alpha, \beta, B, \sigma ^{2}_{I}).
 \end{equation}
  It follows that for any arbitrary probability distribution of the delay times: $T_{1}, T_{2}$ and $T_{3}$,
there exists a positive real number $\mathfrak{m}_{2}>0$ such that
\begin{eqnarray}
  &&\limsup_{t\rightarrow \infty}\frac{1}{t}\int^{t}_{0}\left[ (S(v)-S^{*}_{1})^{2}+ (E(v)-E^{*}_{1})^{2}+  (I(v)-I^{*}_{1})^{2}\right]dv\nonumber\\
  &&\leq \frac{3\sigma^{2}_{S}(S^{*}_{1})^{2}+ 2\sigma^{2}_{E}(E^{*}_{1})^{2}+\sigma^{2}_{I}(I^{*}_{1})^{2}+\sigma^{2}_{\beta}(S^{*}_{1})^{2}(G^{*})^{2}E(e^{-2\mu (T_{1}+T_{2})})+\sigma^{2}_{\beta}(S^{*}_{1})^{2}(G^{*})^{2}E(e^{-\mu T_{1}})}{\mathfrak{m}_{2}},\nonumber\\
  \label{ch2.sec3.thm3.eq2}
\end{eqnarray}
almost surely.
\end{thm}
Proof:\\
From Lemma~\ref{ch1.sec3.lemma1},  (\ref{ch1.sec3.lemma1.eq5a})-(\ref{ch1.sec3.lemma1.eq5c}),
 \begin{eqnarray}
-\tilde{\phi}_{1}&=&-3\mu+\left[2\mu\lambda{(\mu)}+(2\mu+d+\alpha)\frac{\lambda{(\mu)}}{2}+\alpha\lambda{(\mu)}+\frac{\beta S^{*}_{1}\lambda{(\mu)}}{2}+3\sigma ^{2}_{S}+\left(\frac{\beta (G^{*})^{2}}{2\lambda{(\mu)}}+\sigma^{2}_{\beta}(G^{*})^{2}\right)E(e^{-2\mu T_{1}})\right.\nonumber\\
&&\left.+\left(\frac{\beta \lambda{(\mu)}(G^{*})^{2}}{2}+\sigma^{2}_{\beta}(G^{*})^{2}\right)E(e^{-2\mu (T_{1}+T_{2})})\right]\nonumber\\
&&\leq -3\mu+\left[2\mu\lambda{(\mu)}+(2\mu+d+\alpha)\frac{\lambda{(\mu)}}{2}+\alpha\lambda{(\mu)}+\frac{\beta S^{*}_{1}\lambda{(\mu)}}{2}+3\sigma ^{2}_{S}+\left(\frac{\beta (G^{*})^{2}}{2\lambda{(\mu)}}+\sigma^{2}_{\beta}(G^{*})^{2}\right)\right.\nonumber\\
&&\left.+\left(\frac{\beta \lambda{(\mu)}(G^{*})^{2}}{2}+\sigma^{2}_{\beta}(G^{*})^{2}\right)\right]\nonumber\\
&&=-\left(3\mu-\tilde{a}_{1}(\mu, d, \alpha, \beta, B, \sigma ^{2}_{S}, \sigma^{2}_{\beta})\right)\label{ch1.sec3.thm3.proof.eq1}\\
  -\tilde{\psi}_{1} &=& -2\mu+\left[\frac{\beta }{2\lambda{(\mu)}}+\frac{\beta S^{*}_{1}\lambda{(\mu)}}{2}+ \frac{2\mu}{\lambda{(\mu)}}+(2\mu+d+\alpha)\frac{\lambda{(\mu)}}{2}+\alpha \lambda{(\mu)}+ 2 \sigma^{2}_{E} \right]\nonumber\\
  &&=-\left(2\mu-a_{3}(\mu, d, \alpha, \beta, B, \sigma ^{2}_{E})\right)\label{ch1.sec3.thm3.proof.eq2}\\
    -\tilde{\varphi}_{1}&=& -(\mu + d+\alpha)+\left[(2\mu+d+\alpha)\frac{1}{\lambda{(\mu)}}+ \frac{\alpha\lambda{(\mu)}}{2} + \sigma^{2}_{I}+\frac{3\alpha}{2\lambda{(\mu)}}E(e^{-2\mu T_{3}})\right.\nonumber\\
    &&\left. +\left(\frac{\beta S^{*}_{1}(G'(I^{*}_{1}))^{2}}{\lambda{(\mu)}}
\right)E(e^{-2\mu (T_{1}+T_{2})})\right]\nonumber\\
&\leq& -(\mu + d+\alpha)+\left[(2\mu+d+\alpha)\frac{1}{\lambda{(\mu)}}+ \frac{\alpha\lambda{(\mu)}}{2} + \sigma^{2}_{I}+\frac{3\alpha}{2\lambda{(\mu)}}\right.\nonumber\\
    &&\left. +\left(\frac{\beta S^{*}_{1}(G'(I^{*}_{1}))^{2}}{\lambda{(\mu)}}
\right)\right]\nonumber\\
&&=-\left((\mu + d+\alpha)-\tilde{a}_{2}(\mu, d, \alpha, \beta, B, \sigma ^{2}_{I})\right),\label{ch1.sec3.thm3.proof.eq3}
   \end{eqnarray}
since $0<E(e^{-2\mu (T_{i})})\leq 1, i=1, 2,3$. It follows from (\ref {ch1.sec3.lemma1.eq5})-(\ref{ch1.sec3.lemma1.eq7}) and (\ref{ch1.sec3.thm3.proof.eq1})-(\ref{ch1.sec3.thm3.proof.eq3}) that
\begin{eqnarray}
   dV&=&LV(t)dt + \overrightarrow{g}(S(t), E(t), I(t))d\overrightarrow{w(t)},\nonumber\\
   &\leq&-\min\{\left(3\mu-\tilde{a}_{1}(\mu, d, \alpha, \beta, B)\right), \left(2\mu-a_{3}(\mu, d, \alpha, \beta, B)\right), \left((\mu + d+\alpha)-\tilde{a}_{2}(\mu, d, \alpha, \beta, B)\right)\}\times\nonumber\\
   &&\times \left[ (S(t)-S^{*}_{1})^{2}+ (E(t)-E^{*}_{1})^{2}+  (I(t)-I^{*}_{1})^{2}\right]\nonumber\\
   && +3\sigma^{2}_{S}(S^{*}_{1})^{2}+ 2\sigma^{2}_{E}(E^{*}_{1})^{2}+\sigma^{2}_{I}(I^{*}_{1})^{2}+\sigma^{2}_{\beta}(S^{*}_{1})^{2}(G^{*})^{2}E(e^{-2\mu (T_{1}+T_{2})})+\sigma^{2}_{\beta}(S^{*}_{1})^{2}(G^{*})^{2}E(e^{-\mu T_{1}})\nonumber\\
   &&+ \overrightarrow{g}(S(t), E(t), I(t))d\overrightarrow{w(t)},\label{ch1.sec3.thm3.proof.eq4}
 \end{eqnarray}
 where under the assumptions (\ref{ch1.sec3.thm3.eq1}) in the hypothesis,
\begin{equation}\label{ch2.sec3.thm3.proof.eq5}
\mathfrak{m}_{2}=\min\{\left(3\mu-\tilde{a}_{1}(\mu, d, \alpha, \beta, B)\right), \left(2\mu-a_{3}(\mu, d, \alpha, \beta, B)\right), \left((\mu + d+\alpha)-\tilde{a}_{2}(\mu, d, \alpha, \beta, B)\right)\}>0.
\end{equation}
Integrating both sides of (\ref{ch1.sec3.thm3.proof.eq4}) from 0 to $t$, it follows that
 \begin{eqnarray}
 &&(V(t)-V(0))\leq -\mathfrak{m_{2}}\int^{t}_{0}\left[ (S(v)-S^{*}_{1})^{2}+ (E(v)-E^{*}_{1})^{2}+  (I(v)-I^{*}_{1})^{2}\right]dv\nonumber\\
   && +\left(3\sigma^{2}_{S}(S^{*}_{1})^{2}+ 2\sigma^{2}_{E}(E^{*}_{1})^{2}+\sigma^{2}_{I}(I^{*}_{1})^{2}+\sigma^{2}_{\beta}(S^{*}_{1})^{2}(G^{*})^{2}E(e^{-2\mu (T_{1}+T_{2})})+\sigma^{2}_{\beta}(S^{*}_{1})^{2}(G^{*})^{2}E(e^{-\mu T_{1}})\right)t\nonumber\\
  &&+M(t), \label{ch2.sec3.thm3.proof.eq6}
\end{eqnarray}
where $V(0)$ is constant and from (\ref{ch1.sec3.lemma1.eq6}), the function $M(t)$ is defined as follows:-
  \begin{eqnarray}
  M(t)&=&\int^{t}_{0}\overrightarrow{g}(S(v), E(v), I(v))d\overrightarrow{w(v)}\nonumber\\
  &=&\sum^{3}_{i=1}M^{S}_{i}(t)+\sum^{3}_{i=1}M^{E}_{i}(t)+\sum^{3}_{i=1}M^{I}_{i}(t)+\sum^{3}_{i=1}M^{\beta}_{i}(t),\nonumber\\
  \label{ch2.sec3.thm3.proof.eq7}
  \end{eqnarray}
  and
  \begin{eqnarray}
  M^{S}_{1}(t)=-\int^{t}_{0}3\sigma_{S}(S(v)-S^{*}_{1})S(v)dw_{S}(v),\nonumber\\
  M^{S}_{2}(t)=-\int^{t}_{0}2\sigma_{S}(E(v)-E^{*}_{1})S(v)dw_{S}(v),\nonumber\\
  M^{S}_{3}(t)=-\int^{t}_{0}\sigma_{S}(I(v)-I^{*}_{1})S(v)dw_{S}(v).\nonumber\\
  \label{ch2.sec3.thm3.proof.eq8}
  \end{eqnarray}
  Also,
  \begin{eqnarray}
  M^{E}_{1}(t)=-\int^{t}_{0}2\sigma_{E}(S(v)-S^{*}_{1})E(v)dw_{E}(v),\nonumber\\
 M^{E}_{2}(t)=-\int^{t}_{0}2\sigma_{E}(E(v)-E^{*}_{1})E(v)dw_{E}(v),\nonumber\\
  M^{E}_{3}(t)=-\int^{t}_{0}\sigma_{E}(I(v)-I^{*}_{1})E(v)dw_{E}(v),\nonumber\\
  \label{ch2.sec3.thm3.proof.eq9}
  \end{eqnarray}
  and
  \begin{eqnarray}
  M^{I}_{1}(t)=-\int^{t}_{0}2\sigma_{I}(S(v)-S^{*}_{1})I(v)dw_{I}(v),\nonumber\\
 M^{I}_{2}(t)=-\int^{t}_{0}2\sigma_{I}(E(v)-E^{*}_{1})I(v)dw_{I}(v),\nonumber\\
  M^{I}_{3}(t)=-\int^{t}_{0}\sigma_{I}(I(v)-I^{*}_{1})I(v)dw_{I}(v).\nonumber\\
  \label{ch2.sec3.thm3.proof.eq10}
  \end{eqnarray}
  Furthermore,
  \begin{eqnarray}
  M^{\beta}_{1}(t)=-\int^{t}_{0}\sigma_{\beta}(S(v)-S^{*}_{1})S(v)E(e^{-\mu T_{1}}G(I(v-T_{1})))dw_{\beta}(v),\nonumber\\
 M^{\beta}_{2}(t)=-\int^{t}_{0}\sigma_{\beta}(S(v)-S^{*}_{1})E(S(v-T_{2})e^{-\mu (T_{1}+T_{2})}G(I(v-T_{1}-T_{2})))dw_{\beta}(v),\nonumber\\
  M^{\beta}_{3}(t)=-\int^{t}_{0}\sigma_{\beta}(E(v)-E^{*}_{1})E(S(v-T_{2})e^{-\mu (T_{1}+T_{2})}G(I(v-T_{1}-T_{2})))dw_{\beta}(v).\nonumber\\
  \label{ch2.sec3.thm3.proof.eq11}
  \end{eqnarray}
  Applying Lemma~\ref{ch1.sec3.thm3.lemma1}, choose $\delta=\frac{2}{12}, \nu_{k}=\nu$, and $\tau_{k}=k$, then there exists a random number $k_{j}(w)>0, w\in \Omega$, and $j=1,2,\ldots 12$ such that
  \begin{equation}\label{ch2.sec3.thm3.proof.eq12}
  M^{S}_{i}(t)\leq \frac{1}{2}\nu <M^{S}_{i}(t), M^{S}_{i}(t)>+ \frac{\frac{2}{12}}{\nu}\ln{k}, \forall t\in [0,k], i=1,2,3,
  \end{equation}
  \begin{equation}\label{ch2.sec3.thm3.proof.eq13}
  M^{E}_{i}(t)\leq \frac{1}{2}\nu <M^{E}_{i}(t), M^{E}_{i}(t)>+ \frac{\frac{2}{12}}{\nu}\ln{k}, \forall t\in [0,k], i=1,2,3,
  \end{equation}
  \begin{equation}\label{ch2.sec3.thm3.proof.eq14}
  M^{I}_{i}(t)\leq \frac{1}{2}\nu <M^{I}_{i}(t), M^{S}_{i}(t)>+ \frac{\frac{2}{12}}{\nu}\ln{k}, \forall t\in [0,k], i=1,2,3,
  \end{equation}
  \begin{equation}\label{ch2.sec3.thm3.proof.eq15}
  M^{\beta}_{i}(t)\leq \frac{1}{2}\nu <M^{\beta}_{i}(t), M^{\beta}_{i}(t)>+ \frac{\frac{2}{12}}{\nu}\ln{k}, \forall t\in [0,k], i=1,2,3,
  \end{equation}
  where the quadratic variations $<M^{a}_{i}(t), M^{a}_{i}(t)>, \forall i=1,2,3$ and $a\in \{S, E, I, \beta\}$ are computed in the same manner as (\ref{ch2.sec3.thm2.proof.eq6}).

  It follows from (\ref{ch2.sec3.thm3.proof.eq7}) that for all $ k>\max_{j=1,2,\ldots,12}{k_{j}(w)}>0$,
  \begin{eqnarray}
  M(t)&\leq& \frac{1}{2}\nu \sum^{3}_{i=1}<M^{S}_{i}(t), M^{S}_{i}(t)>+\frac{1}{2}\nu \sum^{3}_{i=1}<M^{E}_{i}(t), M^{E}_{i}(t)>\nonumber\\
  &&+\frac{1}{2}\nu \sum^{3}_{i=1}<M^{I}_{i}(t), M^{I}_{i}(t)>+\frac{1}{2}\nu \sum^{3}_{i=1}<M^{\beta}_{i}(t), M^{\beta}_{i}(t)> \nonumber\\
  &&+\frac{2}{\nu}\ln{k}, \forall t\in [0,k].\label{ch2.sec3.thm3.proof.eq16}
  \end{eqnarray}
    Now, rearranging and diving both sides (\ref{ch2.sec3.thm3.proof.eq6}) of by $t$ and $\mathfrak{m}_{2}$,
    it follows that for all $t\in [k-1,k]$,
    \begin{eqnarray}
  &&\frac{1}{t}\int^{t}_{0}\left[ (S(v)-S^{*}_{1})^{2}+ (E(v)-E^{*}_{1})^{2}+  (I(v)-I^{*}_{1})^{2}\right]dv\nonumber\\
  &&\leq \frac{3\sigma^{2}_{S}(S^{*}_{1})^{2}+ 2\sigma^{2}_{E}(E^{*}_{1})^{2}+\sigma^{2}_{I}(I^{*}_{1})^{2}+\sigma^{2}_{\beta}(S^{*}_{1})^{2}(G^{*})^{2}E(e^{-2\mu (T_{1}+T_{2})})+\sigma^{2}_{\beta}(S^{*}_{1})^{2}(G^{*})^{2}E(e^{-\mu T_{1}})}{\mathfrak{m}_{2}}\nonumber\\
  &&+\frac{1}{t}\left(\frac{1}{\mathfrak{m}_{2}}\right)\left(\frac{1}{2}\nu \sum^{3}_{i=1}<M^{S}_{i}(t), M^{S}_{i}(t)>+\frac{1}{2}\nu \sum^{3}_{i=1}<M^{E}_{i}(t), M^{E}_{i}(t)>\right)\nonumber\\
  &&+\frac{1}{t}\left(\frac{1}{\mathfrak{m}_{2}}\right)\left(\frac{1}{2}\nu \sum^{3}_{i=1}<M^{I}_{i}(t), M^{I}_{i}(t)>+\frac{1}{2}\nu \sum^{3}_{i=1}<M^{\beta}_{i}(t), M^{\beta}_{i}(t)> \right)\nonumber\\
  &&+\frac{2}{\nu (k-1)}\ln{k}\left(\frac{1}{\mathfrak{m}_{2}}\right).
  \label{ch2.sec3.thm3.proof.eq17}
  \end{eqnarray}
  Thus letting $k\rightarrow \infty$, then $t\rightarrow \infty$. It follows from (\ref{ch2.sec3.thm3.proof.eq17}) that
   \begin{eqnarray}
  &&\limsup_{t\rightarrow \infty}\frac{1}{t}\int^{t}_{0}\left[ (S(v)-S^{*}_{1})^{2}+ (E(v)-E^{*}_{1})^{2}+  (I(v)-I^{*}_{1})^{2}\right]dv\nonumber\\
  &&\leq \frac{3\sigma^{2}_{S}(S^{*}_{1})^{2}+ 2\sigma^{2}_{E}(E^{*}_{1})^{2}+\sigma^{2}_{I}(I^{*}_{1})^{2}+\sigma^{2}_{\beta}(S^{*}_{1})^{2}(G^{*})^{2}E(e^{-2\mu (T_{1}+T_{2})})+\sigma^{2}_{\beta}(S^{*}_{1})^{2}(G^{*})^{2}E(e^{-\mu T_{1}})}{\mathfrak{m}_{2}}\nonumber\\
  &&+\limsup_{t\rightarrow \infty}\frac{1}{t}\left(\frac{1}{\mathfrak{m}_{2}}\right)\left(\frac{1}{2}\nu \sum^{3}_{i=1}<M^{S}_{i}(t), M^{S}_{i}(t)>+\frac{1}{2}\nu \sum^{3}_{i=1}<M^{E}_{j}(t), M^{E}_{i}(t)>\right)\nonumber\\
  &&+\limsup_{t\rightarrow \infty}\frac{1}{t}\left(\frac{1}{\mathfrak{m}_{2}}\right)\left(\frac{1}{2}\nu \sum^{3}_{i=1}<M^{I}_{i}(t), M^{I}_{i}(t)>+\frac{1}{2}\nu \sum^{3}_{i=1}<M^{\beta}_{i}(t), M^{\beta}_{i}(t)> \right)\nonumber\\.
  \label{ch2.sec3.thm3.proof.eq18}
  \end{eqnarray}
  Finally, by sending $\nu\rightarrow 0$, the result in (\ref{ch2.sec3.thm3.eq2})  follows immediately from (\ref{ch2.sec3.thm3.proof.eq18}).
 \begin{rem}\label{ch1.sec3.rem2}
  When the disease dynamics is perturbed by random fluctuations in the disease transmission or natural death rates, that is,  when at least one of the  intensities $\sigma^{2}_{i}> 0, i= S, E, I, \beta$, it has been noted earlier that the nontrivial steady state, $E_{1}$, of the deterministic system (\ref{ch1.sec0.eq3})-(\ref{ch1.sec0.eq6}) found in Theorem~\ref{ch1.sec3.thm1} no longer exists for the perturbed stochastic system (\ref{ch1.sec0.eq8})-(\ref{ch1.sec0.eq11}).  Nevertheless, the conditions in Theorem~\ref{ch1.sec3.thm2} provide restrictions for the constant delays $T_{1}$, $T_{2}$ and $T_{3}$ in (\ref{ch1.sec3.thm2.eq2})-(\ref{ch1.sec3.thm2.eq3b}) and parametric restrictions in (\ref{ch1.sec3.thm2.eq1}) which are sufficient for the sample path solutions of the stochastic system  (\ref{ch1.sec0.eq8})-(\ref{ch1.sec0.eq11}) to oscillate in the neighborhood of the potential endemic equilibrium $E_{1}$.

   Also, the conditions in Theorem~\ref{ch2.sec3.thm3} provide general restrictions in (\ref{ch1.sec3.thm3.eq1}) irrespective of the probability distribution of the random variable delay times $T_{1}, T_{2}$ and $T_{3}$  that are sufficient for the solutions of the perturbed stochastic system (\ref{ch1.sec0.eq8})-(\ref{ch1.sec0.eq11}) to oscillate near the nontrivial steady state, $E_{1}$, of the deterministic system (\ref{ch1.sec0.eq3})-(\ref{ch1.sec0.eq6}) found in Theorem~\ref{ch1.sec3.thm1}.

    The results in (\ref{ch2.sec3.thm2.eq4}) and (\ref{ch2.sec3.thm3.eq2})  characterize the average distance between the trajectories of the stochastic system (\ref{ch1.sec0.eq8})-(\ref{ch1.sec0.eq11}), and the nontrivial steady state $E_{1}$. Moreover, the results also signify that the size of the oscillations of the trajectories of the stochastic system relative to the state $E_{1}$ depends on the intensities of the noises in the system, that is the sizes of  $\sigma^{2}_{i}> 0, i= S, E, I, \beta$. Indeed, it can be seen easily that the trajectories oscillate much closer to $E_{1}$, whenever the intensities ($\sigma^{2}_{i}> 0, i= S, E, I, \beta$) are small, and vice versa.
   \end{rem}
    \section{Permanence of malaria in the stochastic system\label{ch1.sec5}}
 In this section, the permanence of malaria in the human population is investigated, whenever the system (\ref{ch1.sec0.eq8})-(\ref{ch1.sec0.eq11}) is subject to the influence of  random environmental fluctuations from the disease transmission and natural death rates.

  Permanence in the mean of the disease seeks to determine whether there is always a positive significant average number of individuals of the disease related classes namely- exposed $(E)$, infectious $(I)$,  and removal $(R)$ subclasses in the population over sufficiently long time. That is, seeking to determine whether $\lim_{t\rightarrow\infty}E(||E(t)||)>0$, $\lim_{t\rightarrow\infty}E(||I(t)||)>0$ and $\lim_{t\rightarrow\infty}E(||R(t)||)>0$. In the absence of explicit solutions for the nonlinear stochastic system (\ref{ch1.sec0.eq8})-(\ref{ch1.sec0.eq11}), this information about the asymptotic average number of individuals in the disease related classes is obtained via Lyapunov techniques from the examination of the statistical properties of the paths of the system (\ref{ch1.sec0.eq8})-(\ref{ch1.sec0.eq11}), in particular,  determining the size of all average sample path estimates for the disease related subclasses over sufficiently long time. That is, to determine whether the following hold:  $\liminf_{t\rightarrow\infty} \frac{1}{t}\int^{t}_{0}||E(s)||ds>0$, $\liminf_{t\rightarrow\infty} \frac{1}{t}\int^{t}_{0}||I(s)||ds>0$ and $\liminf_{t\rightarrow\infty} \frac{1}{t}\int^{t}_{0}||R(s)||ds>0$.

  %
  The following definition is given for the stochastic version of strong permanence in the mean of a disease.
 \begin{defn}\label{ch1.sec5.definition1}
 The system (\ref{ch1.sec0.eq8})-(\ref{ch1.sec0.eq11}) is said to be almost surely permanent in the mean\cite{chen-biodyn} (in the strong sense), if
 \begin{eqnarray}
  && \liminf_{t\rightarrow \infty}{\frac{1}{t}\int^{t}_{0}S(s)ds}>0, a.s., \quad \liminf_{t\rightarrow \infty}{\frac{1}{t}\int^{t}_{0}E(s)ds}>0, a.s., \nonumber\\
  && \liminf_{t\rightarrow \infty}{\frac{1}{t}\int^{t}_{0}I(s)ds}>0, a.s., \quad \liminf_{t\rightarrow \infty}{\frac{1}{t}\int^{t}_{0}R(s)ds}>0, a.s.\label{ch1.sec5.definition1.eq1}
 \end{eqnarray}
 where $(S(t), E(t), I(t), R(t))$ is any positive solution of the system (\ref{ch1.sec0.eq8})-(\ref{ch1.sec0.eq11}).
 \end{defn}
The method applied to show the permanence in the mean of the disease in the stochastic system (\ref{ch1.sec0.eq8})-(\ref{ch1.sec0.eq11}) is similar in the cases where (1) all the delays in the systems  $T_{1}, T_{2}$ and $T_{3}$  are constant and finite, and (2) in the case where the delays  $T_{1}, T_{2}$ and $T_{3}$ are random variables. Thus, without loss of generality, the results for the permanence in the mean of the disease in the stochastic system is presented only in the case of random delays in the system. Some ideas in \cite{yanli} can be used to prove this result.
 \begin{thm}\label{ch1.sec5.thm1}
 Assume that the conditions of Theorem~\ref{ch1.sec3.thm1} and Theorem~\ref{ch2.sec3.thm3} are satisfied. Define the following
 \begin{eqnarray}\label{ch1.sec5.thm1.eq1}
   \hat{h}\equiv\hat{h}(S^{*}_{1}, E^{*}_{1}, I^{*}_{1})&=&3(S^{*}_{1})^{2}+ 2(E^{*}_{1})^{2}+(I^{*}_{1})^{2}+(S^{*}_{1})^{2}(G^{*})^{2}E(e^{-2\mu (T_{1}+T_{2})})+(S^{*}_{1})^{2}(G^{*})^{2}E(e^{-\mu T_{1}}),\nonumber\\
   \sigma^{2}_{max}&=& \max{(\sigma^{2}_{S}, \sigma^{2}_{E}, \sigma^{2}_{I}, \sigma^{2}_{\beta})}.
 \end{eqnarray}
 Assume further that the following relationship is satisfied
 \begin{equation}\label{ch1.sec5.thm1.eq2}
   \sigma^{2}_{max}<\frac{\mathfrak{m}_{2}}{\hat{h}}\min{((S^{*}_{1})^{2}, (E^{*}_{1})^{2}, (I^{*}_{1})^{2})}\equiv \hat{\tau},
 \end{equation}
 where $\mathfrak{m}_{2}$ is defined in(\ref{ch2.sec3.thm3.proof.eq5}). Then it follows that
\begin{eqnarray}
  && \liminf_{t\rightarrow \infty}{\frac{1}{t}\int^{t}_{0}S(v)dv}>0,a.s., \quad \liminf_{t\rightarrow \infty}{\frac{1}{t}\int^{t}_{0}E(v)dv}>0,a.s. \nonumber\\
  && and\quad \liminf_{t\rightarrow \infty}{\frac{1}{t}\int^{t}_{0}I(v)dv}>0,a.s.\label{ch1.sec5.thm1.eq3}
 \end{eqnarray}
 In other words, the stochastic system (\ref{ch1.sec0.eq8})-(\ref{ch1.sec0.eq11}) is permanent in the mean.
 \end{thm}
 Proof:\\
 It is easy to see that when Theorem~\ref{ch1.sec3.thm1} and Theorem~\ref{ch2.sec3.thm3} are satisfied, then it follows from (\ref{ch2.sec3.thm3.eq2}) that
 \begin{eqnarray}
  &&\limsup_{t\rightarrow \infty}{\frac{1}{t}\int^{t}_{0} (S(v)-S^{*}_{1})^{2}dv}\leq \sigma^{2}_{max}\frac{\hat{h}}{\mathfrak{m}_{2}},\nonumber\\
  &&\limsup_{t\rightarrow \infty}{\frac{1}{t}\int^{t}_{0} (E(v)-E^{*}_{1})^{2}dv}\leq \sigma^{2}_{max}\frac{\hat{h}}{\mathfrak{m}_{2}},\nonumber\\
 &&\limsup_{t\rightarrow \infty}{\frac{1}{t}\int^{t}_{0} (I(v)-I^{*}_{1})^{2}dv}\leq \sigma^{2}_{max}\frac{\hat{h}}{\mathfrak{m}_{2}}.\quad a.s. \label{ch1.sec5.thm1.proof.eq1}
\end{eqnarray}
For each $a(t)\in \left\{S(t), E(t), I(t)\right\}$  it is easy to see that
\begin{equation}\label{ch1.sec5.thm1.proof.eq2}
  2(a^{*}_{1})^{2}-2a^{*}_{1}a(t)\leq (a^{*}_{1})^{2}+ (a(t)-a^{*}_{1})^{2}.
\end{equation}
It follows from (\ref{ch1.sec5.thm1.proof.eq2}) that
\begin{eqnarray}
  \liminf_{t\rightarrow \infty}{\frac{1}{t}\int^{t}_{0}a(v)dv}&\geq& \frac{a^{*}_{1}}{2}-\limsup_{t\rightarrow \infty}{\frac{1}{t}\int^{t}_{0} \frac{(a(v)-a^{*}_{1})^{2}}{2a^{*}_{1}}dv}\nonumber\\
  &\geq&\frac{a^{*}_{1}}{2}-\sigma^{2}_{max}\frac{\hat{h}}{\mathfrak{m_{2}}}\frac{1}{2a^{*}_{1}}\label{ch1.sec5.thm1.proof.eq3}
\end{eqnarray}
For each $a(t)\in \left\{S(t), E(t), I(t)\right\}$, the inequalities in (\ref{ch1.sec5.thm1.eq3}) follow immediately from (\ref{ch1.sec5.thm1.proof.eq3}), whenever  (\ref{ch1.sec5.thm1.eq2}) is satisfied.\\
The following result about the permanence of the disease in the system in the case where the delays $T_{1}, T_{2}$ and $T_{3}$  are all constant is stated.
 \begin{thm}\label{ch1.sec5.thm2}
 Assume that the conditions of Theorem~\ref{ch1.sec3.thm2} and Theorem~\ref{ch1.sec3.thm1.corrolary1} are satisfied. Define the following
 \begin{eqnarray}
   \hat{h}\equiv\hat{h}(S^{*}_{1}, E^{*}_{1}, I^{*}_{1})&=&(S^{*}_{1})^{2}(G^{*})^{2}e^{-2\mu (T_{1}+T_{2})}+(S^{*}_{1})^{2}(G^{*})^{2}e^{-\mu T_{1}},\label{ch1.sec5.thm2.eq1}
 \end{eqnarray}
 Assume further that the following relationship is satisfied
 \begin{equation}\label{ch1.sec5.thm2.eq2}
   \sigma^{2}_{\beta}<\frac{\mathfrak{m}_{1}}{\hat{h}}\min{((S^{*}_{1})^{2}, (E^{*}_{1})^{2}, (I^{*}_{1})^{2})}\equiv \hat{\tau},
 \end{equation}
 where $\mathfrak{m}_{1}$ is defined in(\ref{ch2.sec3.thm2.proof.eq3}). Then it follows that
\begin{eqnarray}
  && \liminf_{t\rightarrow \infty}{\frac{1}{t}\int^{t}_{0}S(v)dv}>0,a.s. \quad \liminf_{t\rightarrow \infty}{\frac{1}{t}\int^{t}_{0}E(v)dv}>0,a.s. \nonumber\\
  && \liminf_{t\rightarrow \infty}{\frac{1}{t}\int^{t}_{0}I(v)dv}>0,a.s.\label{ch1.sec5.thm2.eq3}
 \end{eqnarray}
 In other words, the stochastic system (\ref{ch1.sec0.eq8})-(\ref{ch1.sec0.eq11}) is permanent in the mean.
 \end{thm}
 Proof:\\
 The proof is similar to the proof of Theorem~\ref{ch1.sec5.thm1} above.
\begin{rem}\label{ch1.sec5.rem1}
Theorems~[\ref{ch1.sec5.thm2} \&\ref{ch1.sec5.thm1}] provide sufficient conditions for the permanence in the mean of the vector-borne disease in the population, where the conditions depend on the intensities of the white noise processes in the system. Indeed, for Theorem~\ref{ch1.sec5.thm1}, the condition in (\ref{ch1.sec5.thm1.eq2}) suggests that there is a bound for the intensities of the white noise processes in the system that allows the disease to persist in the population permanently, provided that the  noise free basic reproduction number of the  disease in (\ref{ch1.sec2.theorem1.corollary1.eq3}) satisfies $R_{0}>1$. Moreover, if one defines the term
\begin{equation}\label{ch1.sec5.rem1.eq1}
  l_{a}(\sigma^{2}_{max})=\frac{a^{*}_{1}}{2}-\sigma^{2}_{max}\frac{\hat{h}}{\mathfrak{m_{2}}}\frac{1}{2a^{*}_{1}}, \forall a(t)\in \left\{S(t), E(t), I(t)\right\},
\end{equation}
then it is easy to see from (\ref{ch1.sec5.thm1.proof.eq3}) that  $ l_{a}(\sigma^{2}_{max})$ is the lower bound for all average sample path estimates for the ensemble  mean of the different states $a(t)\in \left\{S(t), E(t), I(t)\right\}$ of the system  (\ref{ch1.sec0.eq8})-(\ref{ch1.sec0.eq11}) over sufficiently large amount of time, where $a^{*}_{1}\in \left\{S^{*}_{1}, E^{*}_{1}, I^{*}_{1}\right\}$.
  It can  be observed from (\ref{ch1.sec5.rem1.eq1}) and (\ref{ch1.sec5.thm1.eq2}) that
\begin{equation}\label{ch1.sec5.rem1.eq2}
  \lim_{\sigma^{2}_{max}\rightarrow 0^{+}}{l_{a}(\sigma^{2}_{max})}=\frac{1}{2}a^{*}_{1},\quad and \quad  \lim_{\sigma^{2}_{max}\rightarrow \hat{\tau}^{-}}{l_{a}(\sigma^{2}_{max})}=0.
\end{equation}
That is, the presence of noise in the system (\ref{ch1.sec0.eq8})-(\ref{ch1.sec0.eq11}) with significant continuously increasing intensity values (i.e. $\sigma^{2}_{max}\rightarrow \hat{\tau}^{-}$ ) allow for smaller values of the asymptotic average limiting value of all sample paths ($\liminf_{t\rightarrow \infty}{\frac{1}{t}\int^{t}_{0}a(v)dv}$) of the states $a(t)\in \left\{S(t), E(t), I(t)\right\}$ of the system, and vice versa.

 This observation suggests that the occurrence of "stronger" noise (noise with higher intensity) in the system suppresses the permanence of the disease in the population by allowing smaller average values for the sample paths of the disease related states $E, I, R$ in the population over sufficiently large amount of time. However, it should be noted that the smaller average values for the disease related classes $E, I, R$ over sufficiently long time are also matched with smaller average values for the susceptible class $S$ as the intensities of the noises in the system rise. This observation suggests that there is a general decrease in the population size over sufficiently long time as the intensities of the noises in the system increase in magnitude.

 Therefore, one can conclude that for small to moderate values for the intensities of the noises in the system, the disease persists permanently with higher average values for the disease related classes. Furthermore, as the magnitude of the intensities of the noises increase to higher values, the human population may become extinct over sufficiently long time. This observation is illustrated in the Figures~\ref{ch1.sec4.subsec1.1.figure 1}-\ref{ch1.sec4.subsec1.1.figure 3} and Figures~\ref{ch1.sec4.subsec1.1.figure 4}-\ref{ch1.sec4.subsec1.1.figure 6}, and much more in Figures~\ref{ch1.sec4.figure 1}-\ref{ch1.sec4.figure 3}.

The statistical significance of this result is noting that if all the sample paths for a given state $a(t)\in \left\{S(t), E(t), I(t)\right\}$ are bounded from below on average by the same significantly larger positive value $ l_{a}(\sigma^{2}_{max})$ asymptotically, then the ensemble means corresponding to the given states are also bounded by the same value asymptotically. This fact is more apparent as the stationary and ergodic properties of the system (\ref{ch1.sec0.eq8})-(\ref{ch1.sec0.eq11}) are shown in the next section.
\end{rem}
 \section{Stationary distribution and ergodic property}\label{ch1.sec3.sec1}
 The knowledge of the distribution of the solutions of a system of stochastic differential equations holds the key to fully understand the statistical and probabilistic properties of the solutions at any time, and overall to characterize the uncertainties of the states of the stochastic system over time. Furthermore, the ergodicity of the solutions of a stochastic system ensure that one obtains insights about the long-term behavior of the system, that is the statistical properties of the solutions of the system via knowledge of the average behavior of the sample paths or sample realizations of the system over  finite or sufficiently large time interval.

 In this section, the long term distribution and the ergodicity of the positive solutions of the stochastic system (\ref{ch1.sec0.eq8})-(\ref{ch1.sec0.eq11}) are characterized in the neighborhood of a potential endemic steady state $E_{1}$ for the system obtained in Theorem~\ref{ch1.sec3.thm1}. It is shown below that the stochastic system (\ref{ch1.sec0.eq8})-(\ref{ch1.sec0.eq11}) has a stationary endemic distribution for the solutions of the system. First,  the definition of a stationary distribution for a continuous-time and continuous-state Markov process or for the solution of a system of stochastic differential equations is presented in the following:
 \begin{defn}\label{ch1.sec3.sec1.defn1} (see \cite{yongli})
 Denote $\mathbb{P}_{\gamma}$ the corresponding probability distribution of an initial distribution $\gamma$, which describes the initial state of the system (\ref{ch1.sec0.eq8})-(\ref{ch1.sec0.eq11}) at time $t=0$. Suppose that the distribution of $Y(t)=(S(t), E(t), I(t), R(t))$ with initial distribution $\gamma$ converges in some sense to a distribution $\pi=\pi_{\gamma}$ ( a priori $\pi$ may depend on the initial distribution $\gamma$), that is,
 \begin{equation}\label{ch1.sec3.sec1.defn1.eq1}
 \lim_{t\rightarrow \infty}\mathbb{P}_{\gamma}\{X(t)\in F\}=\pi(F)
 \end{equation}
 for all measurable sets $F$, then we say that the stochastic system of differential equations has a stationary distribution $\pi(.)$.
 \end{defn}
 The standard method utilized in \cite{yongli,yongli-2,yzhang} is applied to establish this result.
 The following assumptions are made:-
 let $Y(t)$ be a regular temporary homogeneous Markov process in the $d$-dimensional space $E_{d}\subseteq \mathbb{R}^{d}_{+}$ described by the stochastic differential equation
 \begin{equation}\label{ch1.sec3.sec1.eq1}
 dY(t)=b(Y,t)dt+\sum_{r=1}^{k}g_{r}(Y, t)dB_{r}(t).
 \end{equation}
 Then the diffusion matrix can be defined as follows
 \begin{equation}\label{ch1.sec3.sec1.eq2}
 A(Y)=(a_{ij}(y)), a_{ij}(y)=\sum_{r=1}^{k}g^{i}_{r}(Y, t)g^{j}_{r}(Y, t).
 \end{equation}
 The following lemma describes the existence of a stationary solution for (\ref{ch1.sec3.sec1.eq1}).
 \begin{lemma}\label{ch1.sec3.sec1.lemma1}
 The Markov process $Y(t)$ has a unique ergodic stationary distribution $\pi(.)$ if a bounded domain $D\subset E_{d}$ with regular boundary $\Gamma$ exists and
 \item[H1:] there exists a constant $M>0$ satisfying $\sum^{d}_{ij}a_{ij}(x)\xi_{i}\xi_{j}\geq M|\xi|^{2}$, $x\in D, \xi\in \mathbb{R}^{d}$.
 \item[H2:] there is a $\mathcal{C}^{1,2}$-function $V\geq 0$ such that $LV$ is negative for any $E_{d}\backslash D$. Then
 \begin{equation}\label{ch1.sec3.sec1.lemma1.eq1}
   \mathbb{P}\left\{ \lim_{T\rightarrow \infty}\frac{1}{T}\int_{0}^{T}f(Y(t))dt=\int_{E_{d}}f(y)\pi(dy)\right\}=1,
 \end{equation}
 for all $y\in E_{d}$, where $f(.)$ is an integrable function with respect to the measure $\pi$.
 \end{lemma}
 The result that follows characterizes the existence of stationary distribution for the system (\ref{ch1.sec0.eq8})-(\ref{ch1.sec0.eq11}) in the general case where the delay times $T_{1}, T_{2}$ and $T_{3}$ in the system are distributed.
 \begin{thm}\label{ch1.sec3.sec1.thm1}
 Let the assumptions in Theorem~\ref{ch2.sec3.thm3} be satisfied, and let
 \begin{equation}\label{ch1.sec3.sec1.thm1.eq0}
      R=3\min_{(S, E, I)\in \mathbb{R}^{3}_{+}}{\left(\frac{1}{\left(\frac{1}{3}\Phi_{1}- \sigma^{2}_{\beta}\left(\frac{2}{3}(G^{*})^{2}+2\theta^{2}_{1}\right)\right)},\frac{2}{\Phi_{2}}\left(1+\sigma^{2}_{\beta}(S\theta_{1}+\theta_{2})^{2}\right), \frac{1}{\Phi_{3}}\left(1+2\sigma^{2}_{\beta}\theta^{2}_{2}\right) \right)}<\infty.
    \end{equation}
  Also define the following parameters:-
 \begin{equation}\label{ch1.sec3.sec1.thm1.eq1}
   \Phi=3\sigma^{2}_{S}(S^{*}_{1})^{2}+ 2\sigma^{2}_{E}(E^{*}_{1})^{2}+\sigma^{2}_{I}(I^{*}_{1})^{2}+\sigma^{2}_{\beta}(S^{*}_{1})^{2}(G^{*})^{2}E(e^{-2\mu (T_{1}+T_{2})})+\sigma^{2}_{\beta}(S^{*}_{1})^{2}(G^{*})^{2}E(e^{-\mu T_{1}}),
 \end{equation}
 and
 \begin{equation}\label{ch1.sec3.sec1.thm1.eq1-1}
      R_{min}=3\min{\left(\frac{1}{\left(\frac{1}{3}\Phi_{1}- \sigma^{2}_{\beta}\left(\frac{2}{3}(G^{*})^{2}+2(I^{*}_{1})^{2}\right)\right)},\frac{2}{\Phi_{2}}\left(1+4\sigma^{2}_{\beta}(S^{*}_{1}I^{*}_{1})^{2}\right), \frac{1}{\Phi_{3}}\left(1+2\sigma^{2}_{\beta}(S^{*}_{1}I^{*}_{1})^{2}\right) \right)}<\infty,
    \end{equation}
    where,
 \begin{eqnarray}
   \theta_{1} &=&\int_{t_{0}}^{h_{1}}f_{T_{1}}(s)e^{-\mu s}G(I(t-s))ds,\label{ch1.sec3.sec1.thm1.proof.eq1} \\
    \theta_{2} &=&\int_{t_{0}}^{h_{2}}f_{T_{2}}S(t-u)\int_{t_{0}}^{h_{1}}f_{T_{1}}(s)e^{-\mu( s+u)}G(I(t-s-u))dsdu,\label{ch1.sec3.sec1.thm1.proof.eq2}\\
    \Phi_{1}&=&3\mu-\left[2\mu\lambda{(\mu)}+(2\mu+d+\alpha)\frac{\lambda{(\mu)}}{2}+\alpha\lambda{(\mu)}+\frac{\beta S^{*}_{1}\lambda{(\mu)}}{2}\right.\nonumber\\
&&+\left.\left(\frac{\beta \lambda{(\mu)}(G^{*})^{2}}{2}\right)+\left(\frac{\beta (G^{*})^{2}}{2\lambda{(\mu)}}\right)\right],\label{ch1.sec3.sec1.thm1.proof.eq2a}\\
\Phi_{2}&=&2\mu-\left[\frac{\beta }{2\lambda{(\mu)}}+\frac{\beta S^{*}_{1}\lambda{(\mu)}}{2}+ \frac{2\mu}{\lambda{(\mu)}}+(2\mu+d+\alpha)\frac{\lambda{(\mu)}}{2}+\alpha \lambda{(\mu)}\right],\label{ch1.sec3.sec1.thm1.proof.eq2b}\\
\Phi_{3}&=&(\mu+d+\alpha)-\left[(2\mu+d+\alpha)\frac{1}{\lambda{(\mu)}}+ \frac{\alpha\lambda{(\mu)}}{2}+\frac{3\alpha}{2\lambda(\mu)} \right.\nonumber\\
&&\left.+ \frac{\beta S^{*}_{1}(G'(I^{*}_{1}))^{2}}{\lambda{(\mu)}}\right].\label{ch1.sec3.sec1.thm1.proof.eq2c}
 \end{eqnarray}
 It follows that there is a unique stationary distribution $\pi(.)$ for the solutions of (\ref{ch1.sec0.eq8})-(\ref{ch1.sec0.eq11}), whenever
 \begin{equation}\label{ch1.sec3.sec1.thm1.eq2}
   \Phi<\min{[\Phi_{1}(S^{*}_{1})^{2}, \Phi_{2}(E^{*}_{1})^{2}, \Phi_{3}(I^{*}_{1})^{2}]},
 \end{equation}
  and
 \begin{equation}\label{ch1.sec3.sec1.thm1.eq2-1}
  R<R_{min},\quad and\quad  R_{min}+\Phi<||E_{1}-0||^{2}.
 \end{equation}
 Moreover, the solution of the system (\ref{ch1.sec0.eq8})-(\ref{ch1.sec0.eq11}) is ergodic.
 \end{thm}
 Proof:\\
 The results will be shown for the vector $X=(S, E, I)$ corresponding to the system (\ref{ch1.sec0.eq8})-(\ref{ch1.sec0.eq11}).
Furthermore, the hypothesis $\mathbf{H1}$ in Lemma~\ref{ch1.sec3.sec1.lemma1} is verified in the following. It is easy to see that when the conditions  in (\ref{ch1.sec3.thm3.eq1}) given in Theorem~\ref{ch2.sec3.thm3} are satisfied, it follows from (\ref{ch1.sec3.sec1.thm1.proof.eq2a})-(\ref{ch1.sec3.sec1.thm1.proof.eq2c}) that
 \begin{equation}\label{ch1.sec3.sec1.thm1.proof.eq2d}
   2(G^{*})^{2}\sigma^{2}_{\beta} +3\sigma^{2}_{S}<\Phi_{1},\quad 2\sigma^{2}_{E}<\Phi_{2}\quad and\quad \sigma^{2}_{I}<\Phi_{3}.
 \end{equation}
 In addition, the system (\ref{ch1.sec0.eq8})-(\ref{ch1.sec0.eq11}) can be rewritten in the form (\ref{ch1.sec3.sec1.eq1}), where $d=3$, and the diffusion matrix $ A(X)=(a_{ij}(x))$ is defined as follows:
 \begin{equation}\label{ch1.sec3.sec1.thm1.proof.eq3}
 \left\{
   \begin{array}{ccc}
     a_{11} & = &\sigma^{2}_{S}S^{2}+\sigma^{2}_{\beta}S^{2}\theta^{2}_{1}, \\
     a_{12} & = &\sigma^{2}_{\beta}S\theta_{1}\theta_{2}-\sigma^{2}_{\beta}S^{2}\theta^{2}_{1}, \\
     a_{13} & = &-\sigma^{2}_{\beta}S\theta_{1}\theta_{2},
   \end{array}
   \right.
 \end{equation}
 \begin{equation}\label{ch1.sec3.sec1.thm1.proof.eq4}
 \left\{
   \begin{array}{lll}
     a_{21} & = &a_{12},\\
     a_{22} & = &\sigma^{2}_{E}E^{2}+\sigma^{2}_{\beta}S^{2}\theta^{1}-2\sigma^{2}_{\beta}S\theta_{1}\theta_{2}+\sigma^{2}_{\beta}\theta^{2}_{2}, \\
     a_{23} & = &\sigma^{2}_{\beta}S\theta_{1}\theta_{2}-\sigma^{2}_{\beta}\theta^{2}_{2},
   \end{array}
   \right.
    \end{equation}
    and
    \begin{equation}\label{ch1.sec3.sec1.thm1.proof.eq5}
      a_{31}=a_{13}, a_{32}=a_{32}, a_{33}=\sigma^{2}_{I}I^{2}+\sigma^{2}_{\beta}\theta^{2}_{2}.
    \end{equation}
  Also, define the sets $U_{1}$ and $U_{2}$ as  follows
    \begin{equation}\label{ch1.sec3.sec1.thm1.proof.eq6}
      U_{1}=\left\{(S, E, I)\in \mathbb{R}^{3}_{+}| \mathfrak{m}_{2}(S-S^{*}_{1})^{2}+\mathfrak{m}_{2}(E-E^{*}_{1})^{2}+\mathfrak{m}_{2}(I-I^{*}_{1})^{2}\leq \Phi\right\}
    \end{equation} and
    \begin{eqnarray}
      U_{2}&=&\left\{(S, E, I)\in \mathbb{R}^{3}_{+}|S^{2}(\sigma^{2}_{S}-2\sigma^{2}_{\beta}\theta^{2}_{1})\geq 1,E^{2}\geq \frac{1}{\sigma^{2}_{E}}\left[1+\sigma^{2}_{\beta}(S\theta_{1}+\theta_{2})^{2}\right], I^{2}\geq \frac{1}{\sigma^{2}_{I}}\left(1+2\sigma^{2}_{\beta}\theta^{2}_{2}\right) \right\}.\nonumber\\
      \label{ch1.sec3.sec1.thm1.proof.eq7}
      \end{eqnarray}
      One can see  from (\ref{ch1.sec3.sec1.thm1.proof.eq2d}) that
      \begin{eqnarray}
     U_{2} &\subset&\left\{(S, E, I)\in \mathbb{R}^{3}_{+}| S^{2}>\frac{1}{\left(\frac{1}{3}\Phi_{1}- \sigma^{2}_{\beta}\left(\frac{2}{3}(G^{*})^{2}+2\theta^{2}_{1}\right)\right)},\quad E^{2}>\frac{2}{\Phi_{2}}\left(1+\sigma^{2}_{\beta}(S\theta_{1}+\theta_{2})^{2}\right),\quad\right.\nonumber\\
     &&\left. I^{2}>\frac{1}{\Phi_{3}}\left(1+2\sigma^{2}_{\beta}\theta^{2}_{2}\right)
 \right\},\nonumber\\
  &\subset& \left(\bar{B}_{\mathbb{R}^{3}_{+}}(0; R)\right)^{c},\label{ch1.sec3.sec1.thm1.proof.eq7-1}
    \end{eqnarray}
    where the set $\left(\bar{B}_{\mathbb{R}^{3}_{+}}(0; R)\right)^{c}$ is the complement of the closed ball or sphere in $\mathbb{R}^{3}_{+}$ centered at the origin $(S, E, I)=(0,0,0)$ with radius given by
    \begin{equation}\label{ch1.sec3.sec1.thm1.proof.eq7-2}
      R=3\min_{(S, E, I)\in \mathbb{R}^{3}_{+}}{\left(\frac{1}{\left(\frac{1}{3}\Phi_{1}- \sigma^{2}_{\beta}\left(\frac{2}{3}(G^{*})^{2}+2\theta^{2}_{1}\right)\right)},\frac{2}{\Phi_{2}}\left(1+\sigma^{2}_{\beta}(S\theta_{1}+\theta_{2})^{2}\right), \frac{1}{\Phi_{3}}\left(1+2\sigma^{2}_{\beta}\theta^{2}_{2}\right) \right)}<\infty.
    \end{equation}
    In addition, the symbol "$\subset$" signifies the set operation of proper subset, and  $\Phi$ and $\mathfrak{m}_{2}$ are defined in (\ref{ch1.sec3.sec1.thm1.eq1}), Theorem~\ref{ch2.sec3.thm3} and (\ref{ch2.sec3.thm3.proof.eq5}). Moreover, $ \mathfrak{m}_{2}$ is given as follows:
     \begin{eqnarray}
       \mathfrak{m}_{2}=\min\{\left(3\mu-\tilde{a}_{1}(\mu, d, \alpha, \beta, B)\right), \left(2\mu-a_{3}(\mu, d, \alpha, \beta, B)\right), \left((\mu + d+\alpha)-\tilde{a}_{2}(\mu, d, \alpha, \beta, B)\right)\}.\nonumber\\
       \label{ch1.sec3.sec1.thm1.proof.eq7b}
     \end{eqnarray}
     It is easy to see that when the conditions  in (\ref{ch1.sec3.thm3.eq1}) which are presented in Theorem~\ref{ch2.sec3.thm3} are satisfied, then $\mathfrak{m}_{2}>0$. Also, observe that $U_{1}$ defined in (\ref{ch1.sec3.sec1.thm1.proof.eq6}) represents the interior and boundary of a sphere in $\mathbb{R}^{3}_{+}$ with radius $\Phi$, centered at the endemic equilibrium $E_{1}=(S^{*}_{1},E^{*}_{1}, I^{*}_{1})$. Furthermore,  when (\ref{ch1.sec3.sec1.thm1.eq2}) holds,
 it is easy to see that
      \begin{equation}\label{ch1.sec3.sec1.thm1.proof.eq7c}
       \Phi\ll \Phi_{1}(S^{*}_{1})^{2}+ \Phi_{2}(E^{*}_{1})^{2}+ \Phi_{3}(I^{*}_{1})^{2}<\max{(\Phi_{1}, \Phi_{2}, \Phi_{3})}||E_{1}-0||^{2}.
     \end{equation}
     Thus, from (\ref{ch1.sec3.sec1.thm1.proof.eq7c}), the set $U_{1}$ is non-empty and totally enclosed in $\mathbb{R}^{3}_{+}$.
          Furthermore, the endemic equilibrium $E_{1}=(S^{*}_{1}, E^{*}_{1}, I^{*}_{1})\in U_{2}$. Indeed, it can be seen from the Assumption~\ref{ch1.sec0.assum1} that $\theta_{1}\leq I^{*}_{1}$ and $\theta_{2}\leq S^{*}_{1}I^{*}_{1}$, whenever $X(t)=E_{1}$. Therefore, since $R<R_{min}$ from (\ref{ch1.sec3.sec1.thm1.eq2-1}), it implies that
     \begin{equation}\label{ch1.sec3.sec1.thm1.proof.eq7c-1}
       E_{1}\in \left(\bar{B}_{\mathbb{R}^{3}_{+}}(0; R_{min})\right)^{c}\subset \left(\bar{B}_{\mathbb{R}^{3}_{+}}(0; R)\right)^{c}.
     \end{equation}
    One also notes that the set $U_{1}$ does not overlap or intersect with the closed ball $\bar{B}_{\mathbb{R}^{3}_{+}}(0; R)$, since $R_{min}+\Phi<||E_{1}-0||^{2}$ from (\ref{ch1.sec3.sec1.thm1.eq2-1}). That is, $U_{1}\subset \left(\bar{B}_{\mathbb{R}^{3}_{+}}(0; R)\right)^{c}$.
     Now, define the new set
     \begin{equation}\label{ch1.sec3.sec1.thm1.proof.eq7d}
       U=U_{1}\cap U_{2}.
     \end{equation}
Clearly, $U\neq \emptyset$ since $E_{1}\in U$. Also, from (\ref{ch1.sec3.sec1.thm1.proof.eq7d}) the set $U\subset U_{1}\subset \mathbb{R}^{3}_{+}$ and $U\subset U_{2}\subset \mathbb{R}^{3}_{+}$, thus,  the set $U$ is totally enclosed in $\mathbb{R}^{3}_{+}$, and hence, the set $U$ is well-defined.
        Now, let $(S, E, I)\in U=U_{1}\cap U_{2}$.  It follows from (\ref{ch1.sec3.sec1.eq2}) and (\ref{ch1.sec3.sec1.thm1.proof.eq3})-(\ref{ch1.sec3.sec1.thm1.proof.eq5}), that
    \begin{eqnarray}
       \sum^{3}_{ij}a_{ij}(x)\xi_{i}\xi_{j}=&& \sigma^{2}_{I}S^{2}\xi^{2}_{1} +\sigma^{2}_{E}E^{2}\xi^{2}_{2} +\sigma^{2}_{I}I^{2}\xi^{2}_{3}\nonumber\\
       &&+\sigma^{2}_{\beta}S^{2}\theta^{2}_{1}(\xi_{1}-\xi_{2})^{2}+\sigma^{2}_{\beta}\theta^{2}_{2}(\xi_{2}-\xi_{3})^{2}-2\sigma^{2}_{\beta}S\theta_{1}\theta_{2}\xi^{2}_{2}\nonumber\\
      &&+2\sigma^{2}_{\beta}S\theta_{1}\theta_{2}\xi_{1}\xi_{2}+2\sigma^{2}_{\beta}S\theta_{1}\theta_{2}\xi_{2}\xi_{3}-2\sigma^{2}_{\beta}S\theta_{1}\theta_{2}\xi_{1}\xi_{3}.\label{ch1.sec3.sec1.thm1.proof.eq8}
    \end{eqnarray}
    By applying the following  algebraic inequalities $\min_{a,b}{(-2ab,2ab)}>-a^{2}-b^{2}$, to  $\sum^{3}_{ij}a_{ij}(x)\xi_{i}\xi_{j}, \forall \xi=(\xi_{1},\xi_{2})$, it is easy to see that (\ref{ch1.sec3.sec1.thm1.proof.eq8}) becomes the following:
     \begin{eqnarray}
       \sum^{3}_{ij}a_{ij}(x)\xi_{i}\xi_{j}\geq&& (\sigma^{2}_{S}S^{2}-2\sigma^{2}_{\beta}S^{2}\theta^{2}_{1})\xi^{2}_{1} +(\sigma^{2}_{E}E^{2}-\sigma^{2}_{\beta}(S\theta_{1}+\theta_{2})^{2})\xi^{2}_{2}\nonumber\\
        &&+(\sigma^{2}_{I}I^{2}-2\sigma^{2}_{\beta}\theta^{2}_{2})\xi^{2}_{3}+\sigma^{2}_{\beta}S^{2}\theta^{2}_{1}(\xi_{1}-\xi_{2})^{2}\nonumber\\
        &&+\sigma^{2}_{\beta}\theta^{2}_{2}(\xi_{2}-\xi_{3})^{2}.\label{ch1.sec3.sec1.thm1.proof.eq9}
       \end{eqnarray}
       Define the constant $M$ as follows
       \begin{equation}\label{ch1.sec3.sec1.thm1.proof.eq10}
         M=\min{[(\sigma^{2}_{S}S^{2}-2\sigma^{2}_{\beta}S^{2}\theta^{2}_{1}),(\sigma^{2}_{E}E^{2}-\sigma^{2}_{\beta}(S\theta_{1}+\theta_{2})^{2}),(\sigma^{2}_{I}I^{2}-2\sigma^{2}_{\beta}\theta^{2}_{2}) ]}.
       \end{equation}
       Clearly, from (\ref{ch1.sec3.sec1.thm1.proof.eq7}), one can see that $M\geq 1>0, \forall (S,E,I)\in U$.        It is also seen that by taking $D$ to be a neighborhood of $U$, then for $\forall (S, E, I)\in \bar{D}$, it follows from (\ref{ch1.sec3.sec1.thm1.proof.eq9})-(\ref{ch1.sec3.sec1.thm1.proof.eq10}) that
       \begin{equation}\label{ch1.sec3.sec1.thm1.proof.eq10a}
         \sum^{3}_{ij}a_{ij}(x)\xi_{i}\xi_{j}\geq M|\xi|^{2},
       \end{equation}
       where $\bar{D}$ is the closure of the set $D$.

       The hypothesis $\mathbf{H2}$ in Lemma~\ref{ch1.sec3.sec1.lemma1} follows from (\ref{ch1.sec3.lemma1.eq7}), where it can be seen that
       \begin{equation}\label{ch1.sec3.sec1.thm1.proof.eq10b}
         LV(t)<-\mathfrak{m}_{2}(S-S^{*}_{1})^{2}-\mathfrak{m}_{2}(E-E^{*}_{1})^{2}-\mathfrak{m}_{2}(I-I^{*}_{1})^{2}+ \Phi.
       \end{equation}
       Furthermore, from (\ref{ch1.sec3.sec1.thm1.proof.eq6})-(\ref{ch1.sec3.sec1.thm1.proof.eq7d}), it follows that $LV(t)<0$, for all $X=(S, E, I)\in \mathbb{R}^{3}_{+}\backslash D$. Since the hypotheses $\mathbf{H1}$  and $\mathbf{H2}$ hold, it implies from Lemma~\ref{ch1.sec3.sec1.lemma1} that the system (\ref{ch1.sec0.eq8})-(\ref{ch1.sec0.eq11}) has  stationary Ergodic solutions.
 \begin{rem}\label{ch1.sec3.sec1.thm1.rem1}
 Theorem~\ref{ch1.sec3.sec1.thm1} signifies that the positive solution process $Y(t)=(S(t), E(t), I(t), R(t))\in R_{+}^{4}\times [t_{0}, \infty)$ of the stochastic system (\ref{ch1.sec0.eq8})-(\ref{ch1.sec0.eq11}) converges in distribution to a unique random variable (for instance denoted $Y_{s}=(S_{s}, E_{s}, I_{s}, R_{s})$) that has the stationary distribution $\pi(.)$ in the neighborhood of the endemic equilibrium $E_{1}=(S^{*}_{1}, E^{*}_{1}, I^{*}_{1}, R^{*}_{1})$, whenever the conditions of Theorem~\ref{ch1.sec3.thm1},  Theorem~\ref{ch2.sec3.thm3} and (\ref{ch1.sec3.sec1.thm1.eq2})-(\ref{ch1.sec3.sec1.thm1.eq2-1}) are satisfied.

 The stationary feature of the solutions process $\{Y(t),t\geq t_{0}\}$ ensures that all the statistical properties such as the mean, variance, and moments etc. remain the same over time for every random vector $Y(t)$, whenever $t\geq t_{0}$ is sufficiently large and fixed. In other words, the ensemble mean, variance and moments etc. for the solution process $\{Y(t),t\geq t_{0}\}$ exist for sufficiently large time, and they are constant, and also correspond to the mean, variance and moments etc. of the stationary distribution $\pi(.)$.

 The ergodic feature of the solution process $\{Y(t),t\geq t_{0}\}$ also allows for insights about the statistical properties of the entire process via knowledge of the sample paths over sufficiently large amount of time. For example, over sufficiently large time, the average value of the sample path given by $\hat{\mu}_{f}=\lim_{T\rightarrow \infty}\frac{1}{T}\int_{0}^{T}f(Y(t))dt$ accurately estimates the corresponding ensemble mean given by  $\mu_{f}=E(f(Y(t)))=\int_{E_{d}}f(y)\pi(dy)$, where $f(.)$ is an integrable function with respect to the measure $\pi$. In particular, $\hat{\mu}_{S}=\lim_{T\rightarrow \infty}\frac{1}{T}\int_{0}^{T}S(t)dt$   estimates the ensemble mean  $\mu_{S}=E(S(t))=\int_{E_{d}}S\pi(dS)$, and  $\hat{\mu}_{I}=\lim_{T\rightarrow \infty}\frac{1}{T}\int_{0}^{T}I(t)dt$   estimates the ensemble mean  $\mu_{S}=E(I(t))=\int_{E_{d}}I\pi(dI)$  etc. over sufficiently large values of $t\geq t_{0}$.

 Note that in the absence of explicit sample path solutions of the nonlinear stochastic system  (\ref{ch1.sec0.eq8})-(\ref{ch1.sec0.eq11}), the stationary distribution $\pi(.)$ is numerically approximated in Section~\ref{ch1.sec4} for specified sets of system parameters of the stochastic system.
  \end{rem}
  \section{Example}\label{ch1.sec4}
In this study, the examples exhibited in this section are used to facilitate understanding about the influence of the intensity or "strength" of the noise in the system on the persistence of the disease in the population, and also to illustrate the existence of a stationary distribution $\pi(.)$ for the states of the system. These objectives are achieved in a simplistic manner by examining the behavior of the sample paths of  the different states ($S, E, I, R$) of the stochastic system (\ref{ch1.sec0.eq8})-(\ref{ch1.sec0.eq11}) in the neighborhood of  the potential endemic equilibrium $E_{1}=(S^{*}_{1}, E^{*}_{1},I^{*}_{1}, R^{*}_{1})$ of the system, and also by generating graphical representations for the distributions of samples of observations of the states ($S, E, I, R$) of the  system at a specified time $t\in [t_{0},\infty)$.

Recall Theorem~\ref{ch1.sec3.thm1.corrolary1} asserts that potential endemic equilibrium $E_{1}$ exists, whenever the basic reproduction number $R^{*}_{0}>1$, where $R^{*}_{0}$ is defined in (\ref{ch1.sec2.lemma2a.corrolary1.eq4}). It follows that when the conditions of Theorem~\ref{ch1.sec3.thm1.corrolary1} are satisfied, then the endemic equilibrium $E_{1}=(S^{*}_{1}, E^{*}_{1},I^{*}_{1}, R^{*}_{1})$ satisfies the following system
\begin{equation}\label{ch1.sec4.eq1}
\left\{
\begin{array}{lll}
&&B-\beta Se^{-\mu T_{1}}G(I)-\mu S+\alpha I e^{-\mu T_{3}}=0\\
&&\beta Se^{-\mu T_{1}}G(I)-\mu E -\beta Se^{-\mu (T_{1}+T_{2})}G(I)=0\\
&&\beta Se^{-\mu (T_{1}+T_{2})}G(I)-(\mu+d+\alpha)I=0\\
&&\alpha I-\mu R-\alpha I e^{-\mu T_{3}}=0
\end{array}
\right.
\end{equation}
 \subsection{Example 1: Effect of the intensity of white noise on the persistence of the disease  }\label{ch1.sec4.subsec1}
The following convenient list of parameter values in Table~\ref{ch1.sec4.table2} are used to generate and examine the paths of the different states of the stochastic system (\ref{ch1.sec0.eq8})-(\ref{ch1.sec0.eq11}), whenever $R^{*}_{0}>1$, and the intensities of the white noise processes in the system continuously increase. It is easily seen that for this set of parameter values, $R^{*}_{0}=80.7854>1$. Furthermore, the endemic equilibrium for the system $E_{1}$ is given as follows:- $E_{1}=(S^{*}_{1},E^{*}_{1},I^{*}_{1})=(0.2286216,0.07157075,0.9929282)$.
 \begin{table}[h]
  \centering
  \caption{A list of specific values chosen for the system parameters for the examples in subsection~\ref{ch1.sec4.subsec1}}\label{ch1.sec4.table2}
  \begin{tabular}{l l l}
  Disease transmission rate&$\beta$& 0.6277\\\hline
  Constant Birth rate&$B$&$ \frac{22.39}{1000}$\\\hline
  Recovery rate& $\alpha$& 0.05067\\\hline
  Disease death rate& $d$& 0.01838\\\hline
  Natural death rate& $\mu$& $0.002433696$\\\hline
  Incubation delay time in vector& $T_{1}$& 2 units \\\hline
  Incubation delay time in host& $T_{2}$& 1 unit \\\hline
  Immunity delay time& $T_{3}$& 4 units\\\hline
  \end{tabular}
\end{table}
The Euler-Maruyama stochastic approximation scheme\footnote{A seed is set on the random number generator to reproduce  the same sequence of random numbers for the Brownian motion in order to generate reliable graphs for the trajectories of the system under different intensity values for the white noise processes, so that comparison can be made to identify differences that reflect the effect of intensity values.} is used to generate trajectories for the different states $S(t), E(t), I(t), R(t)$ over the time interval $[0,T]$, where $T=\max(T_{1}+T_{2}, T_{3})=4$. The special nonlinear incidence  functions $G(I)=\frac{aI}{1+I}, a=0.05$ in \cite{gumel} is utilized to generate the numeric results. Furthermore, the following initial conditions are used
\begin{equation}\label{ch1.sec4.eq1}
\left\{
\begin{array}{l l}
S(t)= 10,\\
E(t)= 5,\\
I(t)= 6,\\
R(t)= 2,
\end{array}
\right.
\forall t\in [-T,0], T=\max(T_{1}+T_{2}, T_{3})=4.
\end{equation}
\subsubsection{Effect of the intensity of white noise $\sigma_{\beta}$ on the persistence and permanence of the disease  }\label{ch1.sec4.subsec1.1}
\begin{figure}[H]
\begin{center}
\includegraphics[height=6cm]{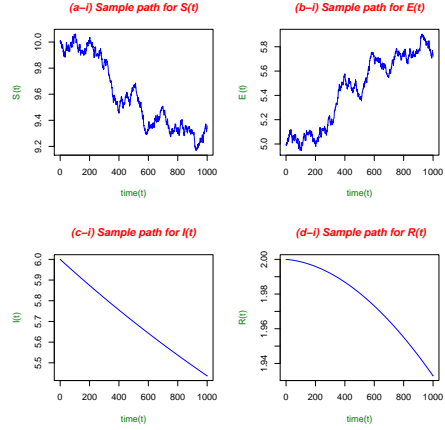}
\caption{(a-i), (b-i), (c-i) and (d-i) show the trajectories of the disease states $(S,E,I,R)$ respectively, whenever the only source of the noise in the system is from the disease transmission rate, and the strength of the noise in the system is relatively small, that is, $\sigma_{i}=0.5, \forall i\in \{S, E, I, R, \beta\} $. The broken lines represent the endemic equilibrium $E_{1}=(S^{*}_{1},E^{*}_{1},I^{*}_{1})=(0.2286216,0.07157075,0.9929282)$. Furthermore, $\min{(S(t))}=9.168214$, $\min{(E(t))}=4.946437$, $\min{(I(t))}=5.43688$ and $\min{(R(t))}=1.93282$.
}\label{ch1.sec4.subsec1.1.figure 1}
\end{center}
\end{figure}
\begin{figure}[H]
\begin{center}
\includegraphics[height=6cm]{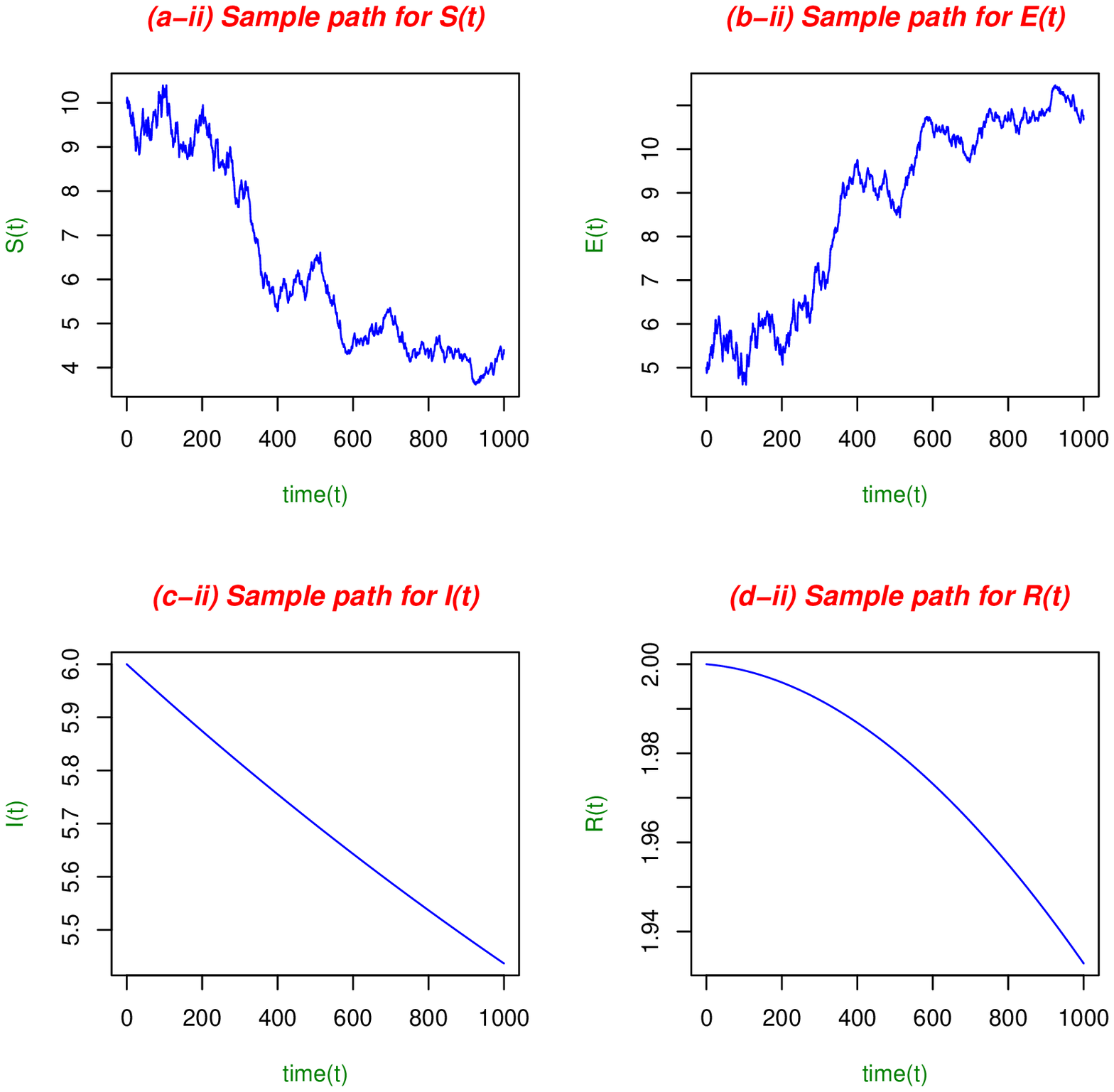}
\caption{$(a-ii)$, $(b-ii)$, $(c-ii)$ and $(d-ii)$ show the trajectories of the disease states $(S,E,I,R)$ respectively, whenever the only source of the noise in the system is from the disease transmission rate, and the strength of the noise in the system is relatively high, that is, $\sigma_{i}=5, \forall i\in \{S, E, I, R, \beta\} $. The broken lines represent the endemic equilibrium $E_{1}=(S^{*}_{1},E^{*}_{1},I^{*}_{1})=(0.2286216,0.07157075,0.9929282)$. Furthermore, $\min{(S(t))}=3.614151$, $\min{(E(t))}=4.611832$, $\min{(I(t))}=5.43688$ and $\min{(R(t))}=1.93282$.
}\label{ch1.sec4.subsec1.1.figure 2}
\end{center}
\end{figure}
\begin{figure}[H]
\begin{center}
\includegraphics[height=6cm]{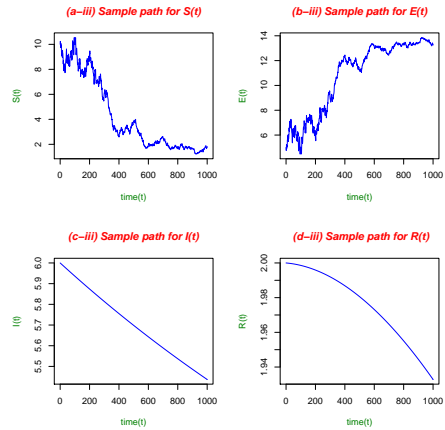}
\caption{$(a-iii)$, $(b-iii)$, $(c-iii)$ and $(d-iii)$ show the trajectories of the disease states $(S,E,I,R)$ respectively, whenever the only source of the noise in the system is from the disease transmission rate, and the strength of the noise in the system is relatively higher, that is, $\sigma_{i}=10, \forall i\in \{S, E, I, R, \beta\} $. The broken lines represent the endemic equilibrium $E_{1}=(S^{*}_{1},E^{*}_{1},I^{*}_{1})=(0.2286216,0.07157075,0.9929282)$. Furthermore, $\min{(S(t))}=1.217144$, $\min{(E(t))}=4.465998$, $\min{(I(t))}=5.43688$ and $\min{(R(t))}=1.93282$.
}\label{ch1.sec4.subsec1.1.figure 3}
\end{center}
\end{figure}
\begin{figure}[H]
\begin{center}
\includegraphics[height=6cm]{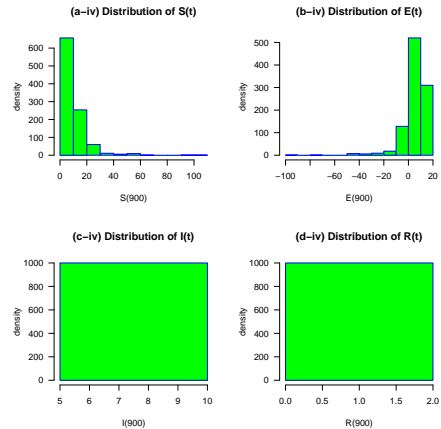}
\caption{(a-iv), (b-iv), (c-iv) and (d-iv) show the approximate distribution of the different disease states $(S,E,I,R)$ respectively,  at the specified time $t=900$, whenever the only source of noise in the system is from the disease transmission rate, and the  intensity of the noise in the system is relatively higher, that is,  $\sigma_{\beta}=10$.
}\label{ch1.sec4.subsec1.1.figure 4}
\end{center}
\end{figure}
\begin{figure}[H]
\begin{center}
\includegraphics[height=6cm]{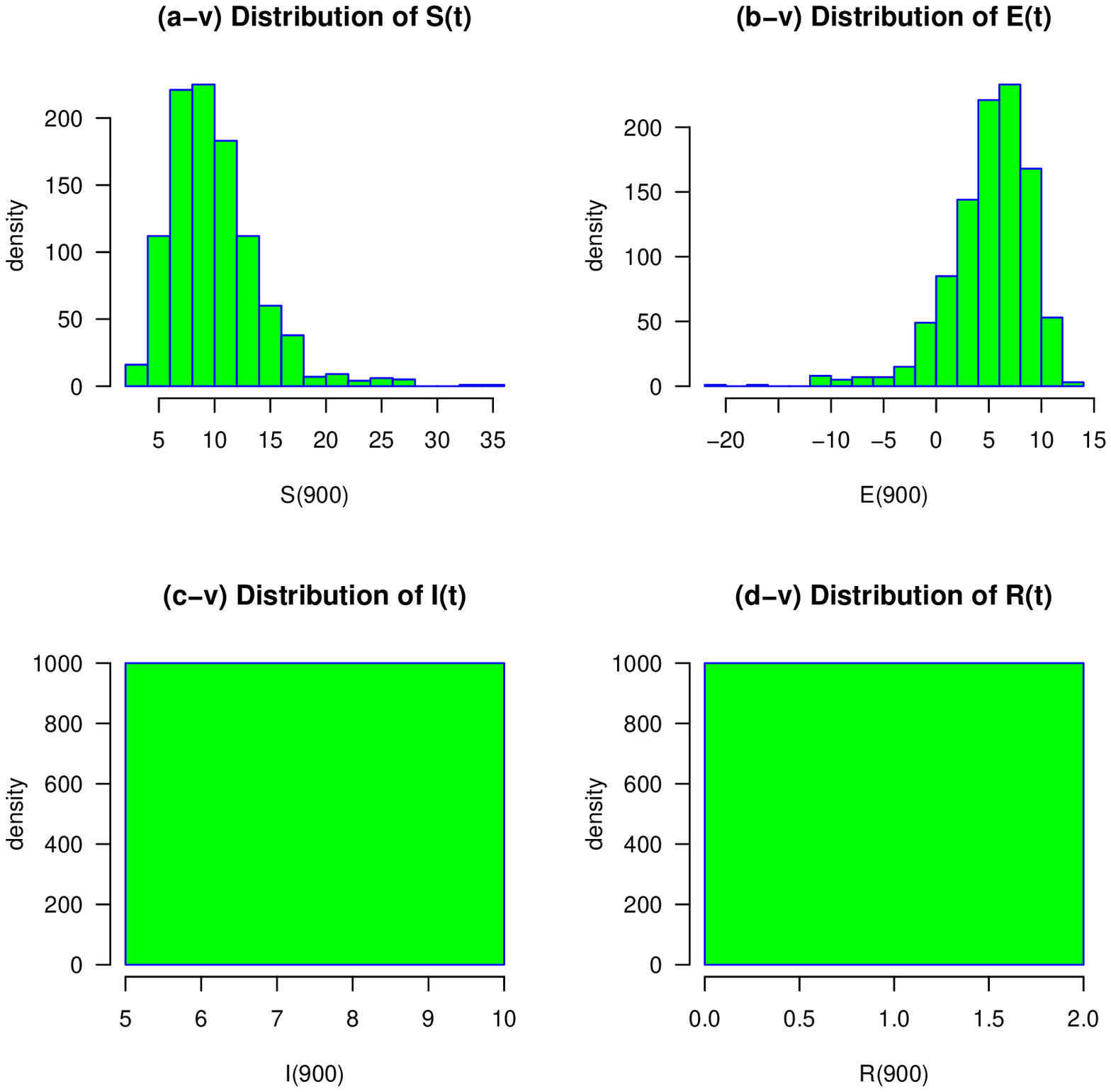}
\caption{(a-v), (b-v), (c-v) and (d-v) show the approximate distribution of the different disease states $(S,E,I,R)$ respectively,  at the specified time $t=900$, whenever the only source of noise in the system is from the disease transmission rate, and the  intensity of the noise in the system is relatively high, that is,  $\sigma_{\beta}=5$.
}\label{ch1.sec4.subsec1.1.figure 5}
\end{center}
\end{figure}
\begin{figure}[H]
\begin{center}
\includegraphics[height=6cm]{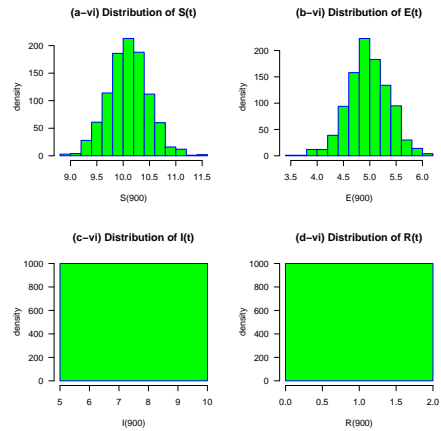}
\caption{(a-vi), (b-vi), (c-vi) and (d-vi) show the approximate distribution of the different disease states $(S,E,I,R)$ respectively,  at the specified time $t=900$, whenever the only source of noise in the system is from the disease transmission rate, and the  intensity of the noise in the system is relatively low, that is,  $\sigma_{\beta}=0.5$.
}\label{ch1.sec4.subsec1.1.figure 6}
\end{center}
\end{figure}

The Figures~\ref{ch1.sec4.subsec1.1.figure 1}-\ref{ch1.sec4.subsec1.1.figure 3} and Figures~\ref{ch1.sec4.subsec1.1.figure 4}-\ref{ch1.sec4.subsec1.1.figure 6} can be used to examine the persistence and permanence of the disease in the human population exhibited in Theorem~\ref{ch1.sec3.thm2} and Theorem~\ref{ch1.sec5.thm2}, respectively. The following observations are made:- (1) the occurrence of noise in the disease transmission rate, that is, $\sigma_{\beta}>0$, results in random fluctuations mainly in the susceptible and exposed states depicted in Figures~\ref{ch1.sec4.subsec1.1.figure 1}-\ref{ch1.sec4.subsec1.1.figure 3}(a-i)-(a-iii), (b-i)-(b-iii). No major oscillations are observed in the infectious and removed states $I(t)$ and $R(t)$. This observation is also more significant in the approximate uniform distributions observed for the $I(t)$ and $R(t)$ states at the time $t=900$ depicted in Figures~\ref{ch1.sec4.subsec1.1.figure 4}-\ref{ch1.sec4.subsec1.1.figure 6}(c-iv)-(c-vi), (d-iv)-(d-vi), based on 1000 sample points for $I(t)$ and $R(t)$ at the fixed time $t=900$ (that is, 1000 sample observations of $I(900)$ and $R(900)$).

(2) Increasing the intensity of the noise from the disease transmission rate, that is, as $\sigma_{\beta}$ increases from $0.5$ to $10$, it results in a rise in infection with many more people becoming exposed to the disease. This fact is depicted in Figures~\ref{ch1.sec4.subsec1.1.figure 1}-\ref{ch1.sec4.subsec1.1.figure 3}(a-i)-(a-iii), (b-i)-(b-iii), where a new higher maximum value for the  trajectories of the exposed state $E(t)$ , and a new lower minimum value for the trajectory of the susceptible state $S(t)$ are attained over the time interval $[0,1000]$, respectively, across the figures as  $\sigma_{\beta}$ increases from $0.5$ to $10$. Therefore, stronger noise in the disease dynamics from the disease transmission rate leads to more persistence of the disease. This observation about the persistence of disease is also significant in the approximate distributions for the susceptible and exposed states, $I(t)$ and $R(t)$, respectively, at the time $t=900$ depicted in Figures~\ref{ch1.sec4.subsec1.1.figure 4}-\ref{ch1.sec4.subsec1.1.figure 6}(a-iv)-(a-vi), (b-iv)-(b-vi), based on 1000 sample points for $I(t)$ and $R(t)$ at the fixed time $t=900$.

Indeed, it can be seen from Figures~\ref{ch1.sec4.subsec1.1.figure 4}-\ref{ch1.sec4.subsec1.1.figure 6}(a-iv)-(a-vi), (b-iv)-(b-vi) that when the intensity of the noise in the system is $\sigma_{\beta}=0.5$, the distributions of $S(900)$ and $E(900)$ are closely symmetric with one peak, with the center for $S(900)$ approximately between $(9.5, 10.5)$, and about $97.3\%$ of the values between $(9.0,11.0)$. The center for $E(900)$ approximately between $(4.5, 5.5)$, and about $97.3\%$ of the values between $(4.0,6.0)$.

Now, when the intensity increases to $\sigma_{\beta}=5$, and to $\sigma_{\beta}=10$,  the distributions of $S(900)$ and $E(900)$ continuously become more sharply skewed, with $S(900)$ skewed the right with center ( utilizing the mode of $S(900)$) shifting to the left with majority of the possible values for $S(900)$ continuously decreasing in magnitude from approximately under the interval $(0, 30)$, to the interval $(0,20)$. These changes of the shape of the distribution, and decrease of the range of values in the support  for the distribution of $S(900)$ as the the intensity rises from $\sigma_{\beta}=5$, and to $\sigma_{\beta}=10$, indicates that more susceptible tend to get infected at the time $t=900$ as the intensity of the noise rises.

Similarly, when the intensity increases from $\sigma_{\beta}=0.5$ to $\sigma_{\beta}=5$, and also from $\sigma_{\beta}=5$ to $\sigma_{\beta}=10$,  the distribution of $E(900)$ is skewed to the left with center ( utilizing the mode of $E(900)$) shifting to the right with majority of the possible values in the support for $E(900)$ continuously increasing in magnitude from under the interval $(0, 10)$, to the interval $(0,20)$. These changes of the shape of the distribution, and increase in the magnitude of the range of values in the the support  for $E(900)$ as the the intensity rises from $\sigma_{\beta}=5$,  to $\sigma_{\beta}=10$, indicates that more susceptible people tend to get infected and become exposed at the time $t=900$.

(3)Finally, the remark about the influence of the strength of the noise on the stochastic permanence in the mean of the disease in Remark~\ref{ch1.sec5.rem1} can be examined using Figures~\ref{ch1.sec4.subsec1.1.figure 1}-\ref{ch1.sec4.subsec1.1.figure 3}, and Figures~\ref{ch1.sec4.subsec1.1.figure 4}-\ref{ch1.sec4.subsec1.1.figure 6}. Recall, (\ref{ch1.sec5.rem1.eq2}) in Remark~\ref{ch1.sec5.rem1} corresponding to Theorem~\ref{ch1.sec5.thm2} asserts that when the intensity of the noise $\sigma_{\beta}$ is infinitesimally small, a larger asymptotic lower bound $l_{a}(\sigma^{2}_{max})=l_{a}(\sigma^{2}_{\beta}),a\in \{S, E, I\}$ is attained for the average in time of the sample path of each state heavily influenced by the random fluctuations in the system, which in this scenario is the $S(t)$ and $E(t)$ states. Across the figures, as the intensity rises from $\sigma_{\beta}=0.5$,  to $\sigma_{\beta}=10$ in Figures~\ref{ch1.sec4.subsec1.1.figure 1}-\ref{ch1.sec4.subsec1.1.figure 3}(a-i)-(a-iii), (b-i)-(b-iii), and Figures~\ref{ch1.sec4.subsec1.1.figure 4}-\ref{ch1.sec4.subsec1.1.figure 6}(a-iv)-(a-vi), (b-iv)-(b-vi),  smaller minimum values for the paths of $S(t)$ and $E(t)$ are observed in Figures~\ref{ch1.sec4.subsec1.1.figure 1}-\ref{ch1.sec4.subsec1.1.figure 3}(a-i)-(a-iii), (b-i)-(b-iii).

 Also, the centers (utilizing the mean) for $S(900)$ and $E(100)$ continuously decrease in magnitude as the intensity rises from $\sigma_{\beta}=0.5$, and to $\sigma_{\beta}=10$. Indeed, this is evident since for the figures that are symmetric with a single peak, that is, Figures~\ref{ch1.sec4.subsec1.1.figure 6}(a-iv)-(b-iv) corresponding to $\sigma_{\beta}=0.5$,  the measures of the center (mean, mode, and median) are all approximately equal and higher in magnitude, while for the figures that are skewed ( to the left, or to the right),  that is, Figures~\ref{ch1.sec4.subsec1.1.figure 4}-\ref{ch1.sec4.subsec1.1.figure 5} ((b-iv) and (b-v) skewed to the left, and (a-iv), (a-v) skewed to the right),  the position of the mean is also pulled further to the direction of the skew, which is relatively lower in magnitude compared to the observation in Figures~\ref{ch1.sec4.subsec1.1.figure 6}. And note that despite the fact that the approximate distribution of $S(900)$ is skewed to the right in Figures~\ref{ch1.sec4.subsec1.1.figure 4}-\ref{ch1.sec4.subsec1.1.figure 5}(a-iv)-(a-v), the mean continuously becomes smaller in magnitude as the intensity rises from $\sigma_{\beta}=0.5$ to $\sigma_{\beta}=10$.
\subsubsection{Joint effect of the intensities of white noise $\sigma_{i},i=S, E, I, \beta$ on the persistence of the disease  }\label{ch1.sec4.subsec1.2}
\begin{figure}[H]
\begin{center}
\includegraphics[height=6cm]{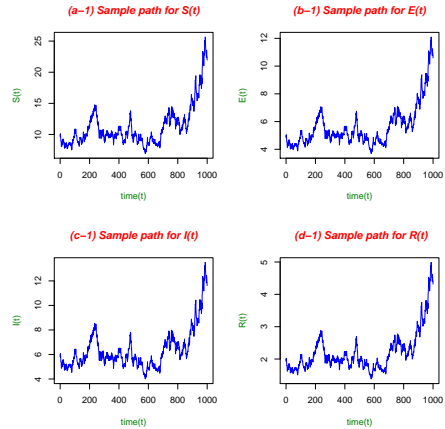}
\caption{(a1), (b1), (c1) and (d1) show the trajectories of the disease states $(S,E,I,R)$ respectively, whenever various sources of the noise in the system (i.e. natural death and disease transmission rates) are assumed to have the same strength or intensity (i.e. $\sigma_{i}=\sigma_{\beta}, i\in \{S, E, I, R\}$), and the strength of the noises in the system are relatively small, that is, $\sigma_{i}=0.5, \forall i\in \{S, E, I, R, \beta\} $. The broken lines represent the endemic equilibrium $E_{1}=(S^{*}_{1},E^{*}_{1},I^{*}_{1})=(0.2286216,0.07157075,0.9929282)$.  Furthermore, $\min{(S(t))}= 6.941153$, $\min{(E(t))}=3.693744$, $\min{(I(t))}=4.004954$ and
$\min{(R(t))}= 1.384626$.
}\label{ch1.sec4.figure 1}
\end{center}
\end{figure}
\begin{figure}[H]
\begin{center}
\includegraphics[height=6cm]{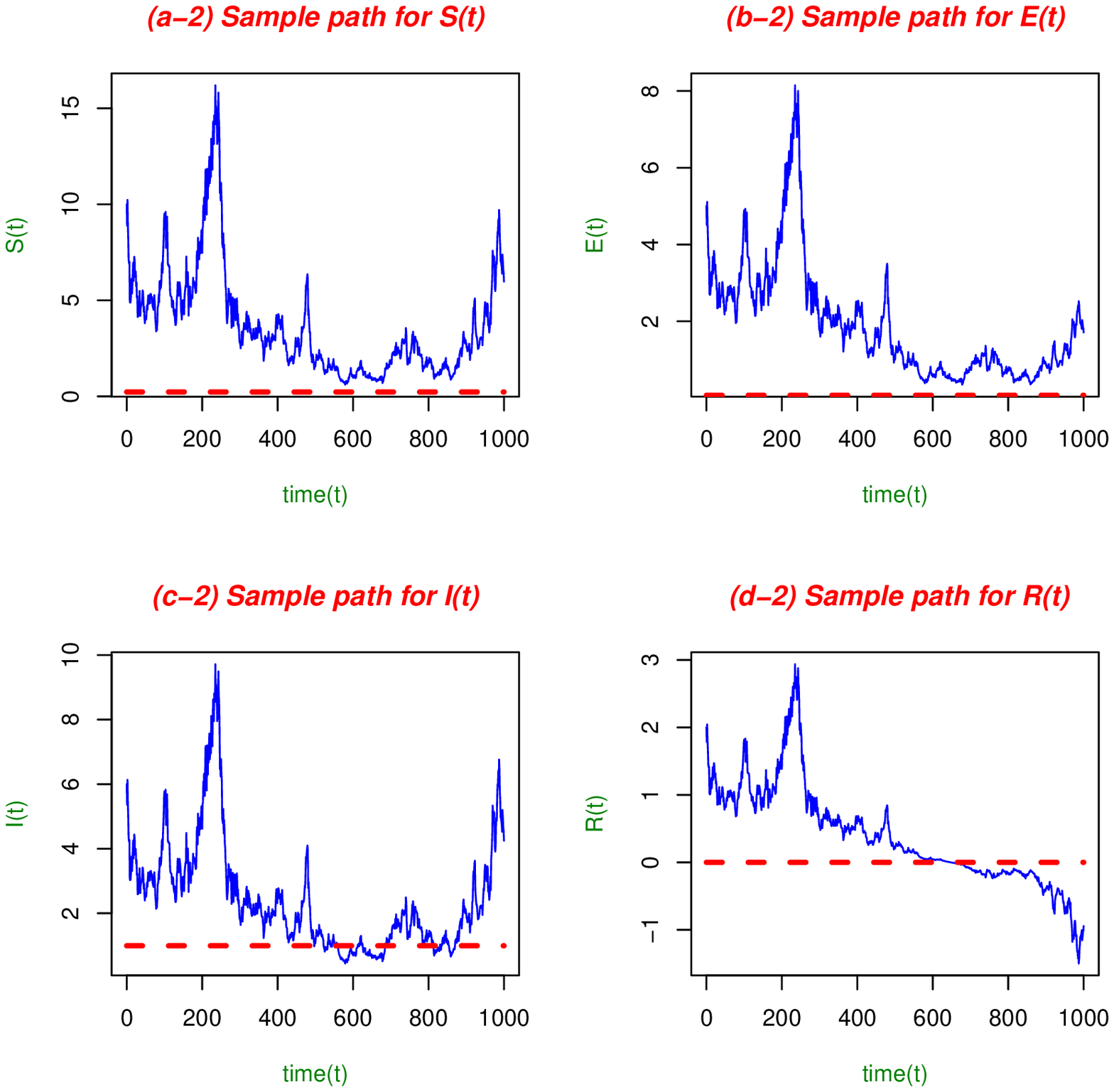}
\caption{(a2), (b2), (c2) and (d2) show the trajectories of the disease states $(S,E,I,R)$ respectively, whenever the various sources of the noise in the system (i.e. natural death and disease transmission rates) are assumed to have the same strength or intensity (i.e. $\sigma_{i}=\sigma_{\beta}, i\in \{S, E, I, R\}$), and the strength of the noises in the system are relatively moderate, that is, $\sigma_{i}=1.5, \forall i\in \{S, E, I, R, \beta\} $. The broken lines represent the endemic equilibrium $E_{1}=(S^{*}_{1},E^{*}_{1},I^{*}_{1})=(0.2286216,0.07157075,0.9929282)$. Furthermore, $\min{(S(t))}= 0.6116444$, $\min{(E(t))}=0.3496159$, $\min{(I(t))}=0.4439692$ and
$\min{(R(t))}= -1.499425$.
}\label{ch1.sec4.figure 2}
\end{center}
\end{figure}
\begin{figure}[H]
\begin{center}
\includegraphics[height=6cm]{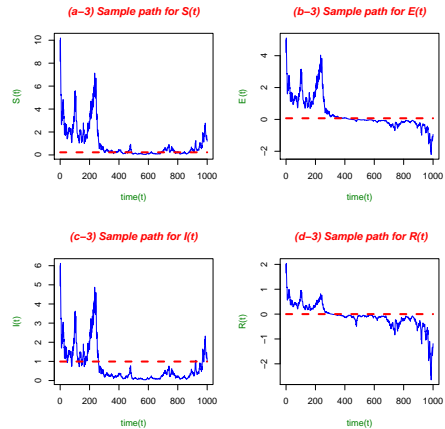}
\caption{(a3), (b3), (c3) and (d3) show the trajectories of the disease states $(S,E,I,R)$ respectively, whenever the various sources of the noise in the system (i.e. natural death and disease transmission rates) are assumed to have the same strength or intensity (i.e. $\sigma_{i}=\sigma_{\beta}, i\in \{S, E, I, R\}$), and the strength of the noises in the system are relatively high, that is, $\sigma_{i}=2.5, \forall i\in \{S, E, I, R, \beta\}$. The broken lines represent the endemic equilibrium $E_{1}=(S^{*}_{1},E^{*}_{1},I^{*}_{1})=(0.2286216,0.07157075,0.9929282)$. Furthermore, $\min{(S(t))}=0.03717095$, $\min{(E(t))}=-2.204265$, $\min{(I(t))}=0.03397191$ and
$\min{(R(t))}= -2.645903$.
 }\label{ch1.sec4.figure 3}
\end{center}
\end{figure}

The Figures~\ref{ch1.sec4.figure 1}-\ref{ch1.sec4.figure 3} can be used to examine the persistence of the disease in the human population as the intensities of the white noises in the system equally and continuously rise in value between $0.5$ to $2.5$, that is, for $\sigma_{i}=0.5, 1.5, 2.5, \forall i\in \{S, E, I, R, \beta\}$.
It can be observed from Figure~\ref{ch1.sec4.figure 1} that when the intensity of noise is relatively small, that is,  $\sigma_{S}=\sigma_{E}=\sigma_{\beta}=\sigma_{I}=\sigma_{R}=0.5$, the disease persists in the population with a significantly higher lower bounds for the disease classes $E, I$, and $R$. The lower bound for these disease classes are seen to continuously decrease in value as the magnitude of the intensities rise from $\sigma_{i}= 0.5$  to $\sigma_{i}= 1.5, \forall i\in \{S, E, I, R, \beta\}$, and also increase to $\sigma_{i}= 2.5, \forall i\in \{S, E, I, R, \beta\}$.
This observation confirms the result in Theorems~[\ref{ch1.sec5.thm1}\&\ref{ch1.sec5.thm2}] and Remark~\ref{ch1.sec5.rem1}, which asserts that an increase in the intensities of the noises in the system tends to lead to  persistence of the disease with smaller lower bounds for the paths of the disease related classes, while a decrease in the strength of the noise in the system allows the disease to persist with a relatively higher lower margin for the paths of the disease related classes. It is important to note that the continuous decrease in the lower bounds for the paths of the disease related states $E, I, R$ are also matched with continuous decrease in the lower margin for the susceptible class exhibited in Figures~\ref{ch1.sec4.figure 1}-\ref{ch1.sec4.figure 3}~[(a1), (a2), (a3)]. This observation suggests that the population gets extinct overtime with continuous rise in the intensity of the noises in the system.

\begin{figure}[H]
\begin{center}
\includegraphics[height=6cm]{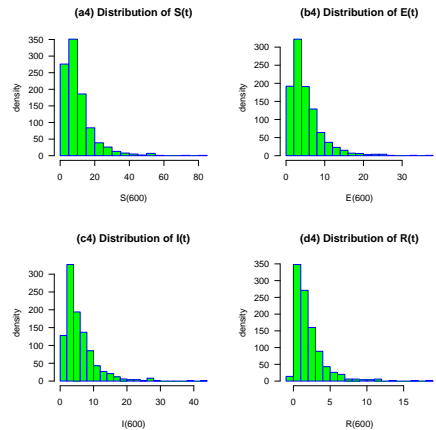}
\caption{(a4), (b4), (c4) and (d4) show the approximate distribution of the different disease states $(S,E,I,R)$ respectively,  at the specified time $t=600$, whenever the intensities of the noises in the system are relatively small, that is,  $\sigma_{E}=\sigma_{I}=\sigma_{R}=0.5$.
}\label{ch1.sec4.figure 4}
\end{center}
\end{figure}
\begin{figure}[H]
\begin{center}
\includegraphics[height=6cm]{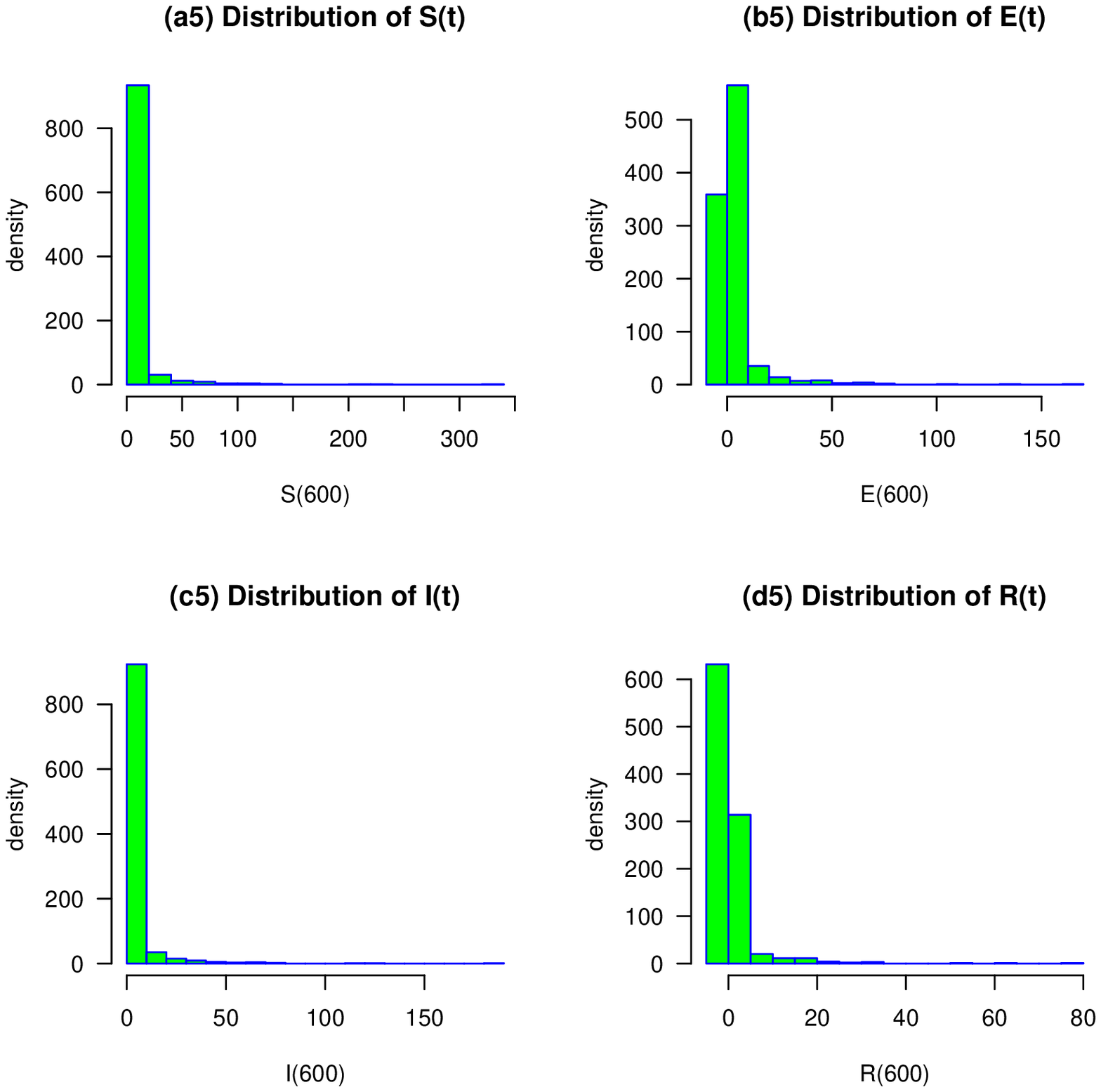}
\caption{(a5), (b5), (c5) and (d5) show the approximate distribution of the different disease states $(S,E,I,R)$ respectively,  at the specified time $t=600$, whenever the intensities of the noises in the system are relatively moderate, that is,  $\sigma_{E}=\sigma_{I}=\sigma_{R}=1.5$.
}\label{ch1.sec4.figure 5}
\end{center}
\end{figure}
\begin{figure}[H]
\begin{center}
\includegraphics[height=6cm]{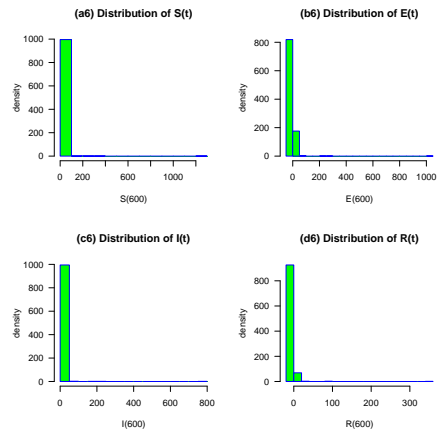}
\caption{(a6), (b6), (c6) and (d6) show the approximate distribution of the different disease states $(S,E,I,R)$ respectively,  at the specified time $t=600$, whenever  the intensities of the noises in the system are relatively high, that is,  $\sigma_{E}=\sigma_{I}=\sigma_{R}=2.5$.
}\label{ch1.sec4.figure 6}
\end{center}
\end{figure}
Figure~\ref{ch1.sec4.figure 4}-Figure~\ref{ch1.sec4.figure 6} provides a clearer picture about the effect of the rise in the intensity of the noises in the system on the persistence of the disease at any  given time, for example, when the time is $t=600$. The statistical graphs in Figure~\ref{ch1.sec4.figure 4}-Figure~\ref{ch1.sec4.figure 6} are based on samples of 1000 simulation observations for the different states in the system $S, E, I$ and $R$ at the time $t=600$.  For the susceptible population in Figure~\ref{ch1.sec4.figure 4}-Figure~\ref{ch1.sec4.figure 6}[(a4)-a(6)], it can be seen that the majority of possible values in the support for $S(600)$ occur in $(0,60)$. However, the frequency of these values dwindle with the rise in the intensity of the noises in the system from $\sigma_{i}= 0.5, \forall i\in \{S, E, I, R, \beta\}$ to $\sigma_{i}= 2.5, \forall i\in \{S, E, I, R, \beta\}$. Moreover, the much smaller values in the support for $S(600)$ tend to occur more frequently as is depicted in Figure~\ref{ch1.sec4.figure 5} and Figure~\ref{ch1.sec4.figure 6}. This observation suggests that the  rise in the intensity of the noise in the system increases the probability of occurrence for smaller values in the support of the susceptible population state $S$ at time $t=200$, and this further suggests that more susceptible people tend to converted out of the susceptible state, either as a result of infection or  natural death.

 Similar observations can be made for the disease related classes:- exposed($E$), infectious ($I$) and removal($R$) populations in Figure~\ref{ch1.sec4.figure 4}-Figure~\ref{ch1.sec4.figure 6}[(b4)-b(6)], [(c4)-c(6)] and [(d4)-d(6)] respectively.  It can be seen that the majority of possible values in the support for $E(600)\leq 50$, $0\leq I(600)\leq 30$, and $R(600)\leq 20$. However, the frequency of these values also dwindle with the rise in the intensity of the noises in the system from $\sigma_{i}= 0.5, \forall i\in \{S, E, I, R, \beta\}$ to $\sigma_{i}= 2.5, \forall i\in \{S, E, I, R, \beta\}$. Moreover, much smaller values in the support for $I(600)$ tend to occur more frequently as is seen in Figure~\ref{ch1.sec4.figure 5} and Figure~\ref{ch1.sec4.figure 6}, while negative values for $E(600)$ and $R(600)$ tend to occur the most for these disease related classes. This observation suggests that the  rise in the intensity of the noise in the system increases the probability for smaller values in the support of  the disease related states $E, I, R$ in the population at time $t=200$ to occur.

 This observation further suggests that more exposed people tend to be converted from the exposed state  either as a result of more exposed people turning into  full blown infectious individuals or  natural death. For the infectious population, this observation suggests that more infectious people tend to be converted from the infectious state either because more infectious people tend to recover from the disease or die naturally, or die from disease related causes. Also, for the recovery class, the high probability of occurrence of smaller values in the support for $R(200)$ suggests that more removed individuals tend to be converted out of the state either because more naturally immune persons tend to lose their naturally acquired immunity and become susceptible again to the disease, or because they tend to die naturally. It is important to note that the occurrence of the negative values in the support for $E(600)$ and $R(600)$ with high probability values signifies that in many occasions, the exposed and removal populations become extinct.
 \begin{rem}
 The numerical simulation results in the subsections~[\ref{ch1.sec4.subsec1.1}-\ref{ch1.sec4.subsec1.2}] suggest and highlight an important fact that the stochastic system (\ref{ch1.sec0.eq8})-(\ref{ch1.sec0.eq11}) exhibits higher sensitivity (such as high persistence of disease, tendency for the population to become extinct etc.) to changes in the magnitude of the intensities of the noises from the natural deathrates (i.e. $\sigma_{i}=0.5, 1.5, 2.5, \forall i\{S, E, I, R\})$ observed in Figures~\ref{ch1.sec4.figure 1}-\ref{ch1.sec4.figure 6}, compared to  changes in the magnitude of the intensity of the noise from the disease transmission rate (i.e. $\sigma_{\beta}=0.5, 5, 10$) exhibited in Figures~\ref{ch1.sec4.subsec1.1.figure 1}-\ref{ch1.sec4.subsec1.1.figure 6}. This fact can be easily seen in the occurrence and size of the oscillations in the sample paths of the different states $S, E, I, R$ represented in the  figures, as the intensities continuously change values from small to high values.  For example, the continuous changes in the intensity  $\sigma_{\beta}=0.5, 5, 10$ result to a possibility for extinction of the human population only for very high intensity values, while small changes in the intensities $\sigma_{i}=0.5, 1.5, 2.5, \forall i\{S, E, I, R\})$ result to the possibility of extinction of the human population at disproportionately smaller magnitudes of the intensity values. This suggests that the growth rates of the intensities in the system influence the qualitative behavior of the trajectories of the system, and consequently impact the disease dynamics over time. This fact about growth rates of the intensities of the white noise processes in the stochastic system (\ref{ch1.sec0.eq8})-(\ref{ch1.sec0.eq11}), and various classifications of the growth rates and the qualitative behavior of the disease dynamics is the central theme of the study Wanduku\cite{wanduku-theorBio}.
\end{rem}
\subsection{Evidence of stationary distribution}\label{ch1.sec4.subsec2}
The following new convenient set of parameter values in Table~\ref{ch1.sec4.table3} are used to examine the results of Theorem~\ref{ch1.sec3.sec1.thm1} about the existence of a stationary distribution for the different states of the  stochastic system (\ref{ch1.sec0.eq8})-(\ref{ch1.sec0.eq11}), whenever $R^{*}_{0}>1$, and the delay times $T_{1}, T_{2}$ and $T_{3}$ are constant.

\begin{figure}[H]
\begin{center}
\includegraphics[height=6cm]{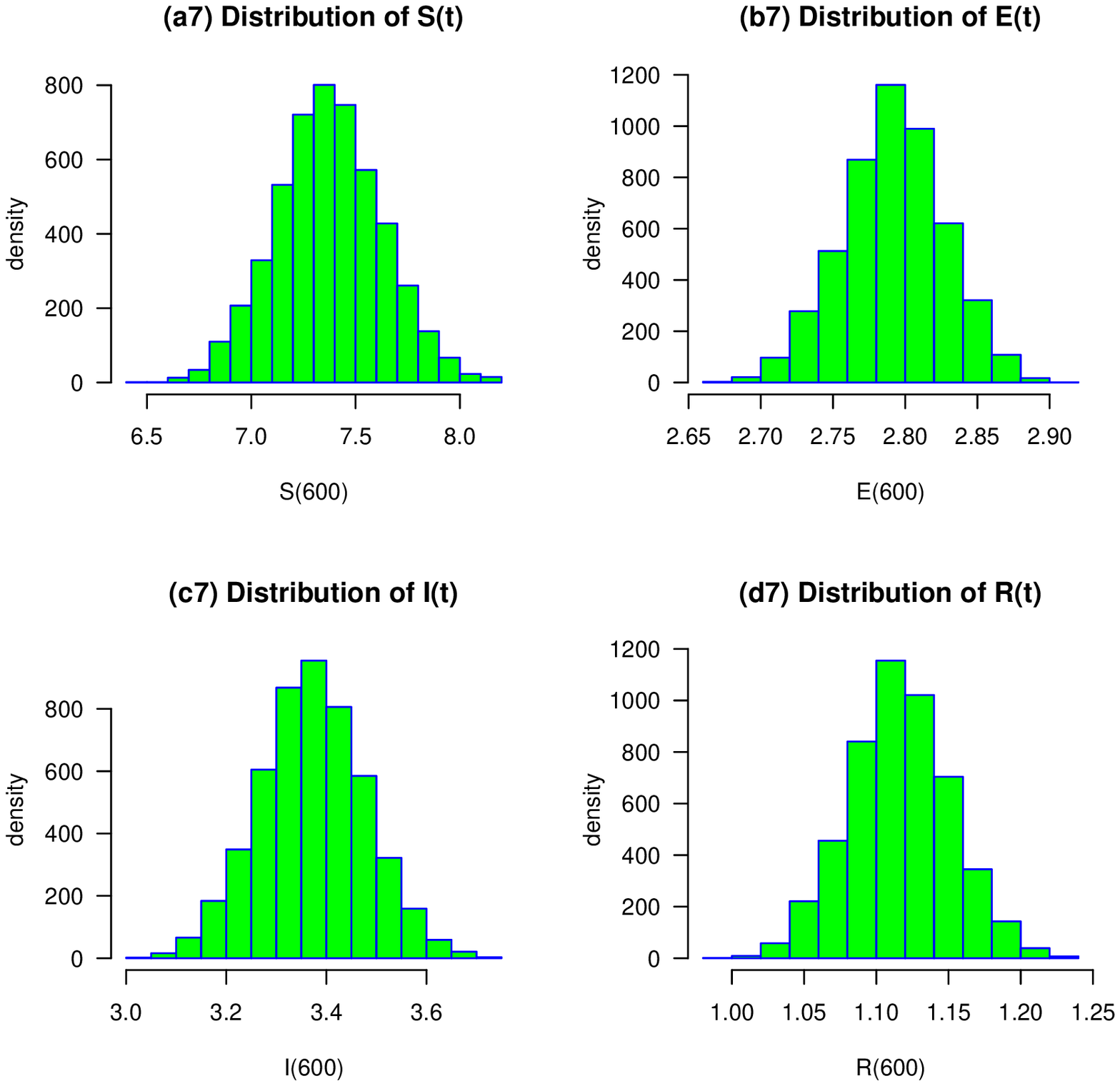}
\caption{(a7), (b7), (c7) and (d7) depict an approximate distribution for the different disease states $(S,E,I,R)$ respectively, at the time when $t=600$.
}\label{ch1.sec4.figure 7}
\end{center}
\end{figure}

\begin{figure}[H]
\begin{center}
\includegraphics[height=6cm]{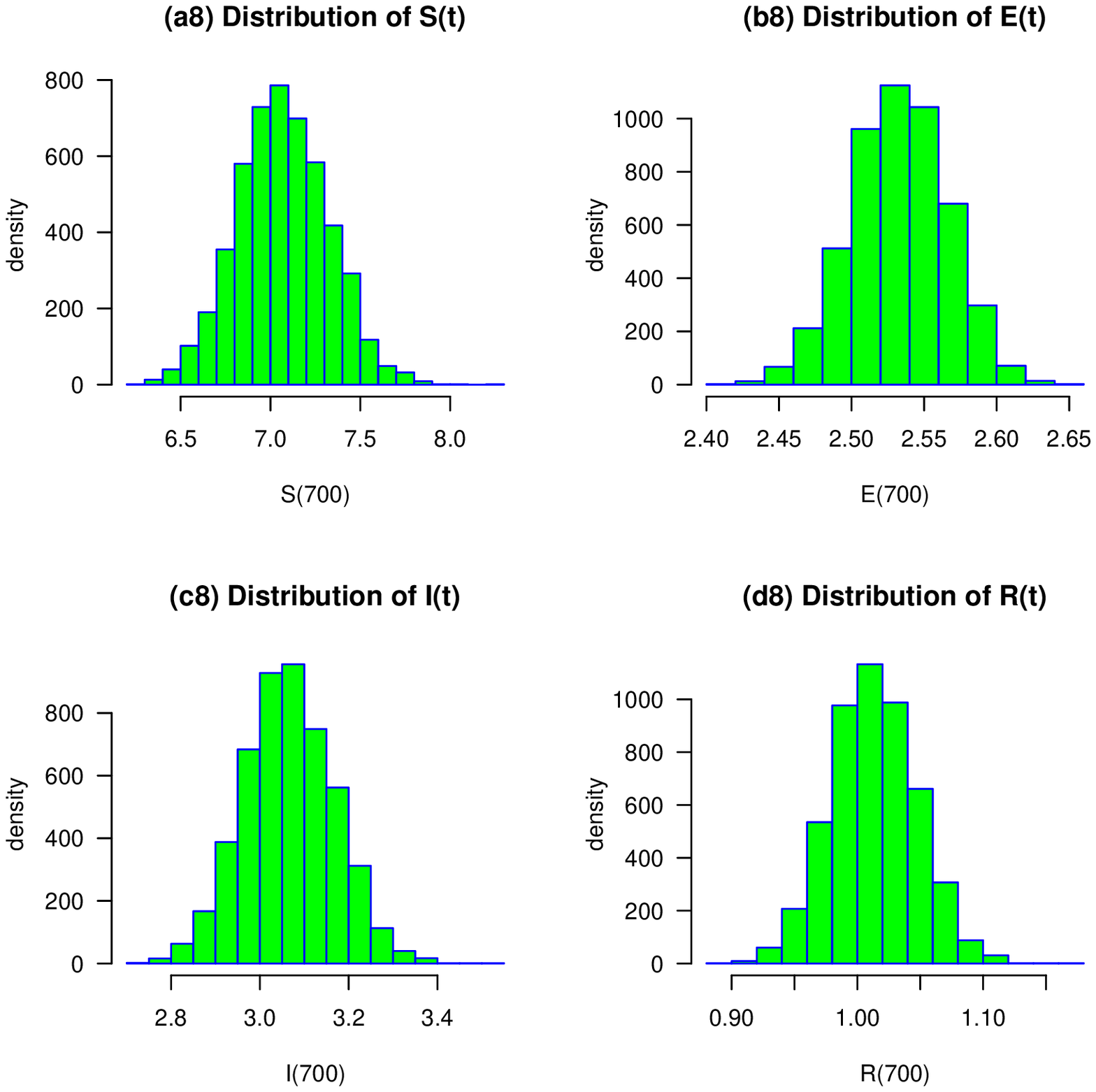}
\caption{
(a8), (b8), (c8) and (d8) depict an approximate distribution for the different disease states $(S,E,I,R)$ respectively, at the time when $t=700$.
}\label{ch1.sec4.figure 8}
\end{center}
\end{figure}

\begin{figure}[H]
\begin{center}
\includegraphics[height=6cm]{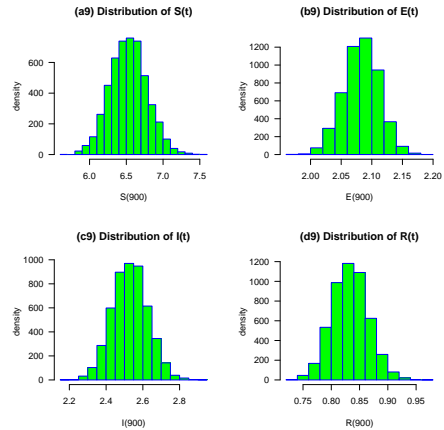}
\caption{
(a9), (b9), (c9) and (d9) depict an approximate distribution for the different disease states $(S,E,I,R)$ respectively, at the time when $t=900$.
}\label{ch1.sec4.figure 9}
\end{center}
\end{figure}

It is easily seen that for this set of parameter values, $R^{*}_{0}=10.59709>1$. Furthermore, the endemic equilibrium for the system $E_{1}$ is given as follows:- $E_{1}=(S^{*}_{1},E^{*}_{1},I^{*}_{1})=(2.83334,0.2755608,2.521028)$. And, the parameter $\lambda(\mu)$ is taken to be $\lambda(\mu)=\mu e^{-\mu(T_{1}+T_{2})}$. Also, the intensities of the white noise processes in the system are $\sigma_{\beta}=2$ and $\sigma_{i}=0.2, \forall i\{S, E, I, R\})$.  In addition, for the condition in (\ref{ch1.sec3.sec1.thm1.eq2}), it is also easy to see that $\Phi=0.01223636$,  and
\[\min{[\Phi_{1}(S^{*}_{1})^{2}, \Phi_{2}(E^{*}_{1})^{2}, \Phi_{3}(I^{*}_{1})^{2}]}=\min{(2.697389, 0.03695982, 1.546866) }.\]
 Thus, (\ref{ch1.sec3.sec1.thm1.eq2}) is satisfied. The approximate distributions of the different disease states $S, E, I, R$ based on  samples of  5000 simulation realizations, at the different times $t=600$, $t=700$ and $t=900$ are given in Figure~\ref{ch1.sec4.figure 7}-Figure~\ref{ch1.sec4.figure 9}, respectively.
 \begin{table}[h]
  \centering
  \caption{A list of specific values chosen for the system parameters for the example in subsection~\ref{ch1.sec4.subsec2}.}\label{ch1.sec4.table3}
  \begin{tabular}{l l l}
  Disease transmission rate&$\beta$& $0.6277e^{-\mu(T_{1}+T_{2})}$\\\hline
  Constant Birth rate&$B$&$ 1$\\\hline
  Recovery rate& $\alpha$& $5.067\times 10^{-8}e^{-\mu(T_{1}+T_{2})}$\\\hline
  Disease death rate& $d$& $1.838\times 10^{-8}e^{-\mu(T_{1}+T_{2})}$\\\hline
  Natural death rate& $\mu$& $0.2433696$\\\hline
  Incubation delay time in vector& $T_{1}$& 2 units \\\hline
  Incubation delay time in host& $T_{2}$& 1 unit \\\hline
  Immunity delay time& $T_{3}$& 4 units\\\hline
  \end{tabular}
\end{table}
It is easy to see from Figure~\ref{ch1.sec4.figure 7}-Figure~\ref{ch1.sec4.figure 9} that the approximate distributions for the different states $S, E, I$ and $R$ have approximately common location, scale and shape parameters.  These same conclusions can be reached about the location, scale and shape parameters of the corresponding distributions for the different states $S, E, I$ and $R$ obtained  for all  bigger times than $t=600$, that is, $\forall t\geq 600$. This clearly proves the existence of a stationary distribution for the states of the stochastic system (\ref{ch1.sec0.eq8})-(\ref{ch1.sec0.eq11}), which is the limit of convergence (in distribution) of the sequence (collection) of distributions of the different states of the system, that is, $S(t), E(t), I(t), R(t)$, indexed by the time $t>0$,  whenever the conditions in Theorem~\ref{ch1.sec3.sec1.thm1} are satisfied, and all possible simulation realizations are generated for the different states in the system at each time $t$.

Furthermore, one can infer from the descriptive statistics of the approximated distributions presented in Figure~\ref{ch1.sec4.figure 7}-Figure~\ref{ch1.sec4.figure 9} that the location or centering parameters (e.g. ensemble mean) of the true stationary distribution for the susceptible population lies approximately in the interval [6.5,8.0]. Moreover, the shape of the distribution is approximately symmetric with a single peak. Similarly, for the disease related classes namely:- exposed, infectious and removal classes, one can infer that the shapes of the stationary distributions for each of these classes are approximately symmetric with a single peak, and the location or centering parameters lie approximately in the intervals $[2.45,2.85]$, $[2.8,3.6]$ and $[0.95,1.90]$, respectively.

It is also important to note that whilst the statistical parameters such as the measures of center, variation and moments of the stationary distribution for the states of the system (\ref{ch1.sec0.eq8})-(\ref{ch1.sec0.eq11}) may change for each given set of values selected for the parameters of the system in Table~\ref{ch1.sec4.table3}, the stationary distribution obtained is unique for that set of system parameter values.

 \section{Conclusion}
 The presented family of stochastic malaria models with nonlinear incidence rates, random delays and environmental perturbations  characterizes the general dynamics of malaria in a highly random environment with variability originating from (1.) the disease transmission rate between mosquitoes and humans, and also from (2.)  the natural death rates of the  susceptible, exposed, infectious and removal individuals of the human population. The random environmental fluctuations are formulated as independent white noise processes, and the malaria dynamics is expressed as a family of  Ito-Doob type stochastic differential equations. Moreover, the family type is determined by a general nonlinear incidence function $G$ with various mathematical properties. The nonlinear incidence function $G$ can be used to describe the nonlinear character of malaria transmission rates in various disease scenarios where the rate of malaria transmission may initially increase or decrease, and become steady or bounded as the number of malaria cases increase in the population.

This work furthers investigation about the malaria project initiated in Wanduku\cite{wanduku-biomath,wanduku-comparative} and focuses on (1) the stochastic permanence of malaria, and (2) the existence of a stationary distribution for the random process describing the states of the disease over time. The investigation about these two aspects above centers on analyzing the behavior of the sample paths of the different states of the stochastic process in the neighborhood of the  potential endemic equilibrium state of the dynamic system. Lyapunov functional techniques, and other local martingale characterizations are applied to characterize the trajectories of the solution process of the stochastic dynamic system.

 In addition, much emphasis is laid on analyzing the impacts of  (a) the sources of the noises-disease transmission or natural death rates, and (b) the intensities of the noises on the permanence of malaria and the existence of a stationary distribution for the disease. Expansive and exhaustive discussions are presented to elucidate the qualitative character of the permanence of malaria and stationary distribution for the disease under the influence of the different sources, and different intensities of the noises in the system.

  The results of this study are illuminated by detailed numerical simulation examples that examine the trajectories of the states of the stochastic process in the neighborhood of the endemic equilibrium, and also under the influence of the different sources, and different intensities of the noises in the system.

  The numerical simulation results suggest that higher intensities of the white processes in the system drive the sample paths of the stochastic system further away from the potential endemic steady state.  Moreover, the intensities of the noises from the natural death rates seem to have stronger consequences on the evolution of the disease in the system compared to the intensity of the noise from the disease transmission rate. Also, there is some evidence of a high chance of the population becoming extinct over time, whenever the intensities of the white noise processes become large.

      Furthermore, in the absence of explicit solution process for the nonlinear stochastic system of differential equations, the stationary distribution is numerically approximated for a given set of parameter values for the stochastic dynamic system of equations.


\newpage
\section{References}
\begin{thebibliography}{300}
\bibitem{yongli-2} Y. Cai, Y. kang et al., a stochastic epidemic model incorporatinmg media coverage, commun. math sci, vol. 14, n0.4,(2016) 893-910
\bibitem{mao-2} A. Gray, D. Greenhalgh, L. Hu, X. Mao, and J. Pan, A Stochastic Differential Equation SIS Epidemic Model, SIAM J. Appl. Math., 71(3), (2011) 876–902
\bibitem{aadil} A. lahrouz, L. Omari, extinction and stationary distribution of a stochastic SIRS epidemic model with non-linear incidence, Statistics \& Probability Letters 83(4):(2013)960–968
\bibitem{wanduku-biomath} D. Wanduku, Threshold conditions for a family of epidemic dynamic models for malaria with distributed delays in a non-random environment, International Journal of Biomathematics Vol. 11, No. 6 (2018) 1850085 (46 pages), DOI: 10.1142/S1793524518500857
    \bibitem{Li2008} Li, Y., On the almost surely asymptotic bounds of a class of ornstein-Uhlenbeck Process in finifte dimensions, Journal of Systems Science and Complexity, 21 (2008), 416-426.
\bibitem{maobook} X. Mao, stochastic differential equations and application, 2nd ed.,  WP, 2007
\bibitem{imf}A. G. Ladde, G.S.Ladde, An introduction to differential equations: stochastic modelling, methods and analysis, vol 2, world scientific publishing, 2013
\bibitem{wanduku-bookchapter}D. Wanduku, Modeling highly random dynamical infectious systems (book chapter), in press, 2017
\bibitem{wanduku-theorBio}D. Wanduku, Analyzing the qualitative properties of white noise on a family of infectious disease models in a highly random environment, available at arXiv:1808.09842 [q-bio.PE]
    \bibitem{wanduku-comparative}D. Wanduku, A comparative stochastic and deterministic study of a class of epidemic dynamic models for malaria: exploring the impacts of noise on eradication and persistence of disease, in press, 2017
\bibitem{yongli} Y. Cai, Y. Kang, W. Wang, a stochastic SIRS epidemic model with nonlinear incidence, applied mathematics and computation 305 (2017)221-240
\bibitem{yzhang}Y. zhang, k. Fan, S. Gao, S. Chen, A remark on stationary distributionof a stochastic SIR epidemic model with double saturated rates, applied mathamtics letters 76 (2018) 46-52
\bibitem{cooke}Cooke, Kenneth L. Stability analysis for a vector disease model. Rocky Mountain Journal of Mathematics 9 (1979), no. 1, 31-42
\bibitem{baretta-takeuchi1}Y. Takeuchi, W. Ma and E. Beretta, Global asymptotic properties of a delay SIR epidemic model with finite incubation times, Nonlinear Anal. 42 (2000), 931-947.
\bibitem{Baretta-kolmanovskii}E. Beretta, V. Kolmanovskii, L. Shaikhet, Stability of epidemic model with time delay influenced by stochastic perturbations, Mathematics and Computers in Simulation 45 (1998) 269-277
\bibitem{capasso}  Vincenzo Capasso, Mathematical Structures of Epidemic Systems, Lecture Notes in Biomathematics,Volume 97 1993
\bibitem{liu}Liu WM, Hethcote HW, Levin SA. Dynamical behavior of epidemiological models with nonlinear incidence rates. J.Math Biol (1987); 25:359
\bibitem{capasso-serio}Capasso V, Serio G.A. A generalization of the Kermack-Mckendrick deterministic epidemic model. Math Biosc 1978; 42:43
\bibitem{xiao}D. Xiao, S. Ruan, Global analysis of an epidemic model with nonmonotone incidence rate. Math Biosci. 208(2):(2007) 419-29
\bibitem{huo} 	Huo, Hai-Feng; Ma, Zhan-Ping, Dynamics of a delayed epidemic model with non-monotonic incidence rate, Communications in Nonlinear Science and Numerical Simulation, Volume 15, Issue 2,(2010) p. 459-468.
    \bibitem{gumel}S.M. Moghadas, A.B. Gumel, Global Statbility of a two-stage epidemic model with generalized nonlinear incidence, Mathematics and computers in simulation 60 (2002), 107-118
\bibitem{kyrychko} Yuliya N. Kyrychko, Konstantin B. Blyussb, Global properties of a delayed SIR model with temporary immunity and nonlinear incidence rate, Nonlinear Analysis: Real World Applications Volume 6, Issue 3, July 2005, Pages 495-507
\bibitem{qun}Qun Liu, Daqing Jiang,  Ningzhong Shi, Tasawar Hayat, Ahmed Alsaedi, Asymptotic behaviors of a stochastic delayed SIR epidemic model with nonlinear incidence, Communications in Nonlinear Science and Numerical Simulation Volume 40, November 2016, Pages 89-99.
    \bibitem{qunliu}Q. Liu, Q. Chen
Analysis of the deterministic and stochastic SIRS epidemic models with nonlinear incidence Physica A, 428 (2015), pp. 140–153
\bibitem{cooke-driessche} Cooke KL, van den Driessche P., Analysis of an SEIRS epidemic model with two delays, J Math Biol. 1996 Dec;35(2):240-60.
\bibitem{nguyen}Nguyen Huu Du, Nguyen Ngoc Nhu, Permanence and extinction of certain stochastic SIR models perturbed by a complex type of noises, Applied Mathematics Letters 64 (2017) 223-230
\bibitem{joaq}Joaquim P. Mateusa, , César M. Silvab,  Existence of periodic solutions of a periodic SEIRS model with general incidence, Nonlinear Analysis: Real World Applications
Volume 34, April 2017, Pages 379-402
\bibitem{sena}M. De la Sena, S. Alonso-Quesadaa, A. Ibeasb, On the stability of an SEIR epidemic model with distributed time-delay and a general class of feedback vaccination rules, Applied Mathematics and Computation Volume 270, 1 November 2015, Pages 953-976
\bibitem{zhica}Zhichao Jianga, b, Wanbiao Mab, Junjie Wei, Global Hopf bifurcation and permanence of a delayed SEIRS epidemic model, Mathematics and Computers in Simulation
Volume 122, April 2016, Pages 35–54
\bibitem{cesar}Joaquim P. Mateus,  César M. Silva, A non-autonomous SEIRS model with general incidence rate , Applied Mathematics and Computation Volume 247, 15 November 2014, Pages 169-189
\bibitem{sen}M. De la Sen, S. Alonso-Quesada, , A. Ibeas, On the stability of an SEIR epidemic model with distributed time-delay and a general class of feedback vaccination rules, Applied Mathematics and Computation Volume 270, 1 November 2015, Pages 953-976
\bibitem{shuj}Shujing Gao, Zhidong Teng, Dehui Xie, The effects of pulse vaccination on SEIR model with two time delays, Applied Mathematics and Computation
Volume 201, Issues 1-2, 15 July 2008, Pages 282-292
\bibitem{Sampath}B. G. Sampath Aruna Pradeep , Wanbiao Ma, Global Stability Analysis for Vector Transmission Disease Dynamic Model with Non-linear Incidence and Two Time Delays,
Journal of Interdisciplinary Mathematics, Volume 18, 2015 - Issue 4
\bibitem{zheng}Zhenguo Bai,  Yicang Zhou, Global dynamics of an SEIRS epidemic model with periodic vaccination and seasonal contact rate, Nonlinear Analysis: Real World Applications
Volume 13, Issue 3, June 2012, Pages 1060-1068
\bibitem{yanli}Yanli Zhou, Weiguo Zhang, Sanling Yuan, Hongxiao Hu, Persistence And Extinction In Stochastic Sirs
Models With General Nonlinear Incidence Rate, Electronic Journal of Differential Equations, Vol. 2014 (2014), No. 42, pp. 1-17.
\bibitem{eric}Eric Avila, ValesBruno Buonomo, Analysis of a mosquito-borne disease transmission model with vector stages and nonlinear forces of infection, Ricerche di Matematica
November 2015, Volume 64, Issue 2, pp 377-390
\bibitem{sya}  S. Syafruddin, Mohd Salmi Md. Noorani, Lyapunov function of SIR and SEIR model for transmission of dengue fever disease, International Journal of Simulation and Process Modelling (IJSPM), Vol. 8, No. 2-3, 2013
    \bibitem{pang}Liuyong Pang, Shigui Ruan, Sanhong Liu , Zhong Zhao , Xinan Zhang, Transmission dynamics and optimal control of measles
epidemics, Applied Mathematics and Computation 256 (2015) 131–147
\bibitem{zhuhu}Ling ZhuHongxiao Hu, A stochastic SIR epidemic model with density dependent birth rate, Advances in Difference Equations December 2015, 2015:330
\bibitem{ladde}G.S. Ladde, V. Lakshmikantham, Random Differential Inequalities, Academic press, New
York, 1980
\bibitem{ross} R. Ross, The Prevention of Malaria, John Murray, London, 1911.
\bibitem{macdonald} Macdonald, G., The analysis of infection rates in diseases in which superinfection occurs. Trop. Dis. Bull. 47, (1950) 907-915
\bibitem{ngwa-shu} G. A. Ngwa, W. Shu, A mathematical model for endemic malaria with variable human and mosquito population, Math. computer. model. 32, (2000)747-763
\bibitem{hyun}Hyun, M.Y., Malaria transmission model for different levels of acquired immunity and temperature dependent parameters (vector). Rev. Saude Publica 2000, 34 (3), 223–231
\bibitem{may}Anderson, R.M., May, R.M., Infectious Diseases of Humans: Dynamics and Control. Oxford University Press, Oxford, 1991.
\bibitem{kazeem} G.A. Ngwa, A.M. Niger, A.B. Gumel, Mathematical assessment of the role of non-linear birth and maturation delay in the population dynamics of the malaria vector, Appl. Math. Comput. 217 (2010) 3286.
\bibitem{gungala}Ngonghala CN, Ngwa GA, Teboh-Ewungkem MI, Periodic oscillations and backward bifurcation in a model for the dynamics of malaria transmission. Math Biosci, 240(1):(2012)45-62
\bibitem{anita}N. Chitnis, J.M. Hyman, J.M. Cushing, Determining important parameters in the spread of malaria through the sensitivity analysis of a mathematical model, Bull. Math. Biol. 70 (2008) 1272.
\bibitem{tabo} M.I. Teboh-Ewungkem, T. Yuster, A within-vector mathematical model of plasmodium falciparum and implications of incomplete fertilization on optimal gametocyte sex ratio, J. Theory Biol. 264 (2010) 273.
\bibitem{Wanduku-2017}D. Wanduku, Complete Global Analysis of a Two-Scale Network SIRS Epidemic Dynamic Model with Distributed Delay and Random Perturbation, Applied Mathematics and Computation Vol. 294 (2017) p. 49 - 76
\bibitem{wanduku-delay}D. Wanduku, G.S. Ladde, Global properties of a two-scale network stochastic delayed human epidemic dynamic model, nonlinear Analysis: Real World Applications 13(2012)794-816
     \bibitem{divine5}D. Wanduku, G.S. Ladde,  The global analysis of a stochastic two-scale Network Human Epidemic Dynamic Model With Varying Immunity Period, Journal of  Applied Mathematics and Physics, 2017, 5,   1150-1173
         \bibitem{wanduku-determ}D. Wanduku and G.S. Ladde Global Stability of Two-Scale Network Human Epidemic Dynamic
Model, Neural, Parallel, and Scientific Computations 19 (2011) 65-90
\bibitem{wanduku-fundamental}D. Wanduku, G.S. Ladde , Fundamental Properties of a Two-scale Network stochastic human epidemic Dynamic model, Neural, Parallel, and Scientific Computations 19(2011) 229-270
    \bibitem{mao} M. Xuerong, Stochastic differential equations and applications Horwood Publishing Ltd.(2008), 2nd ed.
    \bibitem{ladde} G.S., Ladde, Cellular Systems-II. Stability of Campartmental Systems. Math. Biosci.
30(1976), 1-21
\bibitem{malaria}James M. Crutcher, Stephen L. Hoffman, Malaria, Chapter 83-malaria, Medical Microbiology, 4th edition, Galveston (TX): University of Texas Medical Branch at Galveston; 1996.
\bibitem{WHO}http://www.who.int/denguecontrol/human/en/
\bibitem{CDC}https://www.cdc.gov/malaria/about/disease.html
\bibitem{lars}L. Hviid, Naturally acquired immunity to Plasmodium falciparum malaria ,Acta Tropica 95(3): October 2005, 270-5
\bibitem{denise}Denise L. Doolan, Carlota Dobano,J. Kevin Baird, Acquired Immunity to Malaria, clinical microbiology reviews,Vol. 22, No. 1,  Jan. 2009, p. 13–36
\bibitem{yakui}Yakui Xue And Xiafeng Duan, Dynamic Analysis Of An Sir Epidemic Model With Nonlinear Incidence Rate And Double Delays, International Journal Of
Information And Systems SciencesVolume 7, Number 1, (2011) Pages 92–102
\bibitem{muroya}Yoshiaki Muroya , Yoichi Enatsu, Yukihiko Nakata,Global stability of a delayed SIRS epidemic model with a non-monotonic incidence rate, Journal of Mathematical Analysis and Applications Volume 377, Issue 1, 1 May 2011, Pages 1–14
    \bibitem{hethcote}W.M. Liu, H.W. Hethcote, S.A. Levin, Dynamical behavior of epidemiological models with nonlinear incidence rates,
J. Math. Biol. 25 (4) (1987) 359–380
\bibitem{koro}A. Korobeinikov, P.K. Maini, A Lyapunov function and global properties for SIR and SEIR epidemiological models with
nonlinear incidence, Math. Biosci. Eng. 1 (1) (2004) 57–60.
\bibitem{chen-biodyn} Chen, L, Chen, J., Nonlinear Biologiical Dynamical System, beijing, Science Press, 1993.
\bibitem{kuang} Y. Kuang, delay Differential Equations with Applications in population Dynamics, Academic Press, boston. 1993
\bibitem{divine-proceeding1}D. Wanduku, G.S. Ladde, Global stability of a two-scale network SIR delayed epidemic dynamic model
Proceedings of Dynamic Systems and Applications 6 (2012) 437–441
\bibitem{divine-proceeding2}D. Wanduku, Two-Scale Network Epidemic Dynamic  Model for Vector Borne Diseases, Proceedings of Dynamic Systems and Applications 6 (2016) 228–232
\end{thebibliography}
\end{document}